\documentclass[twocolumn,tightenlines,aps,preprintnumbers,superscriptaddress,showpacs,floatfix]{revtex4}
\usepackage{epsf}
\usepackage{amssymb}
\usepackage{graphicx}
\usepackage{amsmath,amssymb,textcomp,array}

\begin{document}

\preprint{LA-UR-09-05888}

\title{Nonparametric Reconstruction of the Dark Energy Equation of State}
\author{Tracy Holsclaw}
\affiliation{Department of Applied Mathematics and Statistics, 
 University of California, Santa Cruz, CA 95064}
\author{Ujjaini Alam}
\affiliation{ISR-1, MS D466, Los Alamos National Laboratory, Los
Alamos, NM 87545}
\author{Bruno Sans\'o}
\affiliation{Department of Applied Mathematics and Statistics, 
 University of California, Santa Cruz, CA 95064}
\author{Herbert Lee}
\affiliation{Department of Applied Mathematics and Statistics, 
 University of California, Santa Cruz, CA 95064}
\author{Katrin Heitmann}
\affiliation{ISR-1, MS D466, Los Alamos National Laboratory, Los
Alamos, NM 87545}
\author{Salman Habib}
\affiliation{T-2, MS B285, Los Alamos National Laboratory, Los
Alamos, NM 87545}
\author{David Higdon}
\affiliation{CCS-6, MS F600, Los
Alamos National Laboratory, Los Alamos, NM 87545}

\date{\today}

\begin{abstract}
 
  A basic aim of ongoing and upcoming cosmological surveys is to
  unravel the mystery of dark energy. In the absence of a compelling
  theory to test, a natural approach is to better characterize the
  properties of dark energy in search of clues that can lead to a
  more fundamental understanding. One way to view this
  characterization is the improved determination of the
  redshift-dependence of the dark energy equation of state parameter,
  $w(z)$. To do this requires a robust and bias-free method for
  reconstructing $w(z)$ from data that does not rely on restrictive
  expansion schemes or assumed functional forms for $w(z)$. We present
  a new nonparametric reconstruction method that solves for $w(z)$ as
  a statistical inverse problem, based on a Gaussian Process
  representation. This method reliably captures nontrivial behavior of
  $w(z)$ and provides controlled error bounds. We demonstrate the
  power of the method on different sets of simulated supernova data;
  the approach can be easily extended to include diverse cosmological
  probes.
 
 \end{abstract}


\maketitle

\section{Introduction}

The discovery of the accelerated expansion of the
Universe~\cite{riess,perlmutter} poses perhaps the greatest puzzle in
fundamental physics today. A solution of this problem will profoundly
impact cosmology and could also provide key insights in reconciling
gravity with quantum theory. Driven by these motivations, the
fundamental aim of ground and space based missions such as the Baryon
Oscillation Spectroscopic Survey (BOSS)~\cite{boss}, the Dark Energy
Survey~\cite{des}, the Joint Dark Energy Mission (JDEM)~\cite{jdem},
the Large Synoptic Survey Telescope (LSST)~\cite{lsst} -- to name just
a few -- is to unravel the secret of cosmic acceleration. In search of
the underlying explanation, theoretical approaches fall into two main
categories: (i) dark energy -- invoking a new cosmic ingredient, the
simplest being a cosmological constant, and (ii) modified gravity --
invoking new dynamics of space-time (for a recent review, see
Ref.~\cite{frieman08}). In this paper we consider only the dark energy
alternative, and, for the moment, ignore possible modifications of
general relativity.

A fundamental difficulty in dark energy investigations is the absence
of any single {\em compelling} theory to test against observations.
Consequently, much of the work in this area has followed the
approach to parameterize dark energy by
its equation of state $w=p/\rho$ (where $p$ is the pressure, and
$\rho$ the density), see, e.g., Ref.~\cite{turnwhite97}; dynamical models of dark energy such as
quintessence fields lead to a time-varying equation of
state~\cite{quint}. Data analysis efforts therefore focus on
characterizing this time-dependence. Current observations are
consistent with the existence of a cosmological constant, $\Lambda$,
($w=-1$), at the 10\% level, the time-variation being unconstrained
(for recent constraints on $w$, see e.g. Refs.~\cite{hicken,amanullah}). The
implied value of $\Lambda$ is, however, in utter disagreement with
simple theoretical estimates of the vacuum energy, being too small by
a factor $>10^{60}$. It is therefore an {\em ad hoc} addition with no
hint of a possible origin, hence the focus on dynamical explanations,
e.g., field theory models or modified gravity. Although a detection of
any time or, equivalently, redshift-dependence in $w(z)$ would
immediately rule out a cosmological constant, such observational
imprints must necessarily be subtle, otherwise they would have been
discovered already. This is the motivation behind constructing a
robust framework with controlled error bounds that allows a reliable
extraction of $w(z)$ from diverse datasets.

Shortly after the discovery of the accelerated expansion, it was
pointed out that a reconstruction program (an inverse analysis of
data) for dark energy working directly with observational supernova
data is computationally possible (see, e.g.,
Refs.~\cite{huterer98,starobinsky98} for early approaches). Soon a
large number of papers followed, suggesting many different ways of
reconstructing diverse properties of dark energy, e.g.
Refs.~\cite{daly03,wang03,huterer04,gerke04,shaf06,zunckel,hojjati09};
a review on dark energy reconstruction methods including a
comprehensive list of references is given in Ref.~\cite{sahni06}.
Broadly speaking, reconstruction techniques fall into two classes, the
first being those based on parameterized forms for $w(z)$ such as
$w=const.$, $w=w_0+w' z$~\cite{coorhut99,maor,weller} or
$w=w_0-w_az/(1+z)$~\cite{chev01,linder03}. These possess the virtue of
simplicity but can have serious shortcomings due to lack of generality
and error control (specifically issues of bias, see, e.g.,
\cite{simpson06}), especially as one goes to higher redshifts. 

The second class consists of nonparametric methods that aim to solve
the inverse problem of determining the actual function $w(z)$ given
observational data, rather than just the parameters specifying some
assumed form of $w(z)$. The hope is to avoid the possible biasing of
results due to specific assumptions regarding the functional form of
$w(z)$, which may turn out to be incorrect. The difficulty with direct
reconstruction methods as applied to supernova data is that extracting
the desired information formally involves taking a second derivative
of the -- unavoidably noisy -- luminosity distance-redshift relation,
and the robustness and error control of the resulting reconstruction
can therefore be suspect.

A separate alternative to the direct reconstruction approach for $w$
from the data, is to falsify classes of dark energy models. For
example, in Ref.~\cite{genovese} different general forms for $w$ are
considered that capture different dynamical dark energy models. A
hypothesis test is then carried out for these models to determine how
likely they are given current data. In the best case scenario, entire
classes of models can be excluded in this way. In
Ref.~\cite{mortonson} classes of dark energy models are falsified by
carrying out a combined analysis of the growth of structure and the
expansion history of the Universe from cosmic microwave background
(CMB) and supernova data. This approach takes advantage of the fact
that a viable dark energy model must be consistent with measurements
of both of these relatively orthogonal probes of dark energy. As
pointed out in Ref.~\cite{mortonson} the falsification of the smooth
dark energy class would be very interesting, and a different paradigm
for explaining the accelerated expansion such as a modification of
gravity on very large scales would be required. Hypothesis testing
therefore provides an interesting alternative to the direct
reconstruction approach. In fact, in order to convincingly exclude a
cosmological constant from future measurements, both approaches should
be employed, with the aim of arriving at a consistent conclusion.

Given finite data sets, there are -- broadly speaking -- two ways in
which one can go wrong in the reconstruction task, (i) errors due to the assumption of the wrong
shape of $w(z)$, as discussed above and (ii) errors due to the complex
nature of the high-dimensional space within which the inverse problem
is being attempted, in particular, problems due to the existence of
degeneracy directions. In this paper, our aim is to address the first
of these problems, i.e., to develop a technique that is sufficiently
flexible, yet not dangerously susceptible to new error sources as a
result of the extra degrees of freedom. The second aspect of the
inverse problem, the difficulty of dealing with degeneracy directions
(as seen in the examples below), is not directly addressed here. This
issue requires sensitivity analyses and a formalism for incorporating
multiple data sources and will be treated elsewhere~\cite{holsclaw}.

In the current paper, we propose a new, nonparametric reconstruction
approach that solves the associated statistical inverse problem by
sampling the posterior distribution using Markov Chain Monte Carlo
(MCMC) methods, while representing $w(z)$ by a Gaussian Process (GP).
Traditionally, GP modeling is a nonparametric regression approach
based on a generalization of the Gaussian probability distribution. It
extends the notion of a Gaussian distribution over scalar or vector
random variables to function spaces. While a Gaussian distribution is
specified by a scalar mean $\mu$ or a mean vector and a covariance
matrix, the GP is specified by a mean function and a covariance
function~\cite{Banerjee,gprefs}. GPs have been successfully applied in
astrophysics and cosmology to construct prediction schemes for the
dark matter power spectrum and the CMB temperature angular power
spectrum~\cite{heitmann06,habib07,heitmann09,lawrence09}, to model
asteroseismic data~\cite{brewer09}, and to derive photometric redshift
predictions~\cite{way09,bonfield09}. Here we will use the GP modeling
approach -- in concert with MCMC -- to reconstruct $w(z)$ from
supernova observations, and not as a data interpolation or regression
tool applied directly to observational or computed data, as is most
often the case.

As of now, supernova datasets hold by far the most information about
possible time dependence of $w(z)$, though baryon acoustic oscillation
(BAO) and CMB measurements contain complementary information (see,
e.g., Ref.~\cite{alam07} for a recent combined reconstruction
analysis). Although the GP approach can be easily extended to
accommodate more than one observational probe, for clarity we will
restrict ourselves in this paper to supernova measurements only. A
more inclusive methodology will be presented in future
work~\cite{holsclaw}.

Since current data quality does not allow placement of strong
constraints on a possible redshift dependence of $w(z)$, we create a
set of simulated data of JDEM-like quality to demonstrate and test our
new method. We consider three models, one with a constant equation of
state and two with varying $w(z)$. Our new approach will be shown to
perform well in capturing nontrivial deviations from a constant
equation of state and in providing reliable error bounds.

The paper is organized as follows. In Section~\ref {sec:sne} we
provide a brief overview of how supernova data are used to constrain
the equation of state of dark energy. We describe the simulated data
sets and their error properties in Section~\ref{sec:data}. In
Section~\ref{sec:recon} we introduce different reconstruction methods
and present our approach in the same section, contrasting our
nonparametric method with results obtained using the popular
parametric forms of Refs.~\cite{chev01,linder03}. We conclude in
Section~\ref{sec:concl}. Details of the implementation of the GP-based
MCMC algorithm are given in an Appendix.

\section{Measuring Expansion History with Supernovae} 
\label{sec:sne}

Type Ia supernova measurements are currently the single best source of
information regarding possible deviations of $w(z)$ from a constant
value. The luminosity distance $d_L$ as measured by supernovae is
directly connected to the expansion history of the Universe described
by the Hubble parameter $H(z)$. For a spatially flat Universe, the
relation is given by
\begin{equation}\label{dl}
d_L(z)=(1+z)\frac{c}{H_0}\int_0^z\frac{ds}{h(s)},
\end{equation}
where $c$ is the speed of light, $H_0$, the current value of the
Hubble parameter ($H(z)=\dot{a}/a$, where $a$ is the scale factor and
the overdot represents a derivative with respect to cosmic time), and
$h(z)=H(z)/H_0$. The assumption of spatial flatness is in effect an
``inflation prior'', although there do exist strong constraints on
spatial flatness when CMB and BAO observations are combined (see,
e.g., Ref.~\cite{flat}). In principle, we can relax this assumption,
but enforce it here to simplify the analysis.

Instead of $d_L(z)$, supernova data are usually specified in terms of the
distance modulus $\mu$ as a function of  redshift. The relation between
$\mu$ and the luminosity distance is 
\begin{eqnarray}\label{muh}
\mu_B(z)&=&m_B-M_B=5\log_{10} \left(\frac{d_L (z)}{1~{\rm Mpc}}\right)+25\\
&=&5\log_{10} \left[(1+z)
  c\int_0^z\frac{ds}{h(s)}\right]-5\log_{10}(H_0)+25,\nonumber
\end{eqnarray} 
where we used Eqn.~(\ref{dl}). $M_B$ is the absolute magnitude of the
object and $m_B$ the ($B$-band) apparent magnitude.
Writing out the expression for the Hubble parameter $h(z)$ in
Eqn.~(\ref{muh}) explicitly in terms of a general dark energy equation
of state for a spatially flat FRW Universe leads to the relation
\begin{eqnarray}
&&\mu_B(z)=25-5\log_{10}(H_0)\label{mu}\\
&&+5\log_{10}\left\{(1+z){c}
\int^z_0 ds\left[\Omega_m(1+s)^{3}\right.\right.\nonumber\\
&&+\left.\left.(1-\Omega_m)(1+s)^3
    \exp\left(3\int_0^{s}\frac{w(u)}{1+u}du\right)\right]^{-1/2}\right\}.
\nonumber
\end{eqnarray}
Note that $H_0$ cannot be determined from supernova measurements in
the absence of an independent distance measurement.  Thus $H_0$
can be treated as unknown and absorbed in a re-definition of
the absolute magnitude:
\begin{equation}\label{calM}
{\cal M}_B=M_B-5\log_{10}H_0+25,
\end{equation}
which accounts for the combined uncertainty in the absolute
calibration of the supernova data, as well as in $H_0$. Using this,
the $B$-band magnitude can be expressed as $m_B=5 \log_{10}{D}_L(z) +
{\cal
  M}_B$ where ${D}_L(z) = H_0 d_L (z)$ is the
``Hubble-constant-free'' luminosity distance. The measurement of $\mu_B$
is only a relative measurement and ${\cal M}_B$ allows for an additive
uncertainty which can be left as a nuisance parameter. To simplify our
notation, we absorb $5\log_{10}(H_0)-25$ into our definition of
the distance modulus, leading to:
\begin{equation}\label{tildemu}
\tilde \mu_B=\mu_B+5\log_{10}(H_0)-25=5\log_{10}[D_L(z)].
\end{equation}
With this definition of the distance modulus we have calibrated the
overall off-set of the data to be zero. To account for uncertainties in
the calibration, we introduce a shift parameter $\Delta_\mu$ with a
broad uniform prior.

Given a set of observations for $\mu_B(z)$ with associated errors, the
task at hand is to solve the statistical inverse problem, i.e., to
extract the corresponding $w(z)$ by inverting the stochastic version
of Eqn.~(\ref{mu}), i.e., inverting a nonlinear smoothing operator,
which can be viewed formally as requiring taking two derivatives of
the (noisy) data, the key difficulty to be overcome in reconstruction.

As previously stated, the present quality of supernova data is not
good enough to determine the equation of state beyond a cosmological
constant (i.e., use of the parameterized form $w=const.$). To do
better than this, both systematic and statistical errors need to be
brought under further control. Such systematic errors can occur due
to, e.g., uncertainties in luminosity corrections and therefore in
distance estimates, or the fitting procedure for the supernova light
curves; for a recent discussion of these issues, see Ref.~\cite{sdss}.
Larger numbers of supernovae, especially at high redshifts, are needed
to get firm constraints on a possible variation in $w$ (see, e.g.,
Refs.~\cite{lh03,genovese}). Future supernova surveys, especially
space-based, hold the promise to remedy this situation. We therefore
explain our method for reconstructing $w(z)$ with simulated data that
mimics the expected quality of future space-based observations. We
turn now to a description of the simulated datasets.

\section{Synthetic datasets}
\label{sec:data}

\begin{figure}[b]
\centerline{
  \includegraphics[width=2.7in]{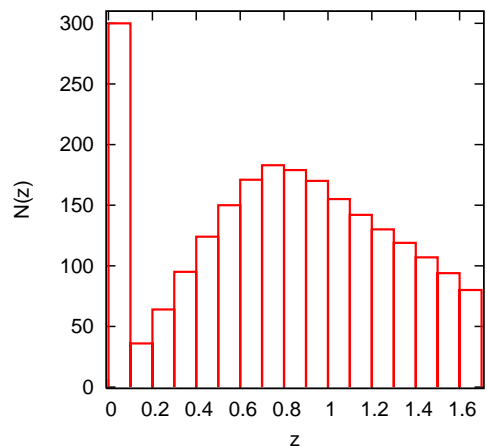}}
\caption{\label{snap}Redshift distribution of supernovae from the
three simulated datasets investigated. In addition to JDEM
measurements of supernovae, we assume a low redshift sample of 300
supernovae for $z\le0.1$.  The bin width is $\Delta z=0.1$.}
\end{figure}

\begin{figure*}[t]	
\centerline{
  \includegraphics[width=2.2in]{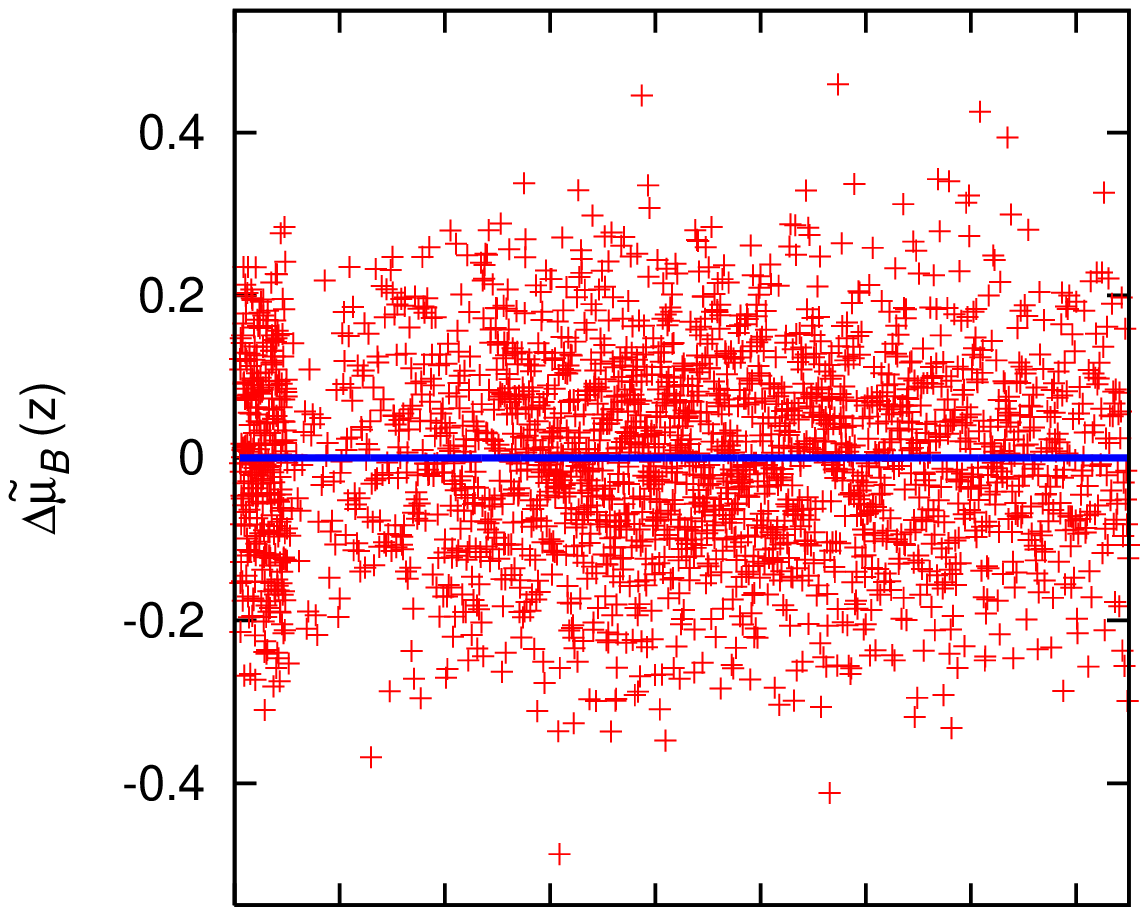}
  \includegraphics[width=2.2in]{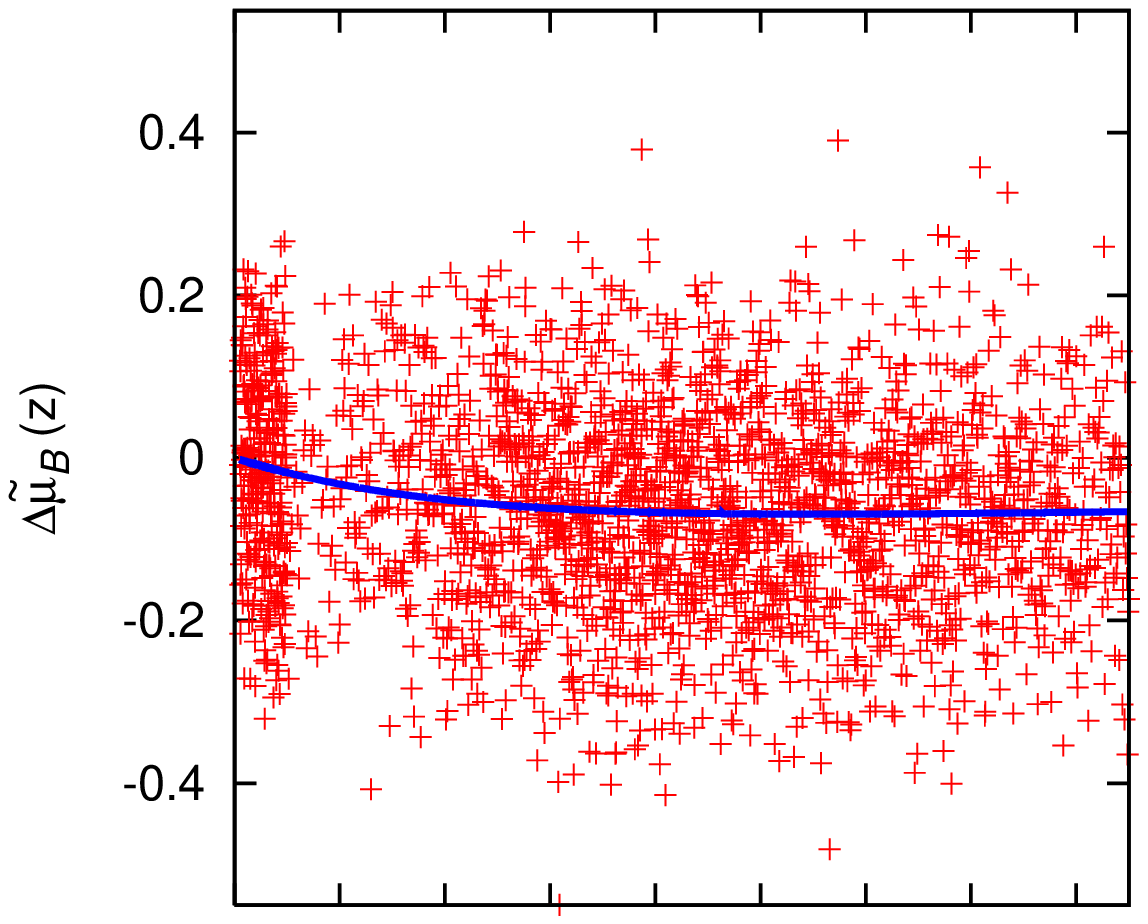}
  \includegraphics[width=2.2in]{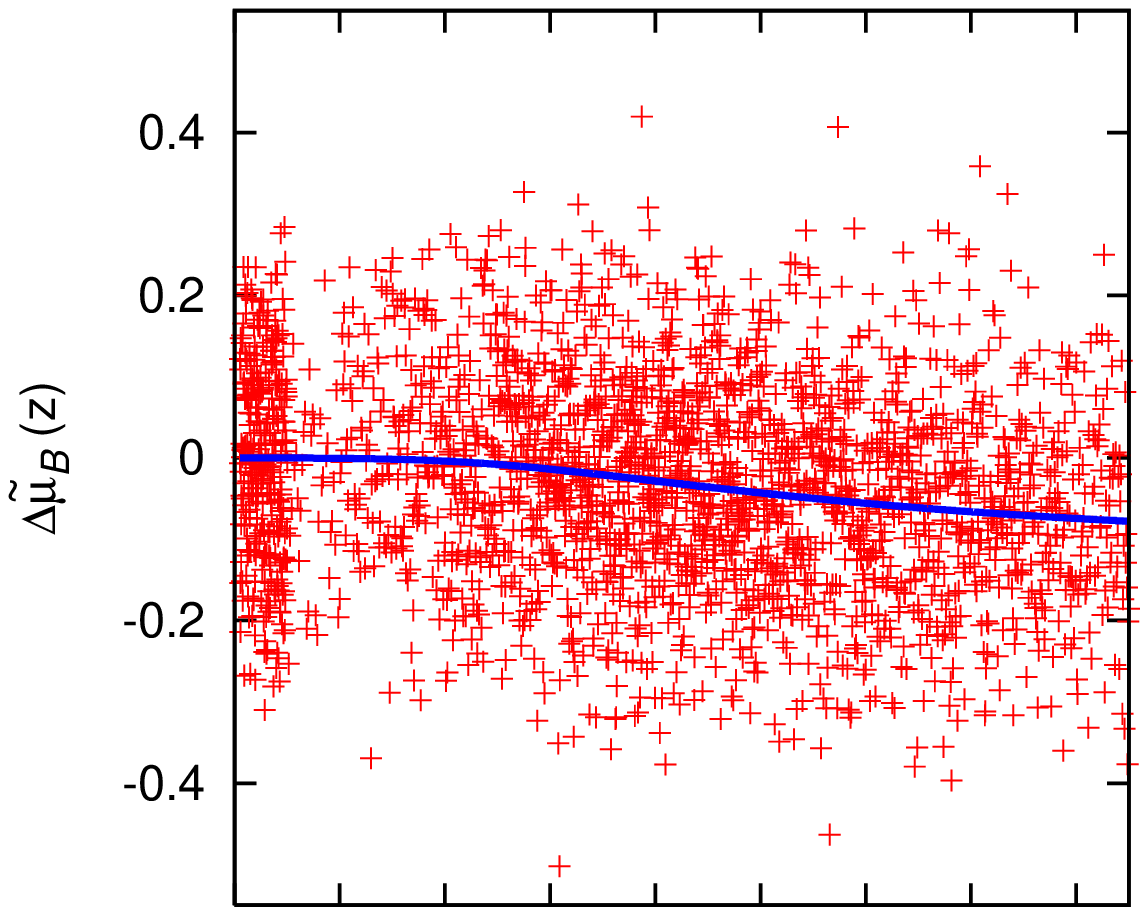}} 

  \vspace{-0.53cm} 

  \centerline{ \includegraphics[width=2.2in]{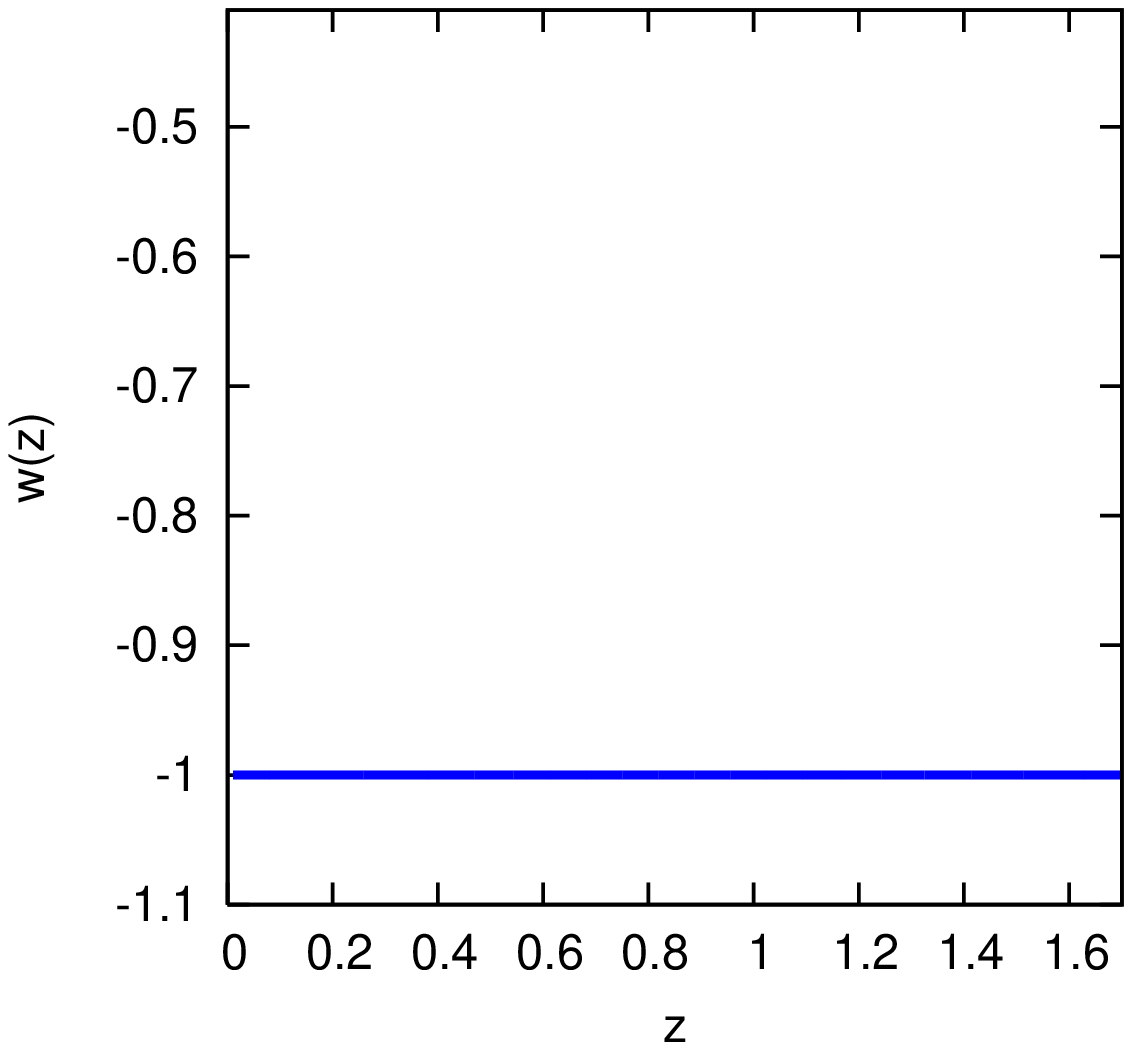}
  \includegraphics[width=2.2in]{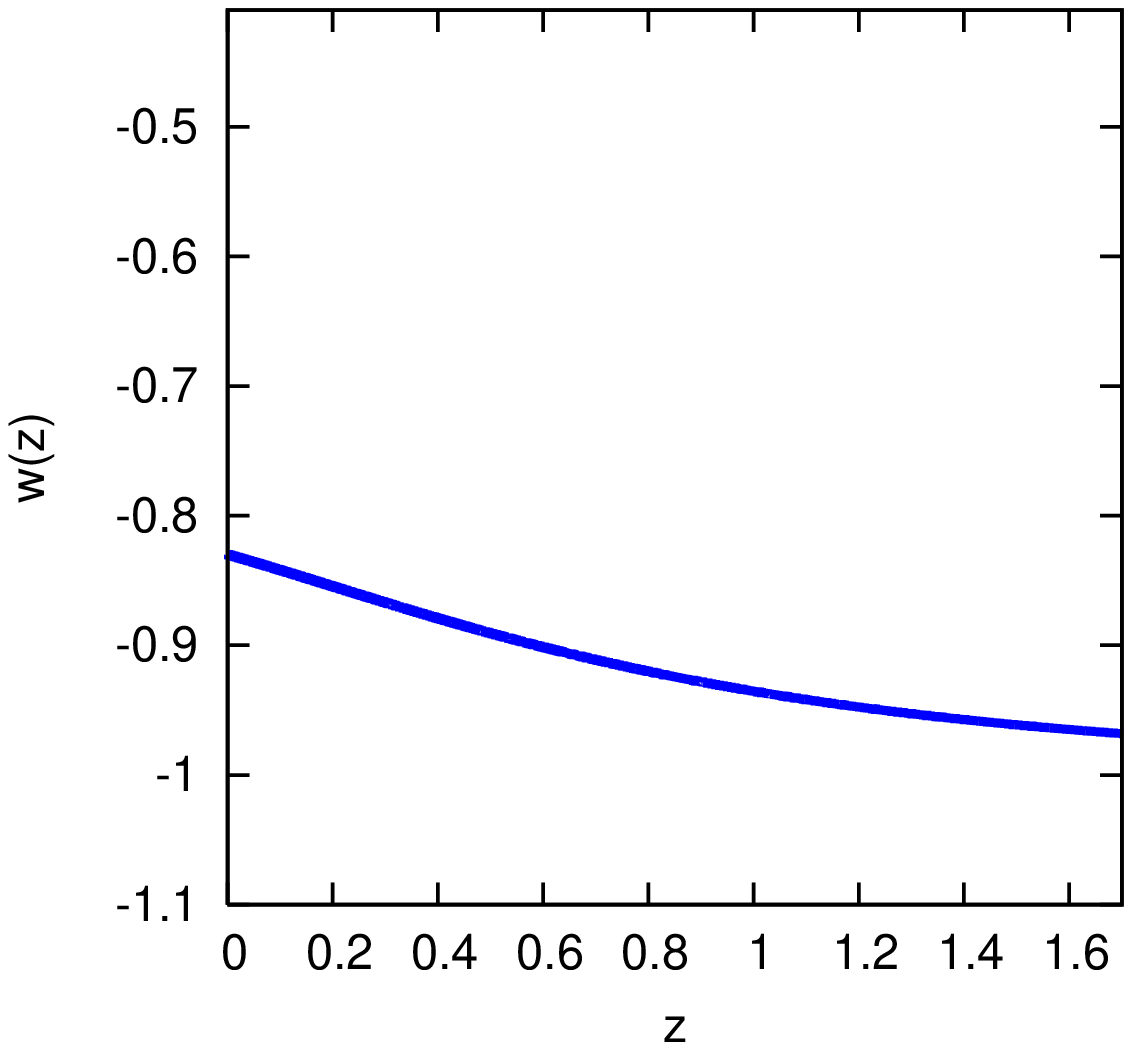}
  \includegraphics[width=2.2in]{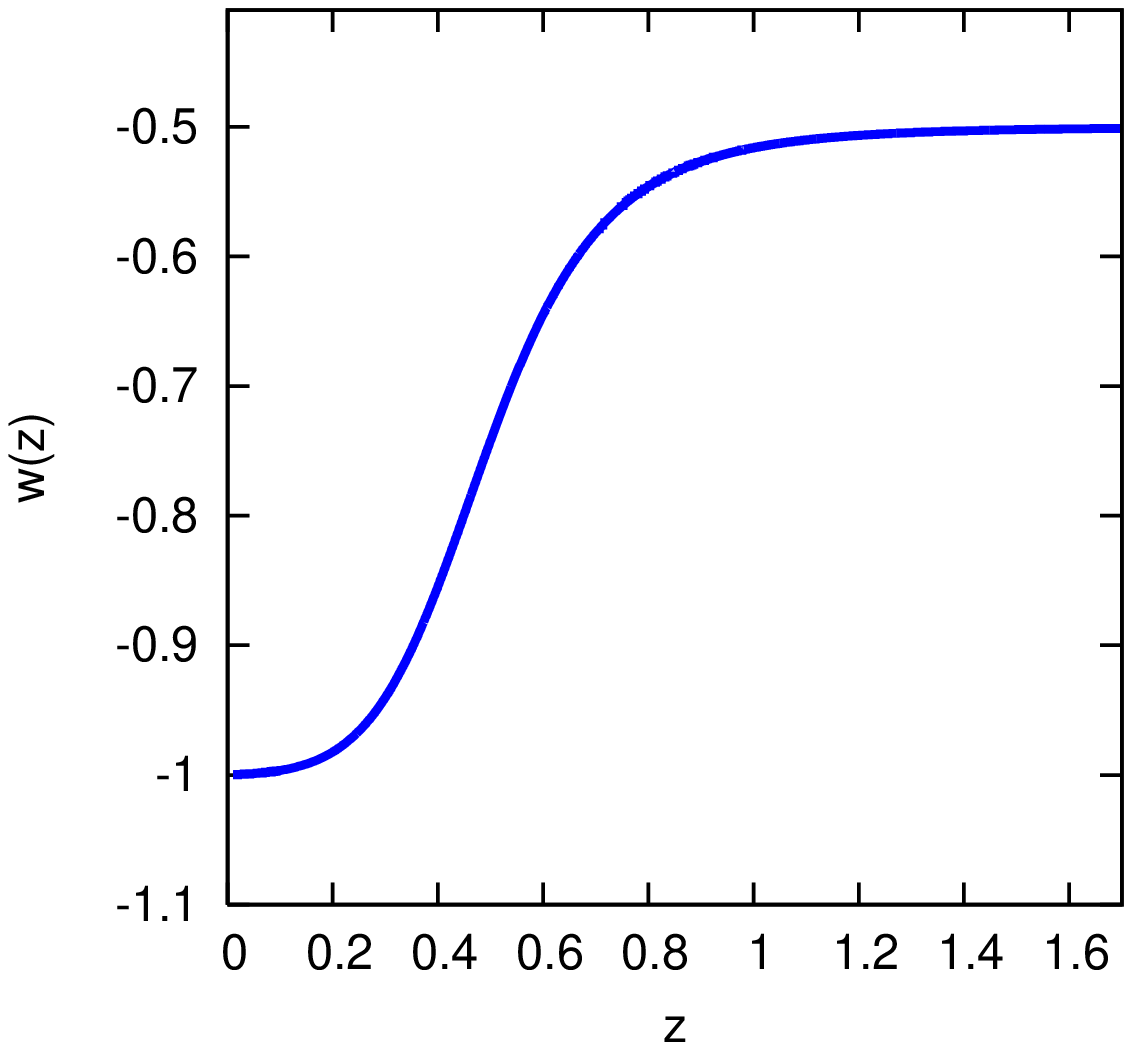}} \caption{\label{data}Three
  simulated datasets. The upper row shows $\Delta\tilde\mu_B$ (the data
  itself with the corresponding value for a $\Lambda$CDM model
  subtracted, red crosses) as a function of redshift, $z$, and the
  exact $\Delta\tilde\mu_B$ for the corresponding model (blue line).
  The lower panels show the behavior of the equation of state $w(z)$
  as a function of redshift. The underlying model for the first
  dataset is a cosmological constant. The second and third datasets
  are based on quintessence models. The third dataset has been chosen
  to test our reconstruction method on a nontrivial equation of
  state.}
\end{figure*}

In this section we introduce three synthetic datasets which we will
use to compare the GP approach to parameterized methods for estimating
$w(z)$. Synthetic datasets have three important attributes: (i) The
underlying ``truth'' is known and one can therefore impose a
quantitative measure on how well each method performs. (ii) The data
quality can be controlled, e.g., we mimic the expected data quality
from future space-based supernova surveys. (iii) Dark energy models
with very different equations of state $w(z)$ can be synthesized.

We assume the measurement of $n\simeq$ 2300 supernovae, distributed
over a redshift range of $0<z<1.7$ with larger concentration of
supernovae in the mid-range redshift bins ($0.4<z<1.1$) and at low
redshift ($z<0.1$). Figure~\ref{snap} shows the detailed distribution
of the supernova data with respect to redshift. To create the
simulated data, we begin with points for $\tilde\mu_B$ shifted
off-center according to some error model for the distance modulus
(Gaussian variance). The distance modulus error can be related to that
in $D_L$ by differentiating Eqn.~(\ref{tildemu}) to yield
$\delta\tilde\mu=(5/\ln 10)(\delta_{D_L}/{D_L})$.
For each supernova,
we provide a measurement for the distance modulus $\tilde\mu_i$ and we
assume a statistical error of $\tau_i = 0.13$, as expected from future
surveys such as JDEM~\cite{snap}. For our purposes here, it is
sufficent to use a simplified error model where the errors are the
same for all supernovae and independent of redshift.  We also do not
explicitly introduce systematic errors.  We represent the measured
points in the following form: 
\begin{equation}
\tilde\mu_i=\alpha(z_i)+\epsilon_i.
\end{equation}
In this notation, the observations $\tilde\mu_i$ follow a normal
distribution with mean $\alpha(z_i)$, the standard deviation being set
by the distribution of the error, $\epsilon_i$, representing a
mean-zero normal distribution with standard deviation, $\tau_i\sigma$.
Here, $\tau_i$ is the observed error and $\sigma$ accounts for a
possible rescaling. In addition, we assume that the errors are
independent. The assumption of normal distributed errors in magnitude
space is consistent with the error distribution of real observations
as quoted in Ref.~\cite{hicken}. For each of the datasets we choose
$\Omega_m=0.27$. The three simulated datasets and corresponding
equations of state are shown in Figure~\ref{data}.

{\bf Dataset 1:} The first dataset is that for a cosmological
constant with a constant equation of 
state, $w=-1$.

{\bf Dataset 2:} The second dataset is based on a quintessence model
with a minimally coupled scalar field. The equation of motion for the
homogeneous mean field is
$\ddot\phi+3H\dot\phi+dV/d\phi=0$. The equation of
state parameter is given by 
\begin{equation}
w=\frac{\dot\phi^2/2-V(\phi)}{\dot\phi^2/2+V(\phi)}.
\end{equation} 
The particular choice of potential used here is
$V(\phi)=V_0\phi^{2}$. This model predicts a relatively
small  variation in the equation of state as a function of $z$ as can
be seen in the middle panel in the lower row in Figure~\ref{data}.

{\bf Dataset 3:} The last dataset is based on a quintessence model
described in Ref.~\cite{coras}. This model has a dark energy
equation of state of the form
\begin{eqnarray}
\label{eq:model3}
&&w(z) =
w_0+(w_m-w_0)\frac{1+\exp(\Delta_t^{-1}(1+z_t)^{-1})}{1-\exp(\Delta_t^{-1})}\\ 
&&\times\left[ 1-
\frac
{\exp(\Delta_t^{-1}) + \exp(\Delta_t^{-1}(1+z_t)^{-1})}
{\exp(\Delta_t^{-1}(1+z)^{-1}) +\exp(\Delta_t^{-1}(1+z_t)^{-1})} 
\right],\nonumber
\end{eqnarray}
with the constants having the values $w_0 = -1.0, ~w_m = -0.5,~z_t =
0.5,~\Delta_t = 0.05$. This model has $w \ge -1$ everywhere, therefore
it can in principle be realized by a quintessence field. The time
variability of the equation of state has an S-shaped form as shown in
the right lower panel in Figure~\ref{data}. The parameter choices for
this model lead to a steeper transition in $w(z)$ from $w=-1$ to
$w=-0.5$ than natural for most quintessence models. Therefore,
compared to dataset 2, this scenario is less realistic. Our choice of
this dataset is dictated by the fact that it cannot be easily fit by
any of the currently used parametric reconstruction methods. (It
represents a general class of models with equations of state that can
exhibit rapid changes.)

\begin{figure}
\centerline{
  \includegraphics[width=2.7in]{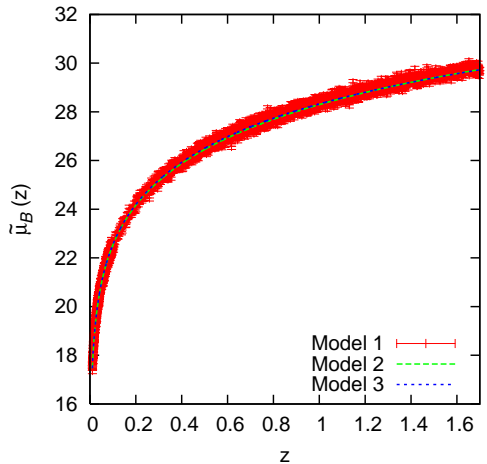}}
\caption{\label{simdata}Simulated data for Model 1 ($w=-1$) with error
  bars in red. The green and blue line show the exact distance modulus $\tilde\mu_B$
  for Model 2 and 3. Note the very small size of the difference.} 
\end{figure}

Figure~\ref{simdata} and the upper panels in Figure~\ref{data} give a
visual impression of the difficulties posed by reconstruction.
Figure~\ref{simdata} shows the simulated data for Model 1 with error
bars and the exact $\tilde\mu_B$ for Models 2 and 3, which are hardly
distinguishable by eye. Figure~\ref{data} shows the differences
$\Delta\tilde\mu_B$ for each dataset with respect to a $\Lambda$CDM
model with $w=-1$; models with nontrivial $w(z)$ show relatively small
deviations from the horizontal line. As we demonstrate below, inverse
modeling using Gaussian processes can successfully discriminate between
these marginal differences and reconstruct the dark energy equation of
state reliably within stated errors.

\section{Reconstruction of the Dark Energy Equation of State}
\label{sec:recon}

As discussed previously, the dark energy equation of state is not
directly measurable from the luminosity distance-redshift relation,
given in Eqn.~(\ref{mu}). The obvious idea of first fitting for
$\mu_B(z)$ and then extracting $w(z)$ by taking two derivatives must
deal with the noise in the data and the filtering required to estimate
the derivatives. Experience with inverse problems has shown that such
approaches can easily yield unsatisfactory results. A detailed
discussion on the shortcomings of this approach can be found in, e.g.,
Ref.~\cite{weller2}.

A simpler alternative is to assume a hopefully well-motivated
parametric form for $w(z)$ and then fit for the parameters (for an
early discussion about the advantages of this approach, see, e.g.,
Ref.~\cite{weller2}). For example, if we assume $w$ to be constant,
the integral over $w(z)$ in Eqn.~(\ref{mu}) can be solved analytically
and the best-fit value for $w$ can then be determined from
measurements of $\mu_B$ via, e.g., maximum likelihood techniques.
Current data are in good agreement with a constant $w$ at the 10\%
level (for a recent analysis see Ref.~\cite{hicken} and references
therein for earlier results). Going beyond this, a weak redshift
dependence of $w(z)$ may be assumed. One way to realize this is a
Taylor expansion of $w(z)$ in its redshift evolution, of the form
$w=w_0+w_az$, as suggested in Refs.~\cite{coorhut99,maor,weller}.
However, this parameterization is not well suited for $z>1$, the
regime that holds the most promise to distinguish different models of
dark energy~\cite{linder03, lh03}. In Ref.~\cite{linder03}, the form
$w=w_0-w_az/(1+z)$ is suggested as a better alternative (also given
previously in Ref.~\cite{chev01}). This parameterization has several
nice features: it is well behaved beyond $z=1$, it has only two
parameters and is therefore relatively easy to constrain, and it
captures the general behavior of different classes of dynamical dark
energy models. The major disadvantage is that the parameterization
will only allow reconstruction of monotonic behaviors of $w(z)$. More
involved parameterizations have been suggested to address this
problem; overviews can be found in Refs.~\cite{sahni06,frieman08}.
Although parameter estimation is technically much easier than
reconstruction, it can have shortcomings due to poor control over
bias~\cite{simpson06}.

Nonparametric reconstruction methods have received less attention, in
part because the current data quality does not fully justify the use
of sophisticated inverse methods. Nevertheless, with future data
quality in mind, nonparametric techniques can be a powerful
alternative for extracting information about $w(z)$. They can capture
more complex behavior in $w(z)$ and -- in principle -- can prevent the
existence of bias due to a restricted parameterization. Early
nonparametric approaches involve a smoothing procedure for either
$d_L$ or related quantities at a characteristic smoothing scale, see,
e.g.~\cite{daly03,shaf06}.

A somewhat intermediate approach is a piecewise constant description
of $w(z)$ (see, e.g., Ref.~\cite{huterer03}) using basis functions
such as top-hat bins or wavelets~\cite{hojjati09}. In the extreme case
of one bin for the whole data range, this method is equivalent to the
$w=const$. parametrization. Determining the optimal number of bins
informed by the data is therefore important though not
straightforward. Too few bins would erase important information, too
many bins would enhance noise to (incorrect) information. In
Ref.~\cite{huterer04}, four redshift bins were used, while
Ref.~\cite{serra09} used five redshift bins over a smaller redshift
range. In order to obtain uncorrelated estimates of the dark energy
parameters in the different bins, a principal component analysis is
carried out first. This method has been used recently by the JDEM
Figure of Merit Science Working Group~\cite{fomswg} to assess the
performance of JDEM with respect to constraining the dark energy
equation of state. In Ref.~\cite{serra09} a combined analysis of
diverse data sets has been performed based on this method and found no
evolution in $w(z)$. In contrast to the piecewise constant description
of the dark energy equation of state, our approach represents $w(z)$
by a continuous Gaussian process, the parameters specifying the
process -- the so-called hyperparameters -- being completely
determined as part of the solution of the inverse problem. It is
important to distinguish the GP hyperparameters from the parameters of
a conventional parametric method. The GP approach is nonparametric,
the hyperparameters specifying aspects of the prior distribution in a
Bayesian approach (such as properties of the allowed classes of
functions). One advantage of this degree of freedom is that one can
explicitly use it to test the sensitivity of the posterior
distribution to assumptions made about the prior, e.g., the order of
differentiability. Here, we make no binning assumptions or assumptions
of the discrete properties of the GPs, favorable when working with a
physical process that is assumed to be continuous in nature.

In this paper we will study the ansatz $w=const.$ and the
parameterization suggested in Refs.~\cite{chev01,linder03} as
reference standards to compare with the GP modeling approach. As a
simplification, in the first step of our analysis, we will assume
knowledge of the value of $\Omega_m$ and assume perfect calibration,
i.e. $\Delta_\mu=0$. In the next step, we will drop these assumptions
and include the parameters as part of the estimation process, as would
be the case in a more realistic scenario (albeit without directly
including non-supernova datasets). To provide a context for the GP
approach we will first present an analysis with parameterized models.

\subsection{Parametric Reconstruction}

\begin{figure*}
\centerline{
 \includegraphics[width=2.2in]{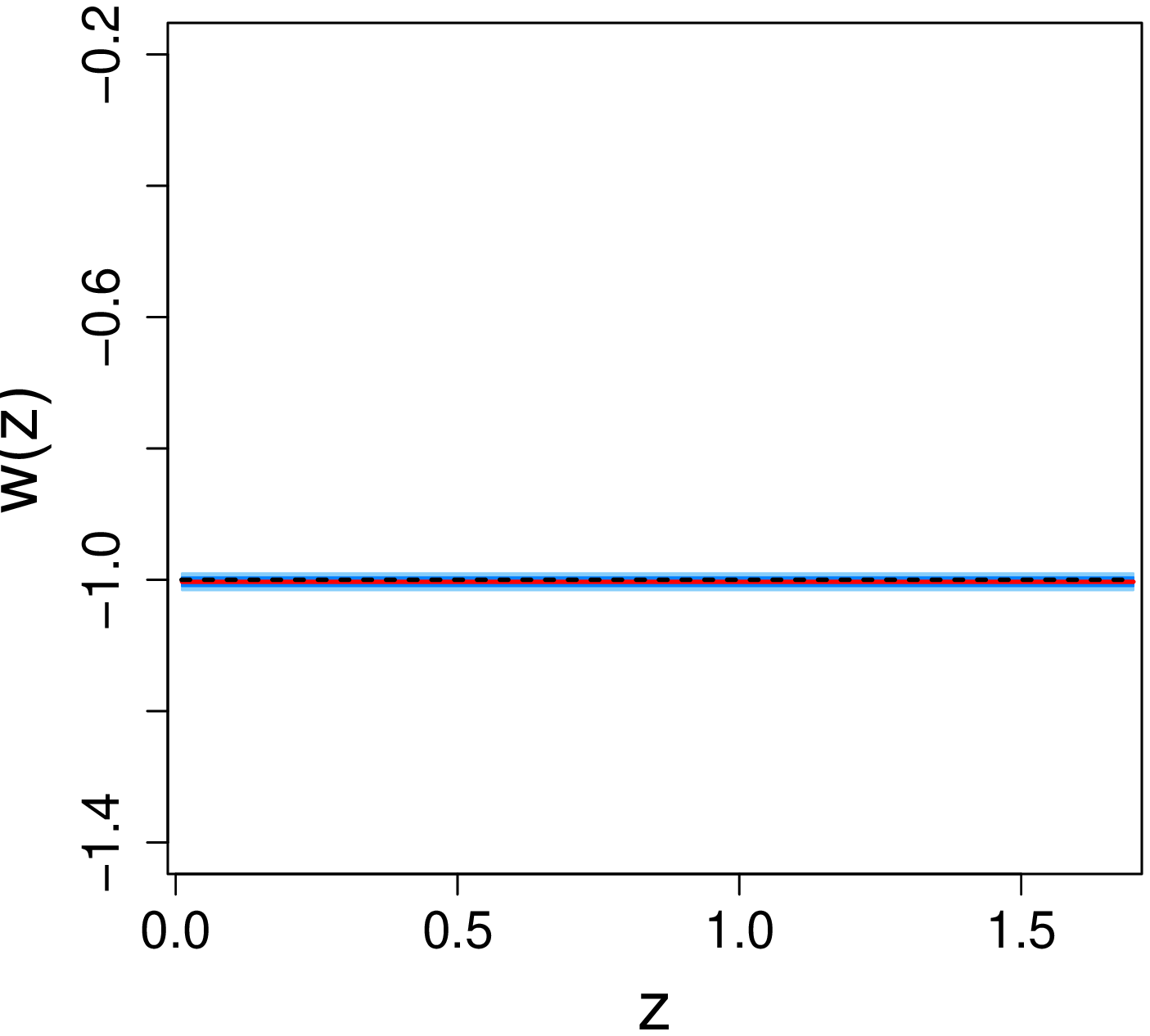}
\includegraphics[width=2.2in]{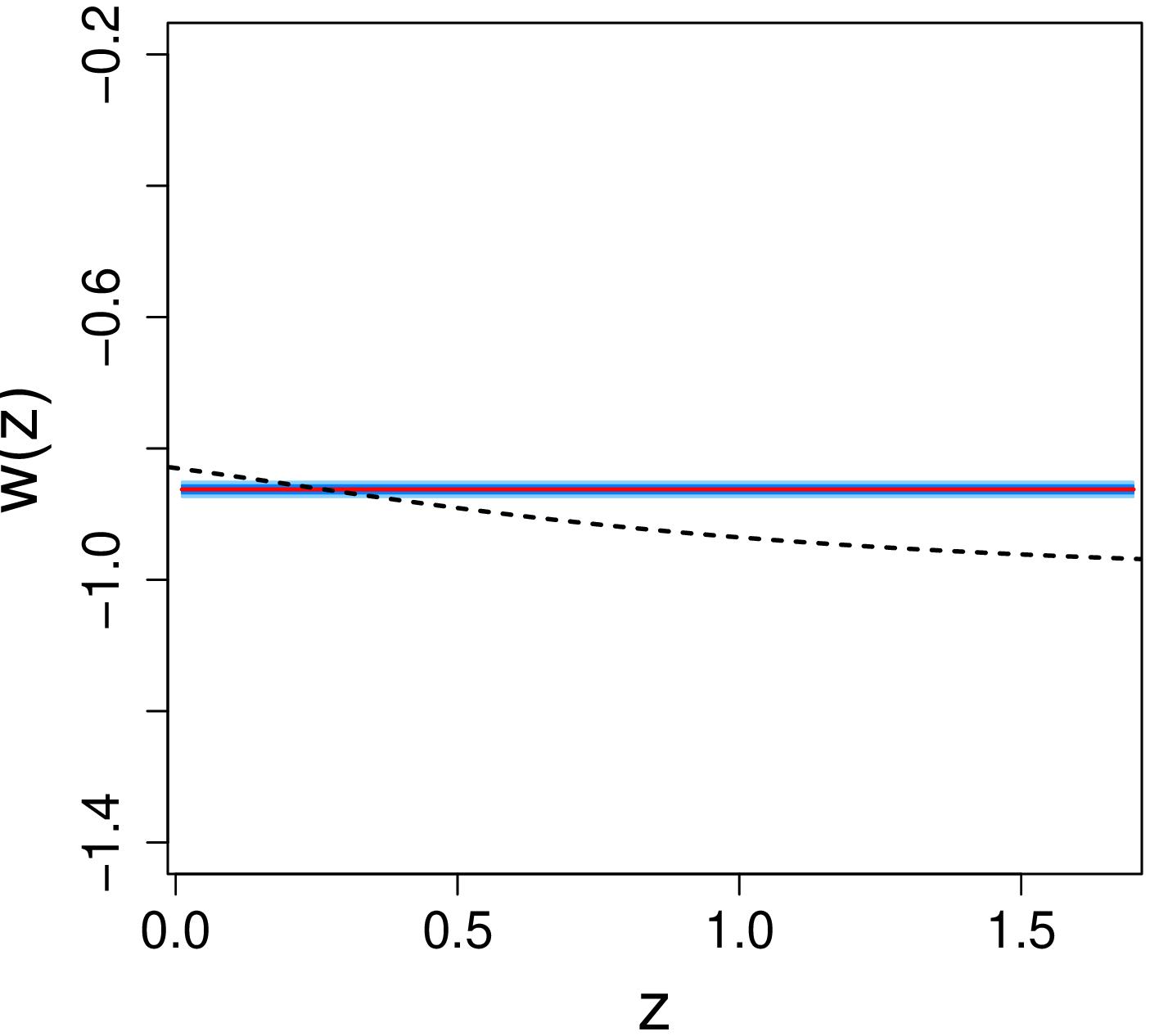}
\includegraphics[width=2.2in]{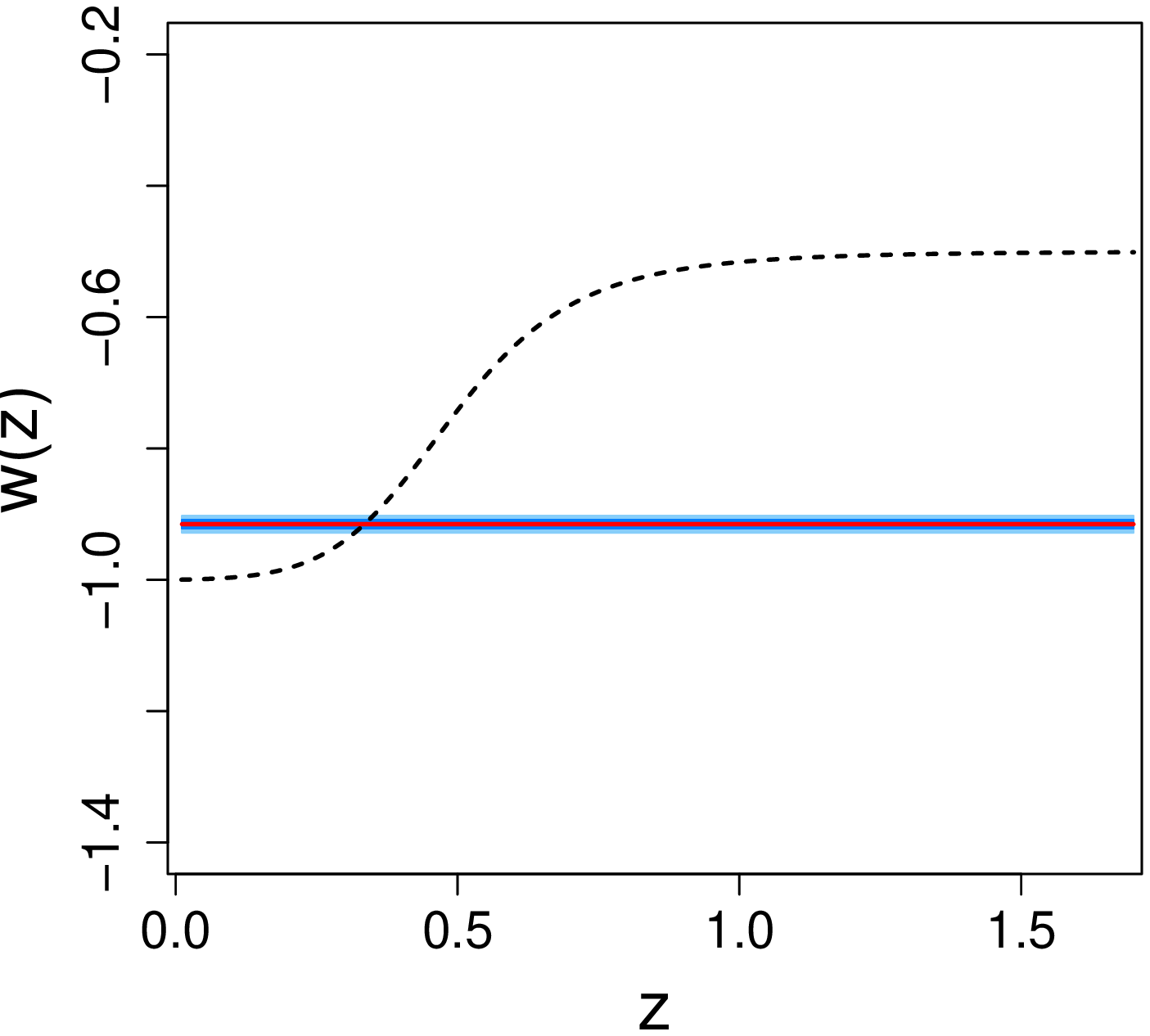}}
\caption{\label{wconst}Reconstruction results for $w$ for datasets
  1-3 (left to right) assuming $w=const.$ and $\Omega_m$ and $\Delta_\mu$
  fixed at their fiducial values. The black dashed curve shows the
  ``truth'' and the red curve, the reconstruction results. The dark
  blue shaded region indicates the 68\% confidence level, while the light
  blue shaded region extends it to 95\%. The assumption $w=const.$ makes
  it impossible to capture the time dependence in datasets 2 and 3.}
\centerline{
\includegraphics[width=2.2in]{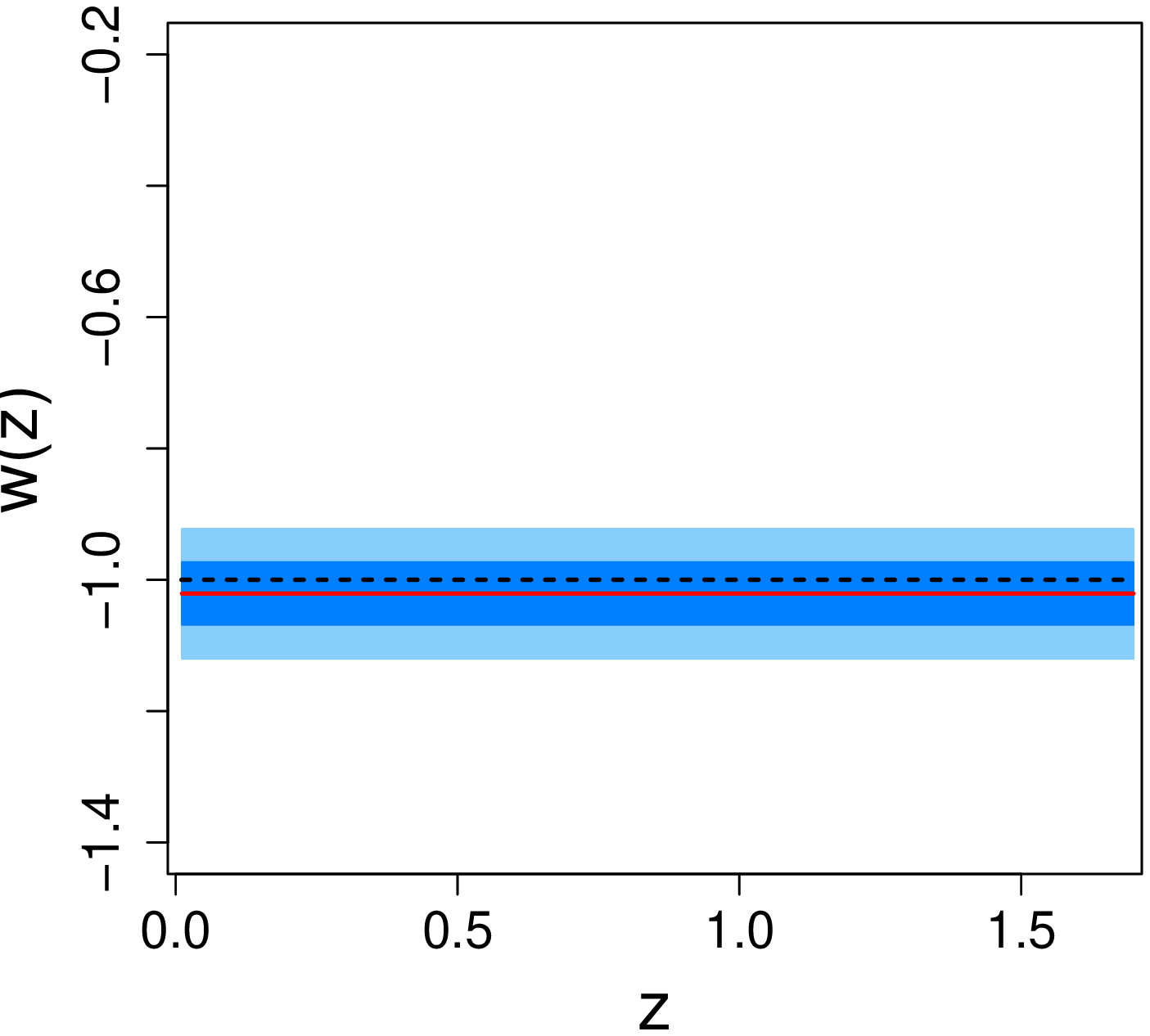}
\includegraphics[width=2.2in]{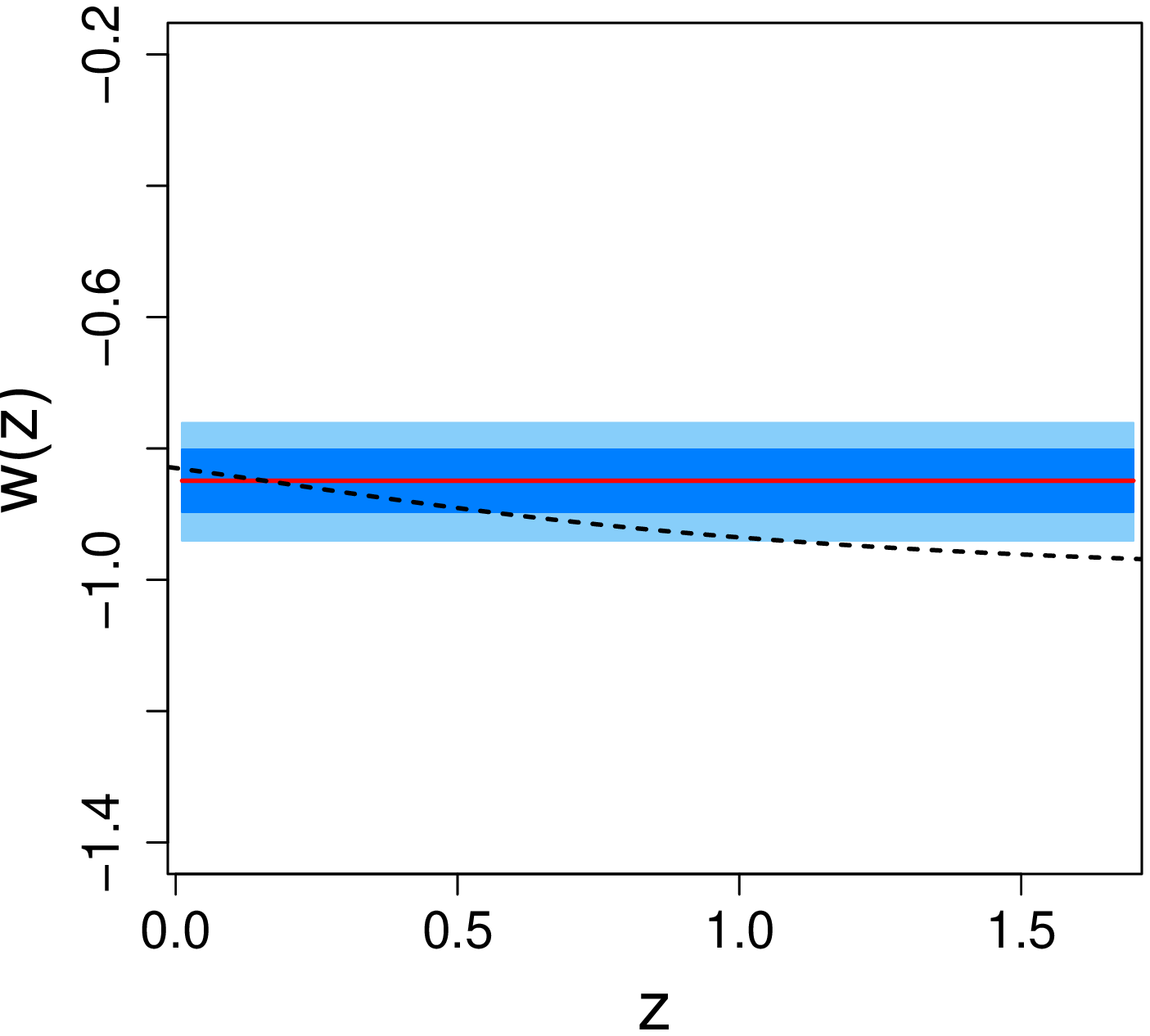}
\includegraphics[width=2.2in]{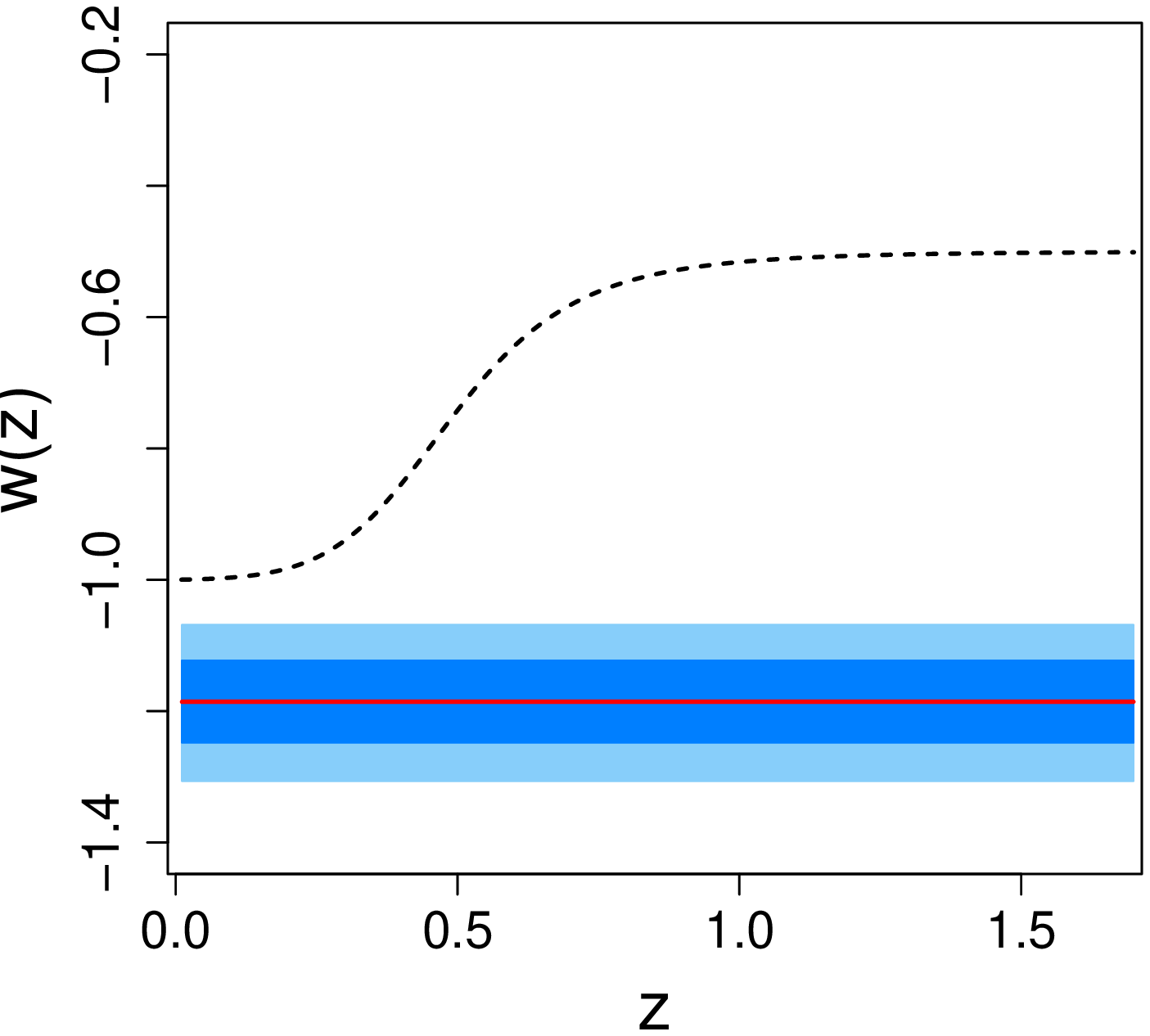}}
\caption{\label{wconst_oh}Results as in Figure~\ref{wconst}, but
  letting $\Omega_m$ and $\Delta_\mu$ vary.  The result for dataset 1 is very
  accurate -- $w_0$ is very close to the true values. The predictions
  for datasets 2 and 3 are poor, not only for $w_0$, but also for the
  incorrect biasing of $\Omega_m$ (see text).}
\end{figure*}

In the study of parametric reconstruction, we follow a Bayesian
analysis approach~\cite{gelman}. We focus the analysis on two of the
previously discussed models: $w=const.=w_0$ and $w(z) = w_0 -
w_az/(1+z)$ and use MCMC algorithms to fit for the model
parameters~\cite{gamerman}, resulting in posterior estimates and
probability intervals for $\Omega_m$, and the parameters that
specify the form of $w(z)$. We have consistent priors in all of our
models (including the GP model described in the next section) so the
results are readily comparable:
\begin{eqnarray}
  \pi(w_0) &\sim& U(-25,1),\\
  \pi(w_a)&\sim& U(-25,25),\\
  \pi(\Omega_{m})& \sim& N(0.27,0.04^2),\label{om}\\
  \pi(\Delta_\mu) &\sim& U(-0.5,0.5),\label{H0}\\
  \pi(\sigma^2) &\propto &\sigma^{-2},
\end{eqnarray}  
and the likelihood
\begin{equation}\label{L}
L(\sigma,\theta)\propto
\left(\frac{1}{\tau_i\sigma}\right)^n
\exp\left(-\frac{1}{2}\sum_{i=1}^n
  \left(\frac{\mu_i-\mu(z_i,\theta)}{\tau_i\sigma}\right)^2\right),    
\end{equation}
where $\theta$ encapsulates the cosmological parameters to be
constrained, i.e., a subset or all of $\{w_0,w_a,\Omega_m\}$, and
$\Delta_\mu$. Here the notation ``$\sim$'' simply means ``distributed
according to''. $U$ is a uniform prior, with the probability density
function $f(x;a,b)=1/(b-a)$ for $x\in [a,b]$ and $0$ otherwise. $N$ is
a Gaussian (or Normal distributed) prior with the probability density
function
$f(x;\mu,\sigma^2)=\exp[-(x-\mu)^2/(2\sigma^2)]/\sqrt{2\pi\sigma^2}$.
The squared notation for the second parameter in $N(\mu,\sigma^2)$ is
used to indicate that $\sigma$ is the standard deviation (to prevent
possible confusion with the variance $\sigma^2$). (The parameters in
the $U$ distribution do not have this same meaning of mean and
standard deviation as in the Normal distribution.) For each case we
study, we confirm that the MCMC chains converged by monitoring the
trace plots and checking for good mixing and stationarity of the
posterior distributions.

The prior for $\Omega_m$ is informed by the 7-year WMAP
analysis~\cite{komatsu} for a $w$CDM model combining CMB, BAO, and
$H_0$ measurement. Since our assumptions on $w$ are less strict than
$w=const.$ we broaden the prior by a factor of two, leading to a
Gaussian prior given in Eqn.~(\ref{om}). As discussed earlier, we also
allow for an uncertainty in the overall calibration of the supernova
data, $\Delta_\mu$. We choose a wide, uniform prior for $\Delta_\mu$
given in Eqn.~(\ref{H0}). We consider two cases in all the analyses
presented in this paper. In the first case we fix $\Omega_m$ to a
fiducial value and reconstruct $w(z)$. This allows us to focus on
biases due to assumed parametric forms (parametric models) or possible
shortcomings due to the ill-posedness of the inverse problem (GP
methodology). In the second case, we let $\Omega_m$ be a free variable
within the specified prior, allowing us to study problems with
degeneracies that are highlighted when $w$ is a nontrivial function of
redshift.

\subsubsection{Constant Equation of State}

The simplest extension beyond a cosmological constant is to assume
that $w(z)$ is redshift independent. In this
case, Eqn.~(\ref{tildemu}) simplifies to
\begin{eqnarray}
\tilde\mu_B(w_0,z)&=&5\log_{10}\left\{(1+z)c
\right.\int^z_0 ds\left[\Omega_m(1+s)^{3}\right.\nonumber\\
&&\hspace{-0.5cm}+\left.\left.(1-\Omega_m)(1+s)^3(1+s)^{3w_0}
\right]^{-1/2}\right\}.
\end{eqnarray}
Current data are in good agreement with this assumption. We will
use the ansatz $w=const.=w_0$ as a first test in attempting to
reconstruct all three datasets. As discussed previously, an MCMC
algorithm is employed with the chain being run about 10,000 times.
Convergence is very quickly attained, within about the first one
hundred iterations.

Figure~\ref{wconst} shows the results for the case where we fix
$\Omega_m=0.27$ and assume perfect calibration. As expected, the
reconstruction works extremely well for the model where in fact
$w=const.$ (left panel). The best fit value for $w_0$ and its
probability intervals (PIs) are given in Table~\ref{table:m1} and
match the chosen value within small errors. Not surprisingly, the
results for the models with time varying $w$ are rather inaccurate.
For dataset 2, the value for $w$ is predicted slightly higher than the
average would be. In general, a larger $w$ leads to a lower
$\Delta\tilde\mu_B$. As can be seen in Figure~\ref{data},
$\Delta\tilde\mu_B(z)$ is slightly below the $\Lambda$CDM model for
this dataset. The one-parameter best fit for $w_0$ therefore has to be
high in order to capture this behavior, if we do not allow any other
parameter to vary. For the third dataset, we find a similar situation.
As can be seen in Figure~\ref{data} in the right panel,
$\Delta\tilde\mu_B$ is below the fiducial model. Capturing this
behavior with only one parameter to vary, $w_0$, leads to a value $w_0
> -1$ in order to fit the behavior in $\Delta\tilde\mu_B(z)$
reasonably well.

\begin{table}
        \caption{$w=const.$ - 95\% Probability Intervals (PIs)}
        \centering
                \begin{tabular}{c c  c c c}
                        \hline
        Set &  $w_0$ &  $\Omega_{m}$&$\Delta_\mu$ & $\sigma^2$ \\ [0.9ex]
 \hline
 1 & $-1.003^{+0.012}_{-0.013}$ &0.27 &0& $0.97^{+0.06}_{-0.05}$ \\
 2 & $-0.862^{+0.011}_{-0.011}$ &0.27& 0& $0.97^{+0.06}_{-0.05}$\\
 3 & $-0.915^{+0.012}_{-0.013}$ &0.27& 0&$0.99^{+0.06}_{-0.06}$\\
 1 & $-1.021^{+0.098}_{-0.100}$&$0.273^{+0.022}_{-0.027}$ & $-0.003^{+0.017}_{-0.017}$& $0.97^{+0.06}_{-0.05}$\\
 2 & $-0.849^{+0.089}_{-0.091}$ &$0.258^{+0.030}_{-0.035}$& $-0.005^{+0.014}_{-0.015}$&$0.97^{+0.06}_{-0.05}$\\
 3 & $-1.186^{+0.117}_{-0.120}$ &$0.347^{+0.018}_{-0.022}$& $-0.006^{+0.016}_{-0.017}$&$0.97^{+0.06}_{-0.05}$\\
 \hline
                \end{tabular}
        \label{table:m1}  
 \end{table}

\begin{figure*}
\centerline{
\includegraphics[width=2.2in]{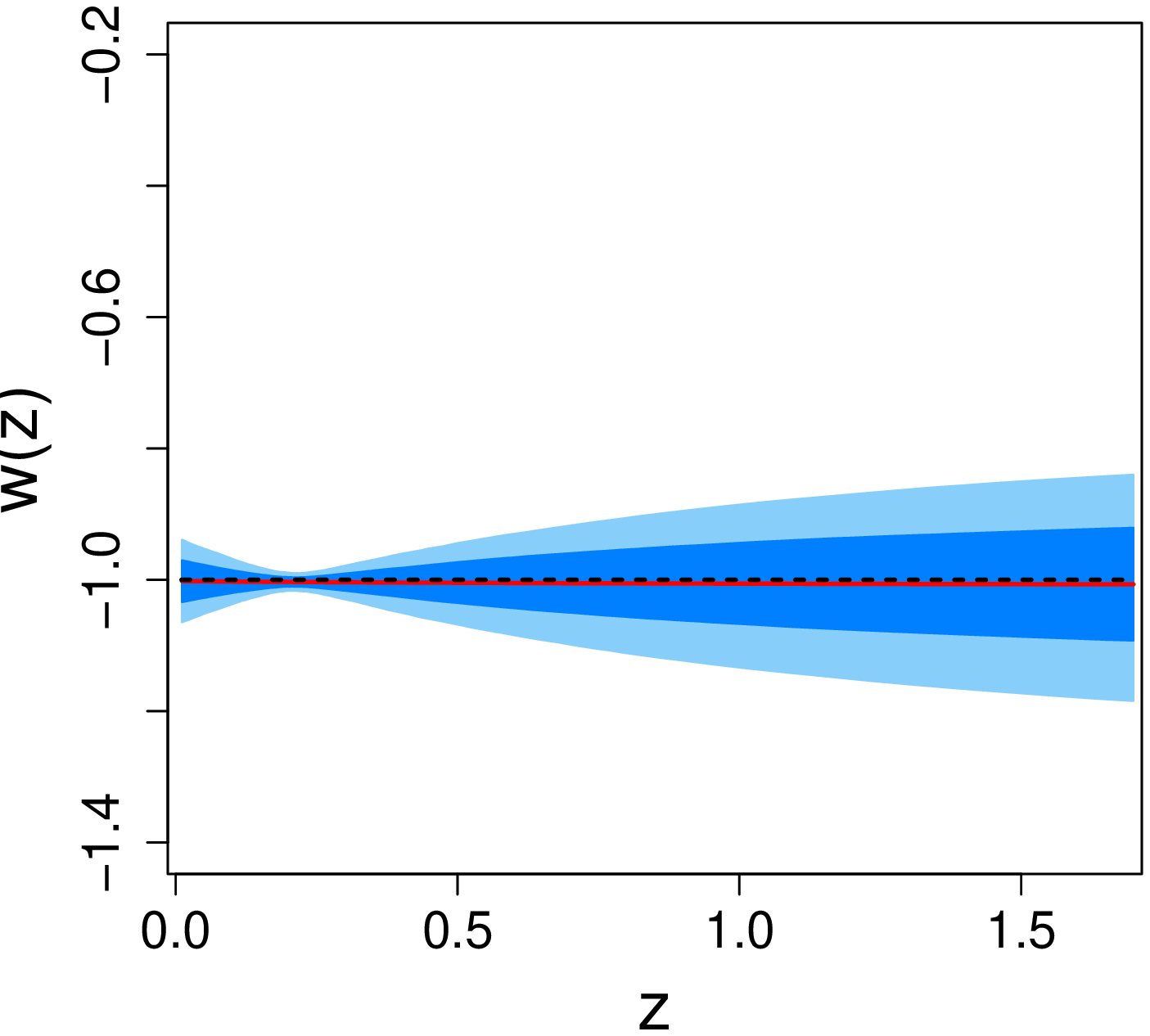}
\includegraphics[width=2.2in]{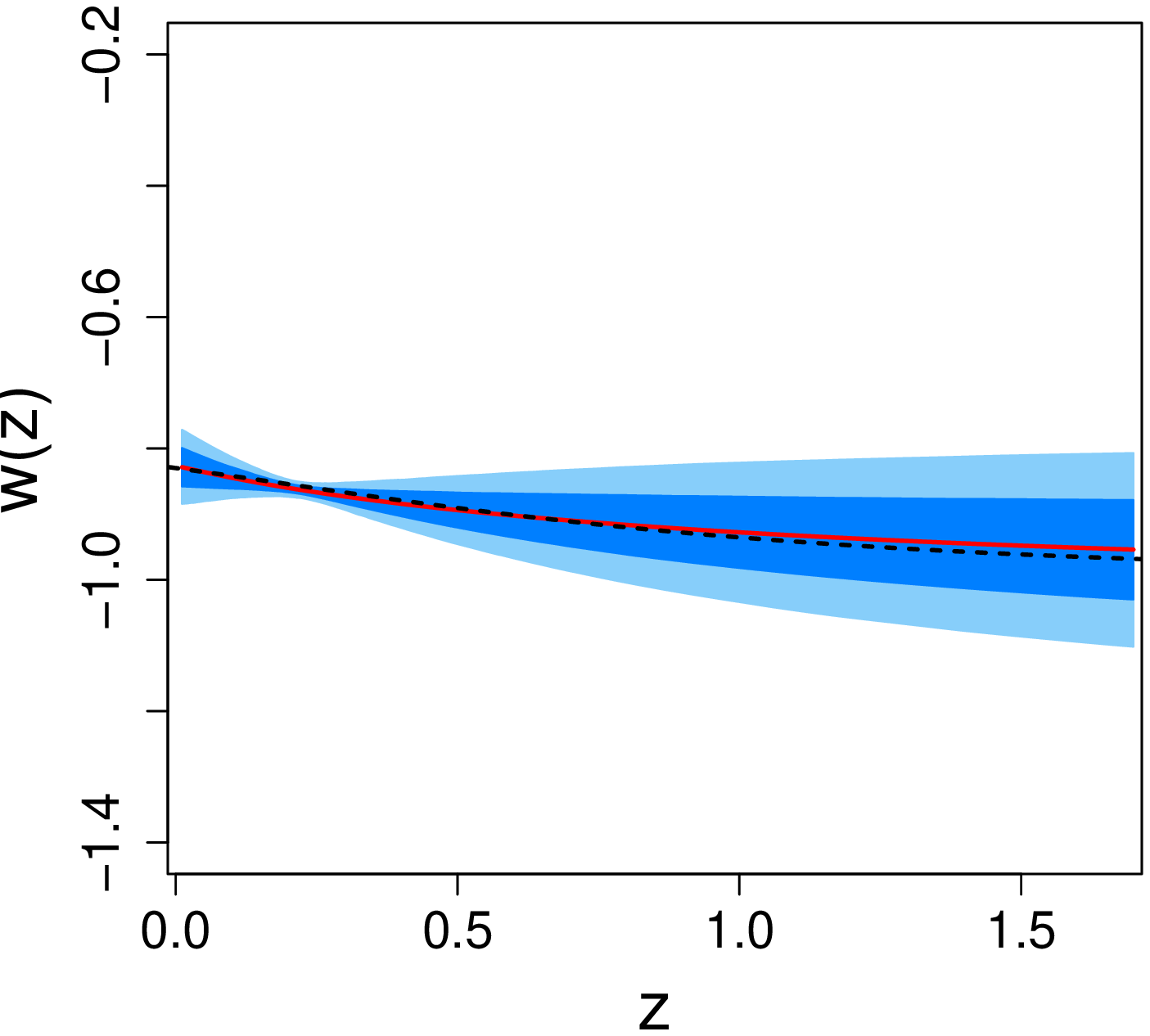}
\includegraphics[width=2.2in]{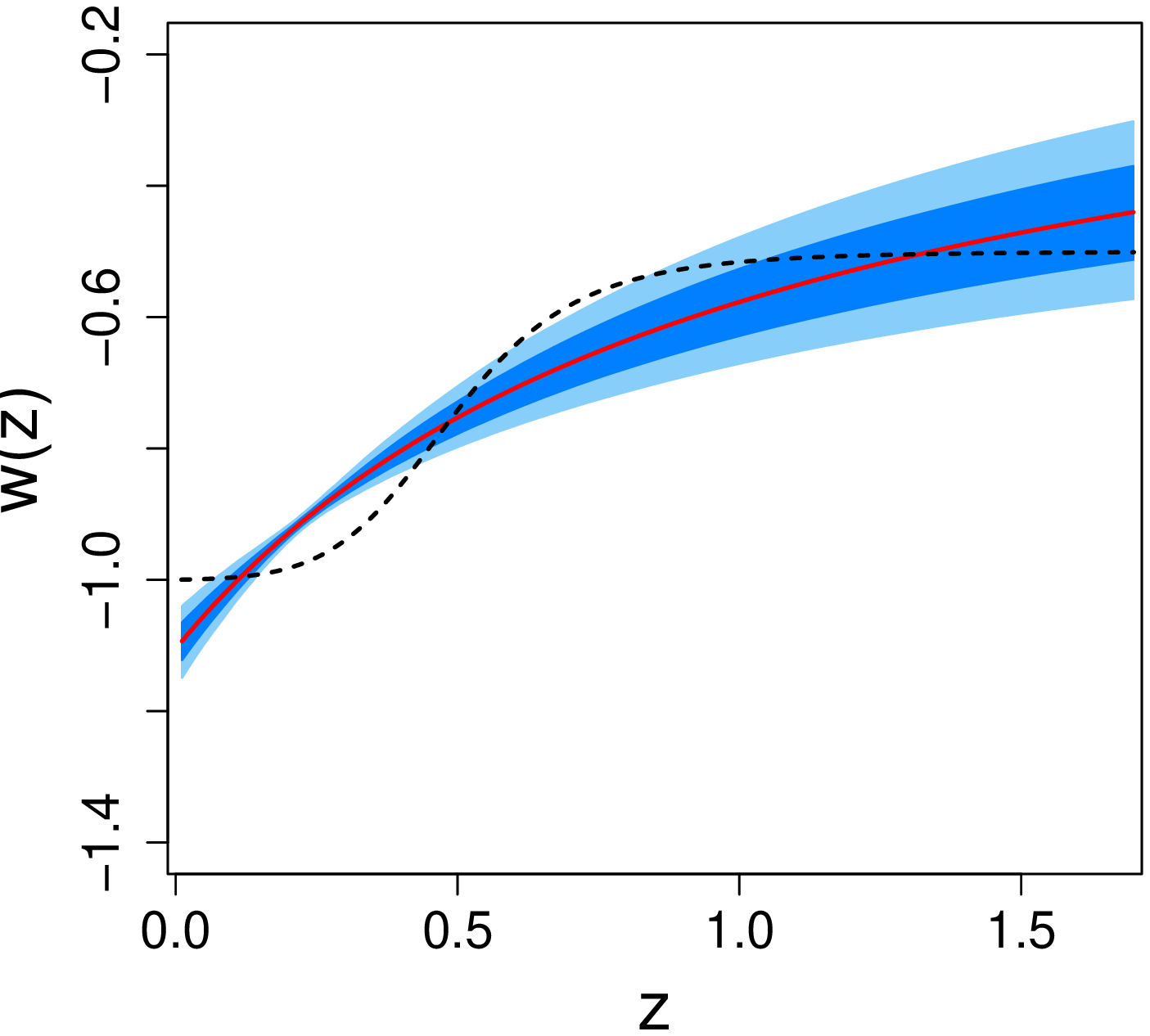}}
\centerline{
\includegraphics[width=2.2in,angle=0]{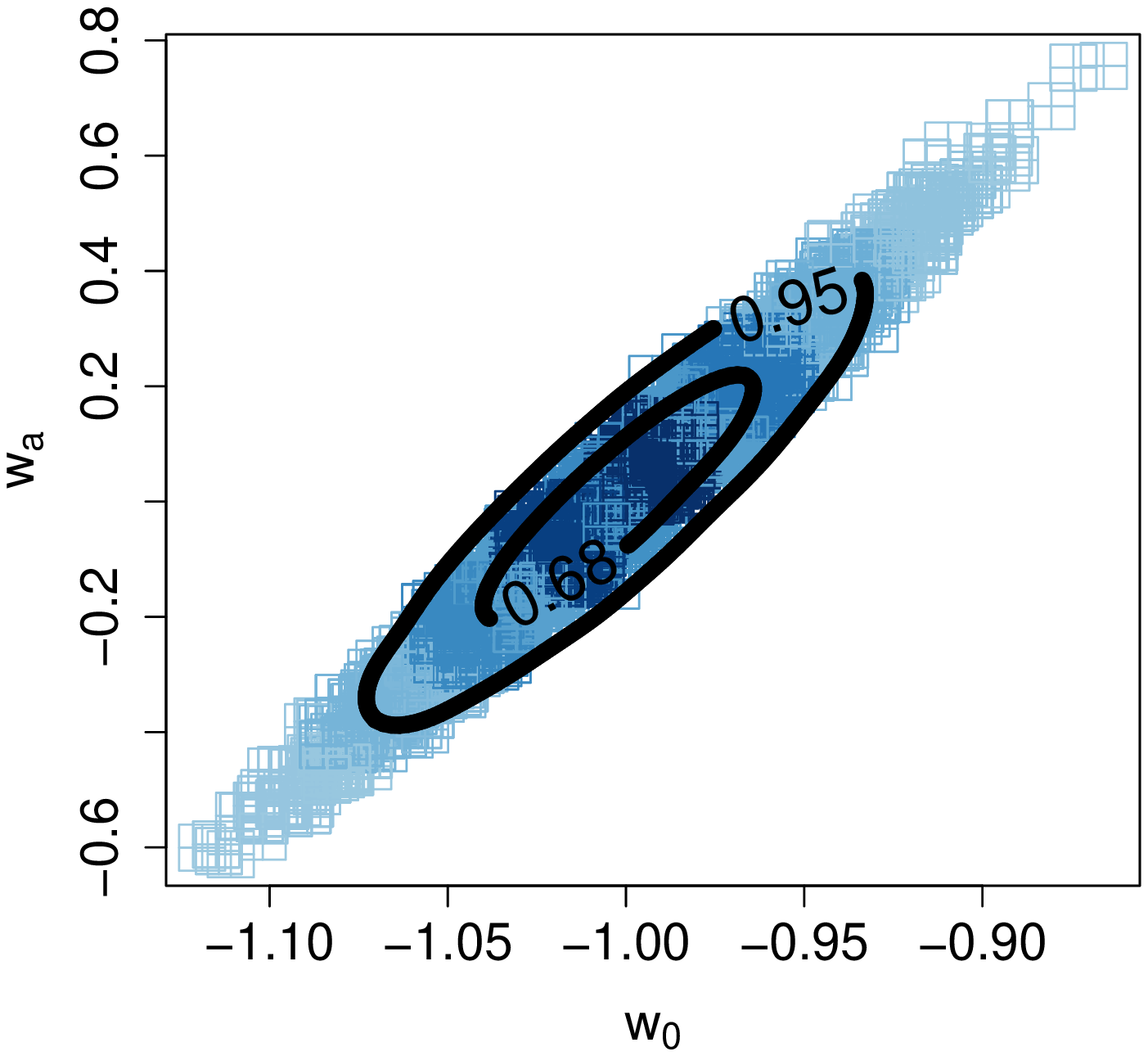}
\includegraphics[width=2.2in,angle=0]{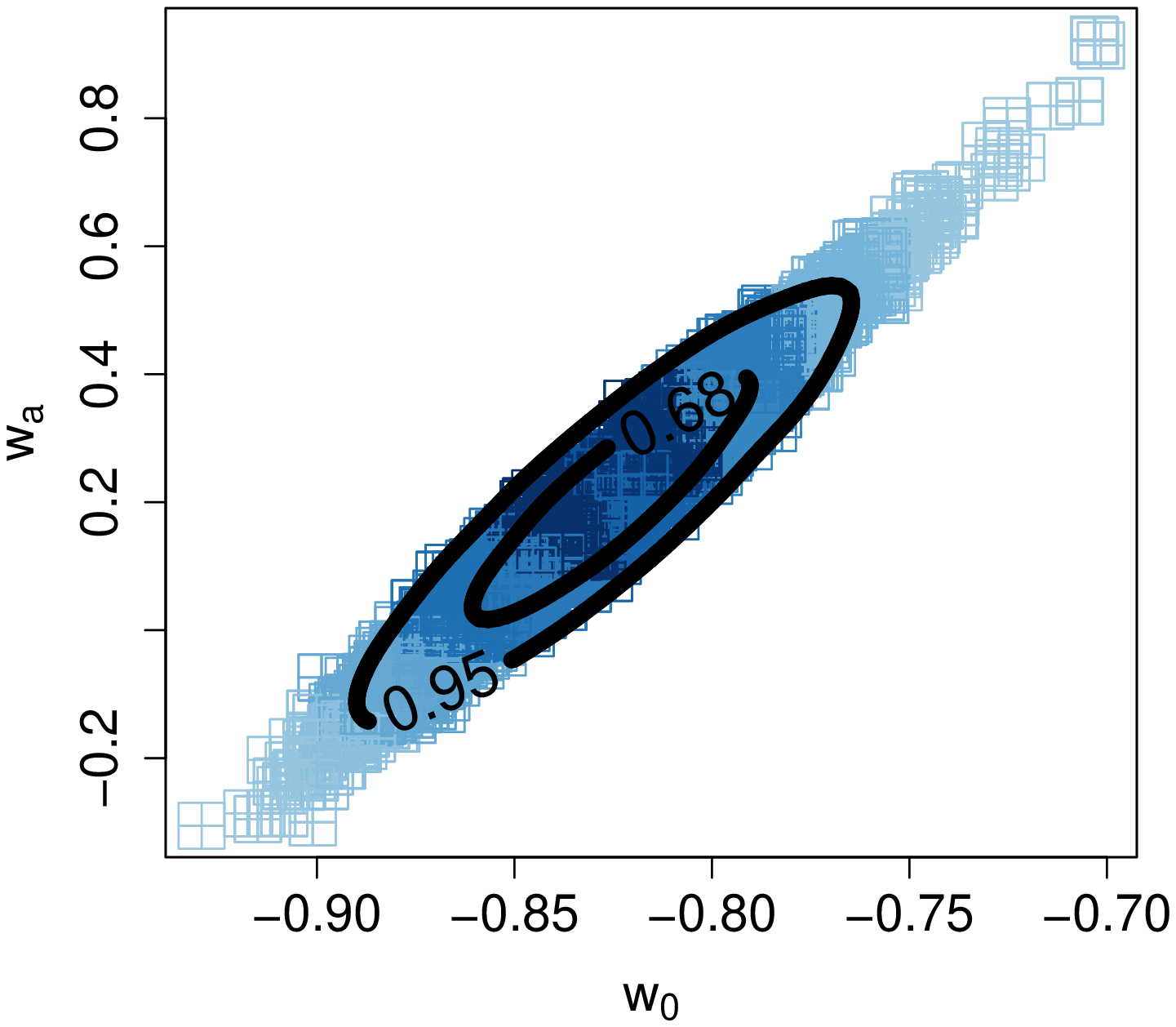}
\includegraphics[width=2.2in,angle=0]{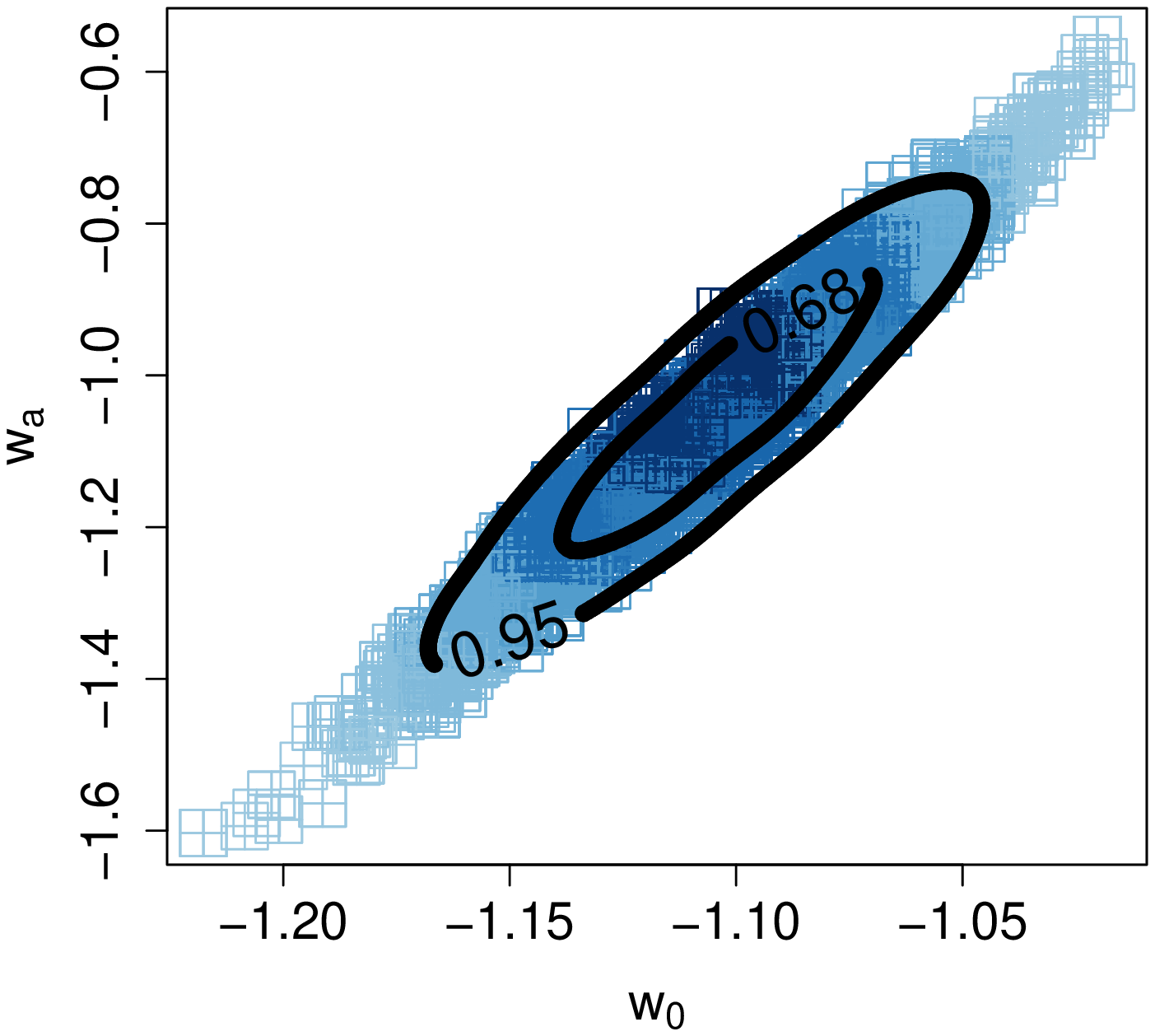}}
\caption{\label{cpl}Upper row: same as in Figure~\ref{wconst}, but
  with the reconstruction based on the parameterization of $w$
  represented by Eqn.~(\ref{cpl_equ}). The parameterization captures
  the variation in dataset 2 reasonably well, but is still not
  flexible enough to reconstruct an equation of state with less smooth
  changes -- as in dataset 3. The lower panels shows the 68\% and 95\%
  confidence contours for the fitting parameters $w_0$ and $w_a$ in
  Eqn.~(\ref{cpl_equ}) for the three datasets. Note that the axes of
  the contour plots have different ranges; the uncertainties for
  dataset 3 are the largest.} 
\end{figure*}

In the next step, we allow $\Omega_m$ and $\Delta_\mu$ to vary within
the assumed priors given in Eqs.~(\ref{om}) and (\ref{H0}). The
results for $w$ (including the truth) are shown in
Figure~\ref{wconst_oh}. The best fit values including error bars are
given in Table~\ref{table:m1}. Since $\Omega_m$ and $w_0$ are highly
correlated they must be sampled jointly with a covariance structure
obtained after running the process for some time. As in the case of
$\Omega_m$ and $\Delta_\mu$ fixed, the analysis is robust and works
well for the case of $w=const.$ Although the error bands increase, the
best fit values for all three parameters are very close to the assumed
model values. In the two cases of variable $w$, the strong degeneracy
between $w_0$ and $\Omega_m$ becomes very apparent, as both $w_0$ and
$\Omega_m$ influence the behavior of $\tilde\mu_B$ in a very similar
way: This bias on $w$ will disappear if other datasets such as CMB
data are included to provide good constraints on $\Omega_m$. For the
second dataset, where $\Delta\tilde\mu_B$ has a downward trend for
higher $z$, the behavior can be captured by a low value for $\Omega_m$
and a high value for $w$, or vice-versa. The best fit values will also
account for the curvature in $\Delta\tilde\mu_B$ and we find that the
best-fit model underpredicts $\Omega_m$ and overpredicts $w_0$. This
degeneracy can only be broken if we have better estimates for
$\Omega_m$. For dataset 3 the situation is even more severe: in order
to capture the slope of $\Delta\tilde\mu_B$, the estimates for both
parameters, $\Omega_m$ and $w_0$, are off, $\Omega_m$ is highly
overestimated, while the value for $w_0$ is underestimated and in fact
does not even go through the true $w(z)$ any more as can be seen in
Figure~\ref{wconst_oh}. This example demonstrates the bias that can be
introduced in the reconstruction of $w(z)$ if the assumed form for $w$
is too restricted and degeneracies are present. Also note that the
true result no longer falls within the predicted error bands.

For both datasets, the prediction for $\Delta_\mu$, which is mainly
anchored by the amplitude of the measurements for $\tilde\mu_B$, is
close to the true value. We also note that the ``truth'' for
$\Omega_m$ and $\Delta_\mu$ is not exact since we are working with one
finite realization for each dataset.

\subsubsection{$w_0-w_a$ Parameterization}

\begin{figure*}
\centerline{
\includegraphics[width=2.2in]{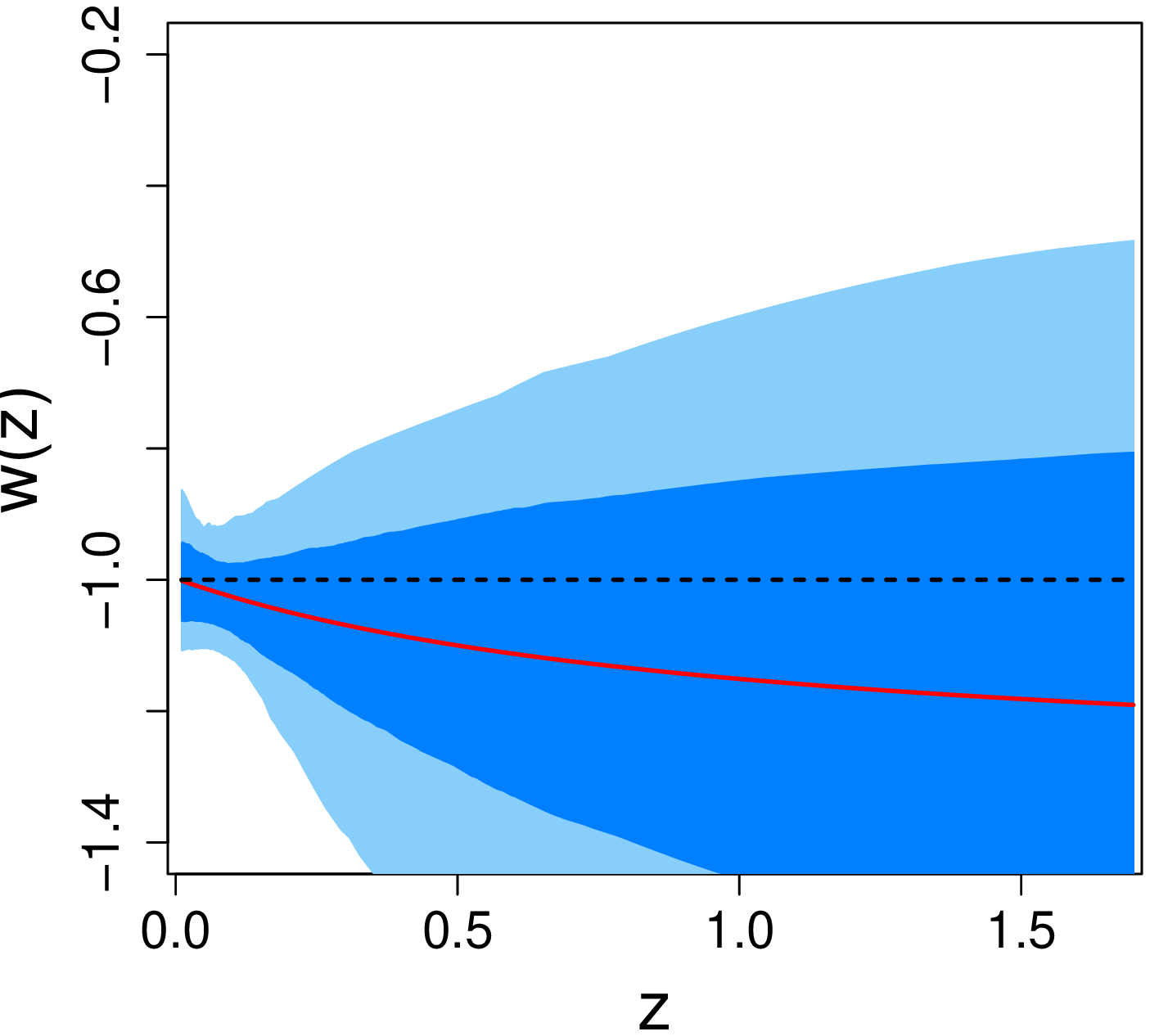}
\includegraphics[width=2.2in]{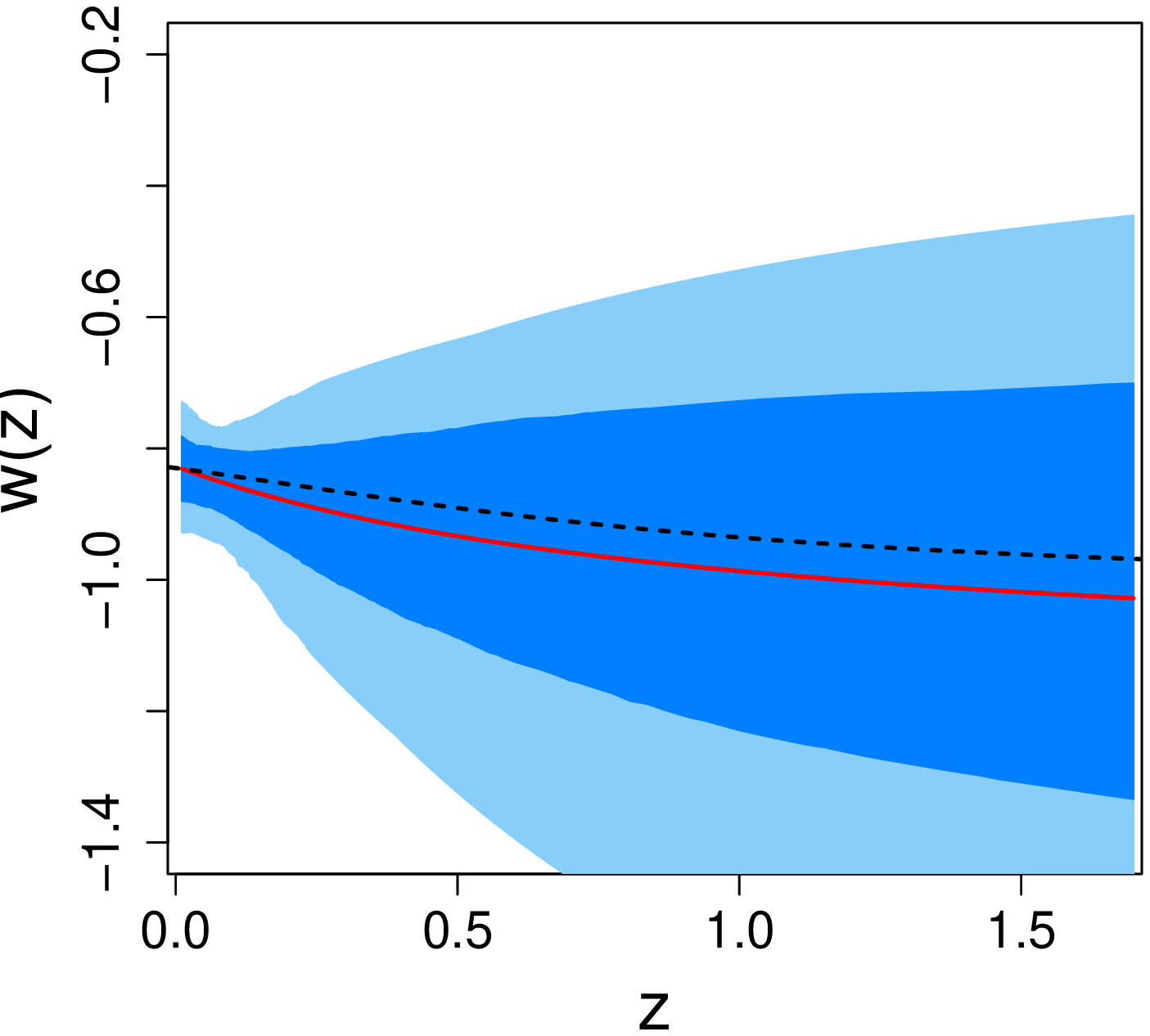}
\includegraphics[width=2.2in]{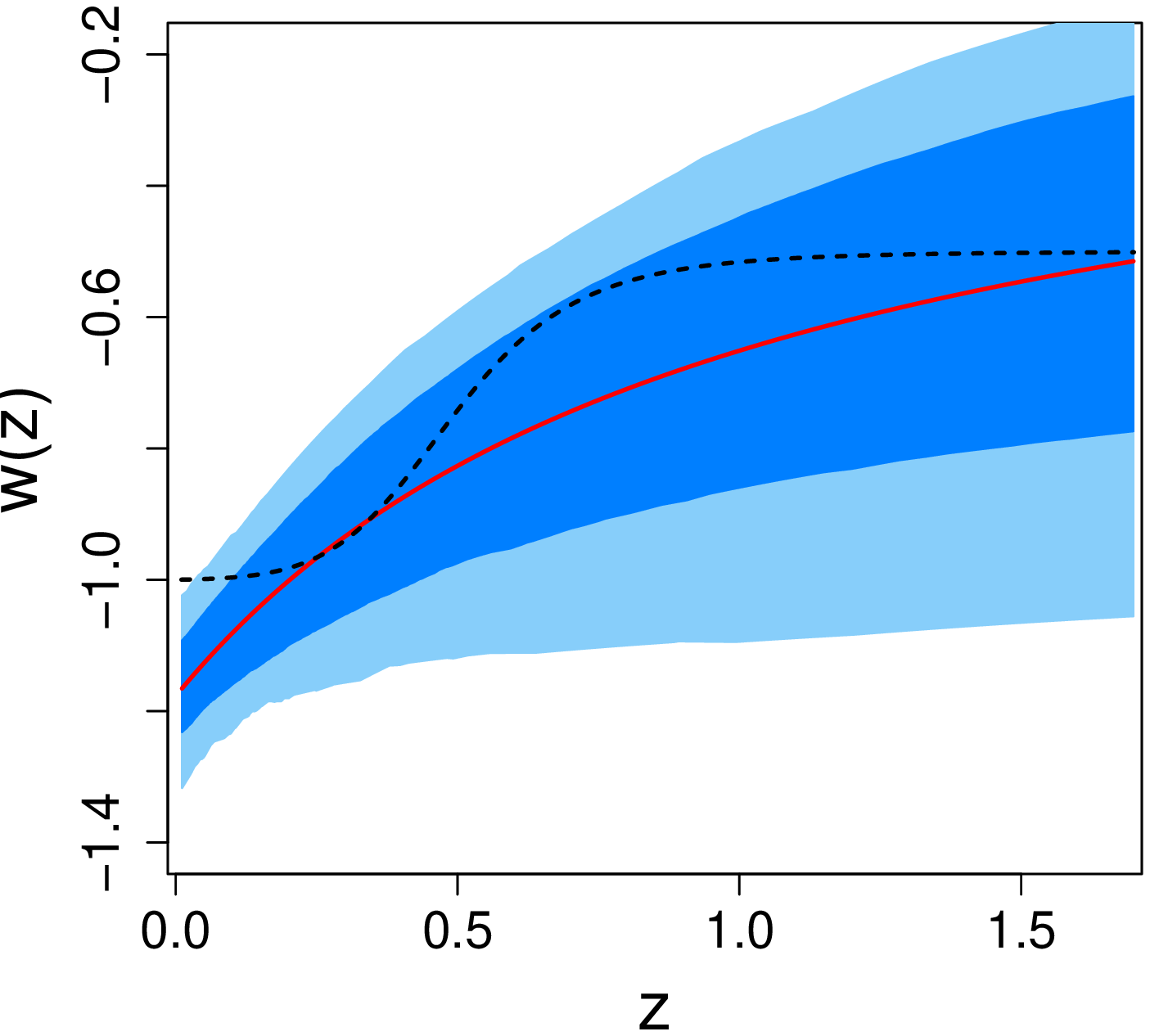}}
\centerline{
\includegraphics[width=2.2in]{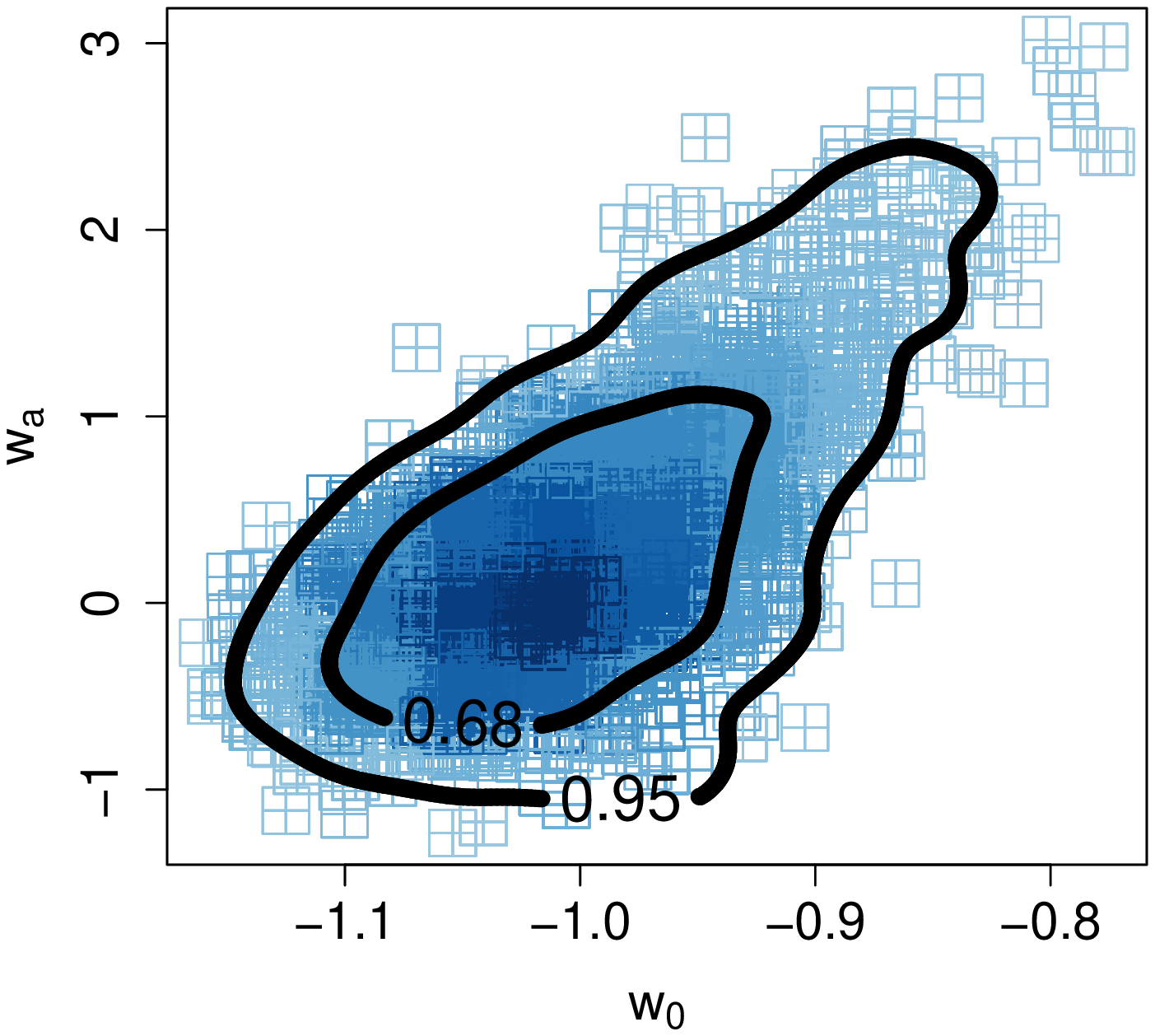}
\includegraphics[width=2.2in]{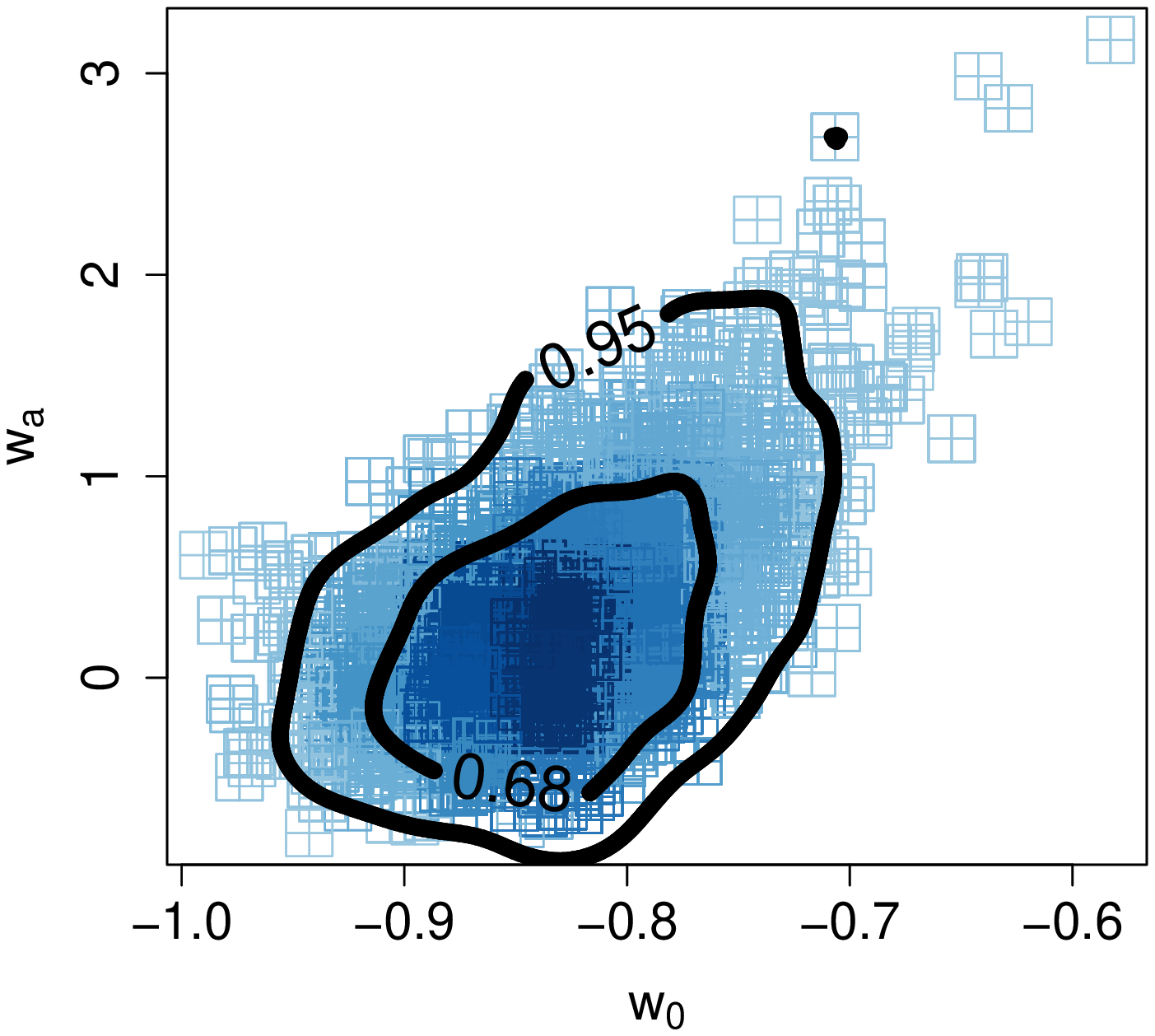}
\includegraphics[width=2.2in]{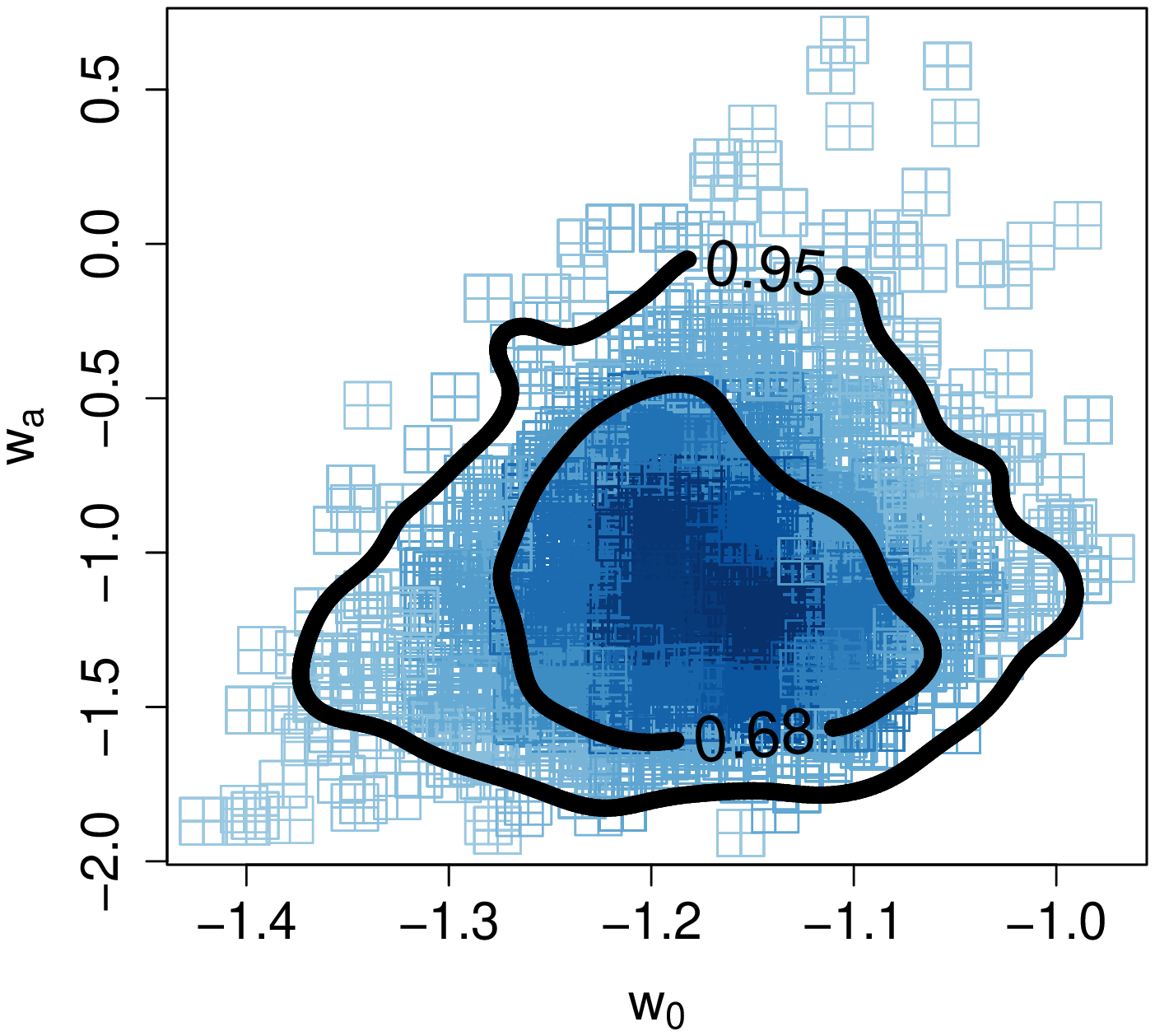}}
\caption{\label{cploh}Results shown as in
  Figure~\ref{cpl} but allowing $\Omega_m$ and $\Delta_\mu$ to vary.}
\end{figure*}

We now turn to the investigation of a commonly used parameterization of
the dark energy equation of state given by Refs.~\cite{chev01, linder03}:
\begin{equation}\label{cpl_equ}
w(z)=w_0-w_a\frac{z}{1+z}.
\end{equation}
As for the case $w=const.$, one integral in Eqn.~(\ref{mu}) can be
solved analytically and the expression for $\tilde\mu_B$ simplifies to:
\begin{eqnarray}
&&\tilde\mu_B(w_0,w_a,z)=\nonumber\\
&&5\log_{10}\left\{(1+z)c
\right.\int^z_0 ds\left[\Omega_m(1+s)^{3}\right.\nonumber\\
&&+\left.\left.(1-\Omega_m)(1+s)^{3(w_0-w_a+1)}e^{3w_a
      s/(1+s)}\right]^{-1/2}\right\}.\nonumber\\
\end{eqnarray}
This parameterization allows for a weak monotonic time dependence in
$w$ and should therefore capture the behavior of our second model
reasonably well.

Following the analysis in the previous case ($w=const.$), we first fix
$\Omega_m$ and $\Delta_\mu$ to their fiducial values. The results are
summarized in Figure~\ref{cpl} and Table~\ref{table:m2}.  For the
$w=const.$ dataset the parameterization picks up a very small
variation in $w$ but the prediction $w=-1$ is well within errors. The
mild variation with $z$ in the second dataset is captured reasonably
well. A rather high value for $w_0$ leads to a pull-down of
$\Delta\tilde\mu_B$. This is then compensated by a large
positive value for $w_a$ which leads to an upturn in
$\Delta\tilde\mu_B$.

For the third dataset the parameterization is not quite flexible
enough. While the overall behavior (the rise at high redshift) is
captured, the S-shape of the underlying equation of state cannot be
extracted.  Moreover, in an attempt to fit the data, the value of
equation of state today is decreased to $w_0 < -1$. This decrease
leads to an upturn of $\Delta\tilde\mu_B(z)$ while the large negative
value for $w_a$ acts in the opposite direction. The parameterization
finds a time dependence in $w$, but not of the correct specific form
as would be required for distinguishing different models of dark
energy.

The results including estimations for $\Omega_m$ and $\Delta_\mu$ are
similar. As for the $w=const.$ parameterization, the parameters are
all sampled jointly because of their strong correlations. These
correlations degrade the accuracy of the $w$ reconstruction. For the
first dataset, the prediction for $\Omega_m$ is slightly high which in
turn amplifies a time dependence in the best fit for $w$ which does
not exist in the original dataset. Again, the error bars are large and
clearly $w=-1$ is well within the error bounds. For the second data
set, the prediction for $\Omega_m$ is rather accurate. For dataset 3,
$\Omega_m$ is overpredicted which leads to a slight degradation in the
prediction for $w$ itself. The values for $w_0$ and $w_a$ are similar
to the case of fixed $\Omega_m$.

 \begin{table}
        \caption{$w=w_0-w_a z/(1+z)$ - 95\% PIs}
                \begin{tabular}{c c c c c }
                        \hline
        Set &  $w_0$ &  $w_a$ & $\Omega_{m}$ &$\Delta_\mu$\\ 
 \hline
 1 & $-1.002^{+0.061}_{-0.066}$ & $0.008^{+0.351}_{-0.365}$ & 0.27 & 0\\
 2 & $-0.826^{+0.056}_{-0.059}$ & $0.203^{+0.309}_{-0.325}$ & 0.27 & 0\\
 3 & $-1.105^{+0.051}_{-0.059}$ & $-1.056^{+0.273}_{-0.307}$ & 0.27 & 0\\
 1 & $-0.998^{+0.134}_{-0.111}$ & $0.306^{+1.705}_{-1.126}$ & 
 $0.281^{+0.053}_{-0.077}$ &  $-0.001^{+0.012}_{-0.018}$\\
 2 & $-0.827^{+0.107}_{-0.101}$ & $0.319^{+1.305}_{-0.909}$ & 
 $0.272^{+0.068}_{-0.094}$ & $-0.003^{+0.018}_{-0.018}$\\
 3 & $-1.177^{+0.129}_{-0.156}$ & $-1.052^{+0.820}_{-0.575}$ & 
 $0.284^{+0.056}_{-0.077}$ & $-0.012^{+0.017}_{-0.019}$\\
 \hline
 \hline
  Set & $\sigma^2$  \\ 
 \hline
 1 & $0.97^{+0.06}_{-0.06}$ \\
 2&  $0.97^{+0.06}_{-0.05}$ \\
 3 & $0.97^{+0.06}_{-0.05}$ \\
 1 & $0.97^{+0.06}_{-0.05}$ \\
 2 & $0.97^{+0.06}_{-0.05}$ \\
 3 & $0.97^{+0.06}_{-0.06}$ \\
 \hline
                \end{tabular}
        \label{table:m2}  
 \end{table}

Overall, the parameterization provides a reasonable description of the
data, especially for moderately varying $w$, as is expected. The
drawback is obvious: rapid changes in $w$ as shown in dataset 3 cannot
be captured.

\subsection{Nonparametric Reconstruction: Gaussian Process Model}

The previous exploration of the standard parametric methods makes it
clear that as long as the data correspond to the models that the
methods are designed for (e.g., if $w$ is in fact constant, the ansatz
$w=w_0$ will obviously lead to the best result), the results are
rather good. However, as soon as datasets are introduced for which the
parameterizations are not flexible enough to track the true behavior
of $w(z)$, the analysis is susceptible to unacceptable levels of bias
in determining modeling and cosmological parameters.

It is common practice to employ a parametric form for $w(z)$ and then
assess the robustness of the result by a goodness of fit test.  For
example, Ref.~\cite{genovese} uses a Bayesian information criterion
(BIC) statistic for this task. We applied the BIC model comparison
criteria for the two parametric reconstruction approaches without much
success -- instead of choosing the model that corresponds to the truth
(which in case of dataset 2 and 3 is the $w_0-w_a$ parametrization
over $w=const.$) the BIC always preferred the parametric form with the
least parameters, in this case $w=const.$ 

A nonparametric form of $w(z)$ can address some of the shortcomings of
parametric reconstruction methods. We now describe a new,
nonparametric method based on GP modeling~\cite{Banerjee,gprefs}. As
mentioned previously, a GP is a stochastic process, which in our case
is indexed by $z$. The defining property of a GP is that the vector
that corresponds to the process at any finite collection of points
follows a multivariate normal (MVN) distribution. Gaussian processes
are elements of an infinite dimensional space, this is the sense in
which they provide a nonparametric method for curve fitting. They are
characterized by a mean and a covariance function, often defined by a
small number of hyperparameters.

We assume that the data errors are Gaussian and use the same
likelihood as in the treatment of the parameterized models. The
use of Bayesian estimation methods (including the MCMC algorithm)
allows us to estimate the hyperparameters of the GP correlation function
together with any other parameters, comprehensively propagating all
estimation uncertainties~\cite{gamerman}. Using the definition of a
GP, we assume that, for any collection $z_1,...,z_n$, $w(z_1),...,
w(z_n)$ follow a multivariate Gaussian distribution with a constant
negative mean and exponential covariance function written as
\begin{equation}
K(z,z')=\kappa^2\rho^{|z-z'|^\alpha}.  
\end{equation}
Here $\rho \in (0,1)$ is a free parameter that, together with $\kappa$
and the parameters defining the likelihood, are fit from the data. The
form of the assumed correlation function implies that, theoretically,
there is non-zero correlation between any two points. $\rho$ controls
the exponential decay of the correlation as a function of distance in
redshift, but it does not provide a bound for the correlation between
two points. This is analogous to the concept of standard deviation --
many, but not all, of the observations are within one standard
deviation of the mean, and most are within two, but there is no
theoretical bound that all observations have to fall within. In
principle, we could include an explicit noise term in the correlation
(a so-called ``nugget''). Instead, we chose to include the noise term
$\sigma^2$ in the likelihood equation (\ref{L}) from which it will
propagate to the GP.

\begin{figure}
\centerline{
  \includegraphics[width=1.2in]{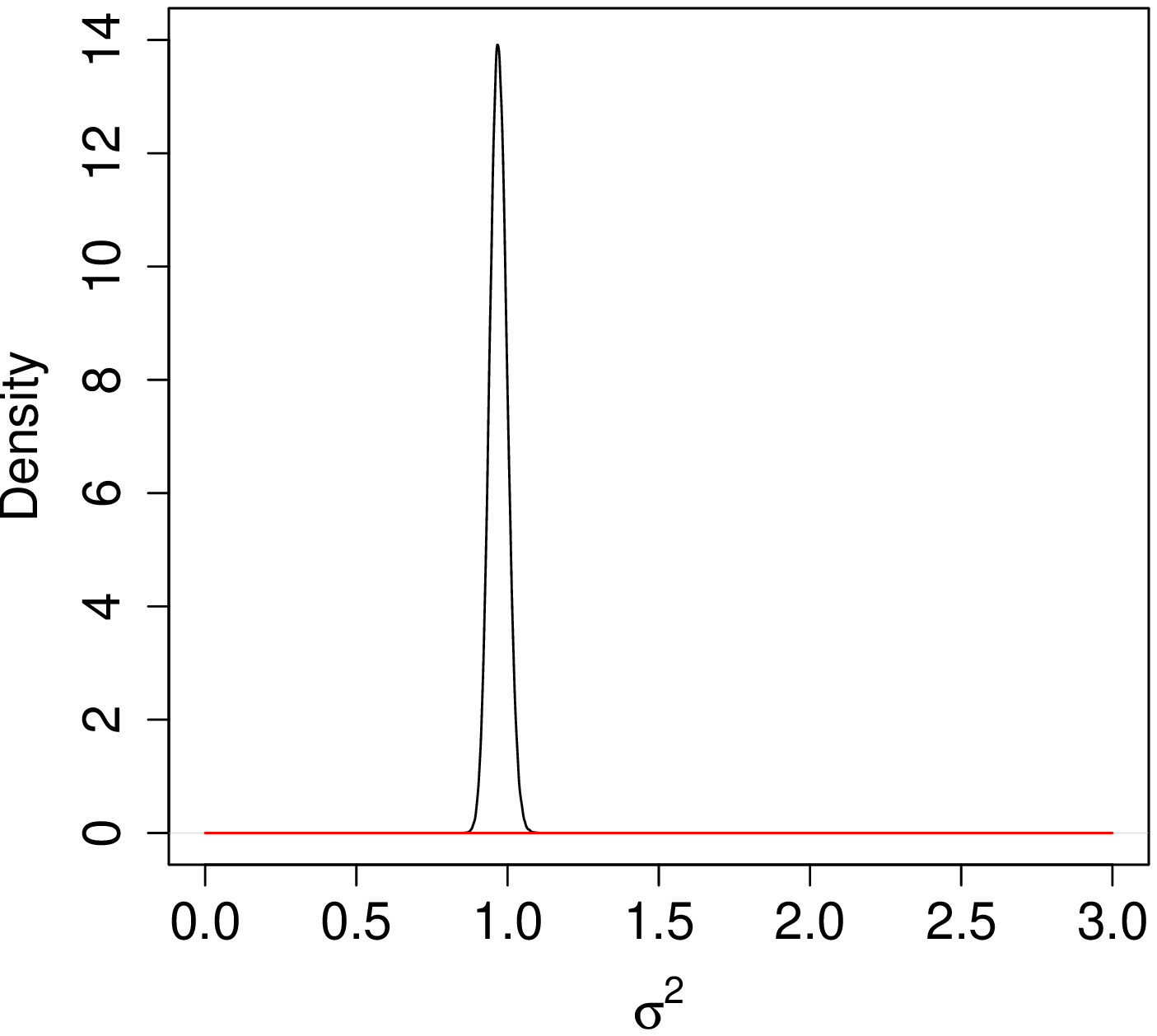}
  \includegraphics[width=1.2in]{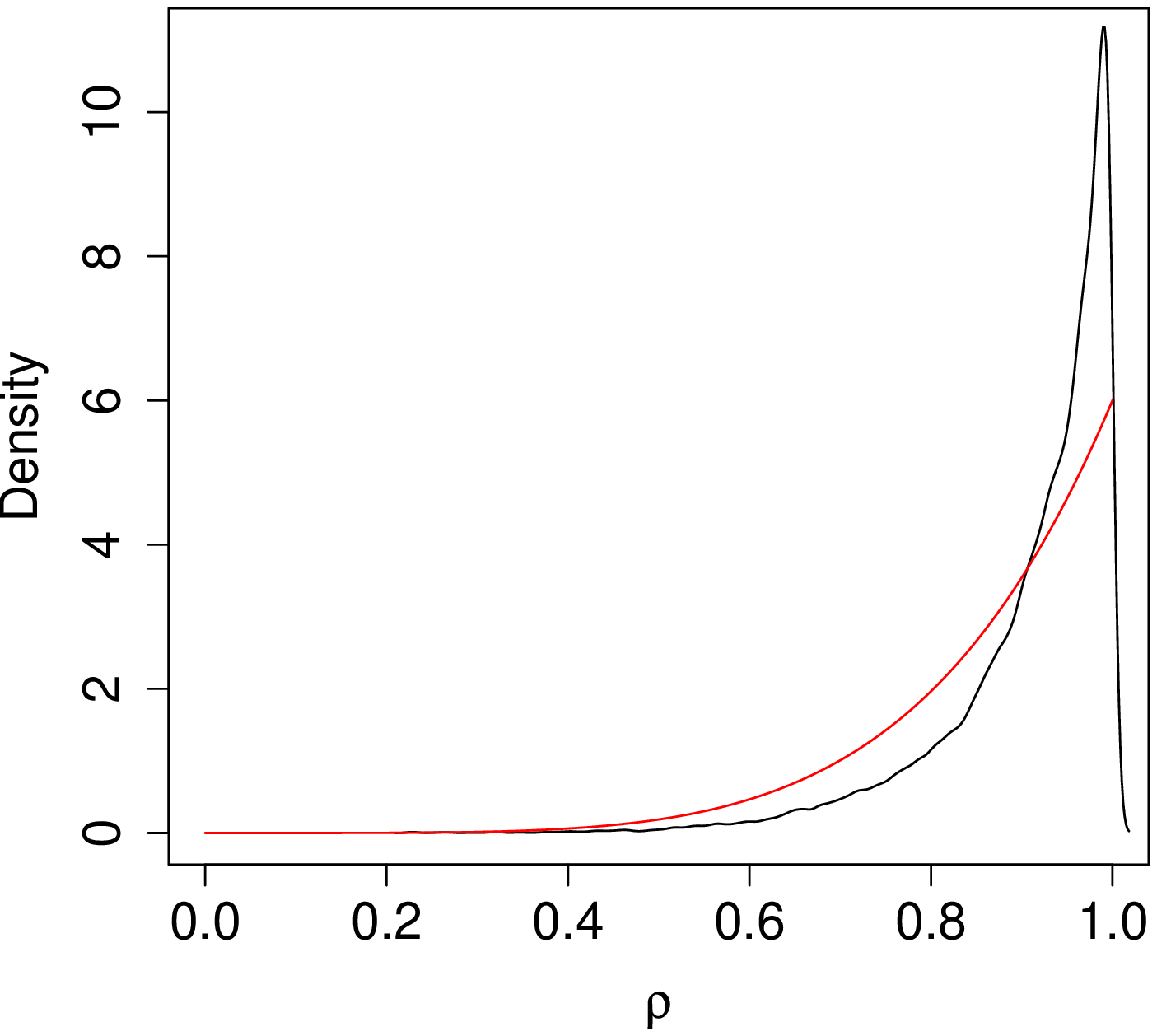}
  \includegraphics[width=1.2in]{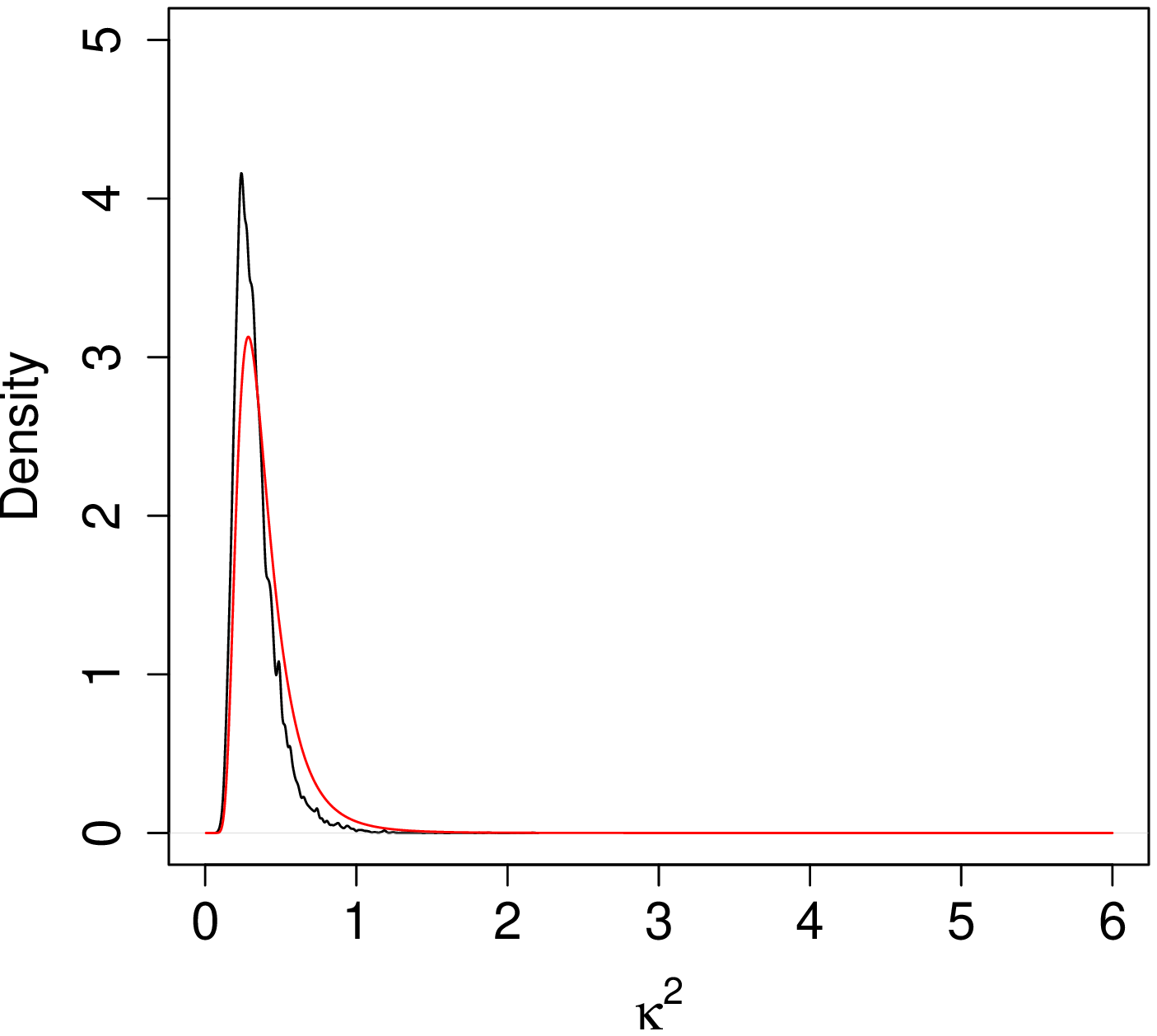}}
  \caption{\label{priors} Example for the priors (red lines) and
  posteriors (black lines) for $\sigma^2$, $\rho$ and $\kappa^2$ for
  dataset 1. The parameters $\sigma^2$ has a non-informative prior,
  $\kappa^2$ has an inverse Gamma prior, and $\rho$ has a $Beta$
  prior. The posteriors for the three different datasets for these
  parameters are very similar.}
\end{figure}

The value of $\alpha \in (0,2]$ influences the smoothness of the GP
realizations: for $\alpha=2$, the realizations are smooth with
infinitely many derivatives, while $\alpha=1$ leads to rougher
realizations suited to modeling continuous non-differentiable
functions. Here we use $\alpha=1$ to allow for maximum flexibility in
reconstructing $w$. (For a comprehensive discussion of different
choices for covariance functions and their properties, see
Ref.~\cite{gprefs}.) The mean of the GP is taken to be fixed. $\rho$
has a prior of $Beta(6,1)$ and $\kappa^2$ has a prior of $IG(6,2)$.
$IG$ is an inverse Gamma distribution prior, with the probability
density function $f(x;\alpha,\beta)=\beta^{\alpha}x^{-\alpha-1}
\Gamma(\alpha)^{-1}\exp(-\beta/x)$, with $x>0$. The probability
distribution of the $Beta$ prior is given by
$f(x;\alpha,\beta)=\Gamma(\alpha+\beta)x^{\alpha-1}(1-x)^{\beta-1}/
[\Gamma(\alpha)\Gamma(\beta)]$. Figure~\ref{priors} shows an example
of the priors for $\sigma^2$ and the two GP model parameters $\rho$
and $\kappa$ for dataset 1. The posteriors indicate that the data are
informative about $\sigma^2$ and reasonably informative about $\rho$.
As for the parametric reconstruction, $\Omega_m$ is given a prior
based on currently available estimates.

\begin{figure*}
\centerline{
 \includegraphics[width=2.2in]{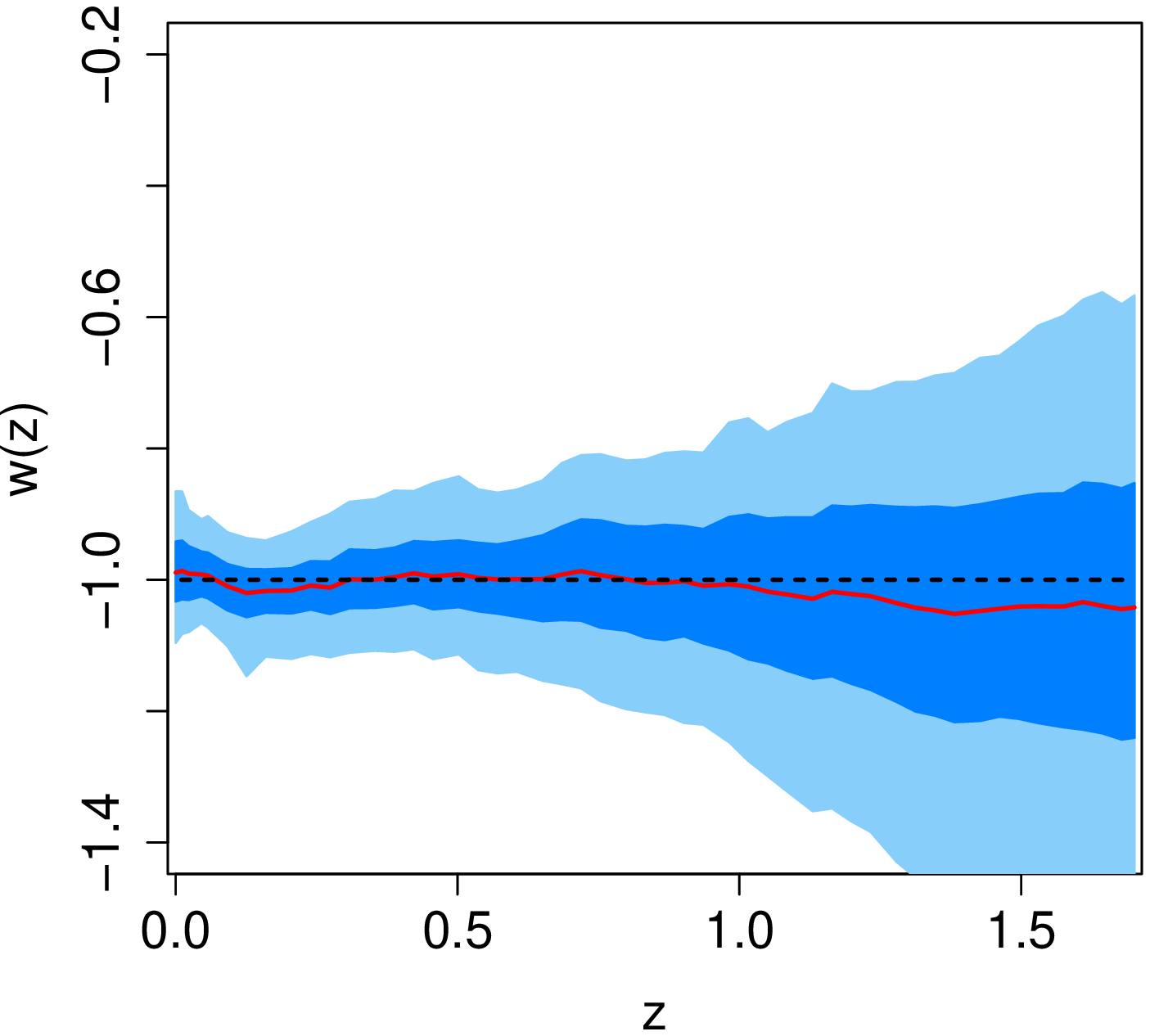}
\includegraphics[width=2.2in]{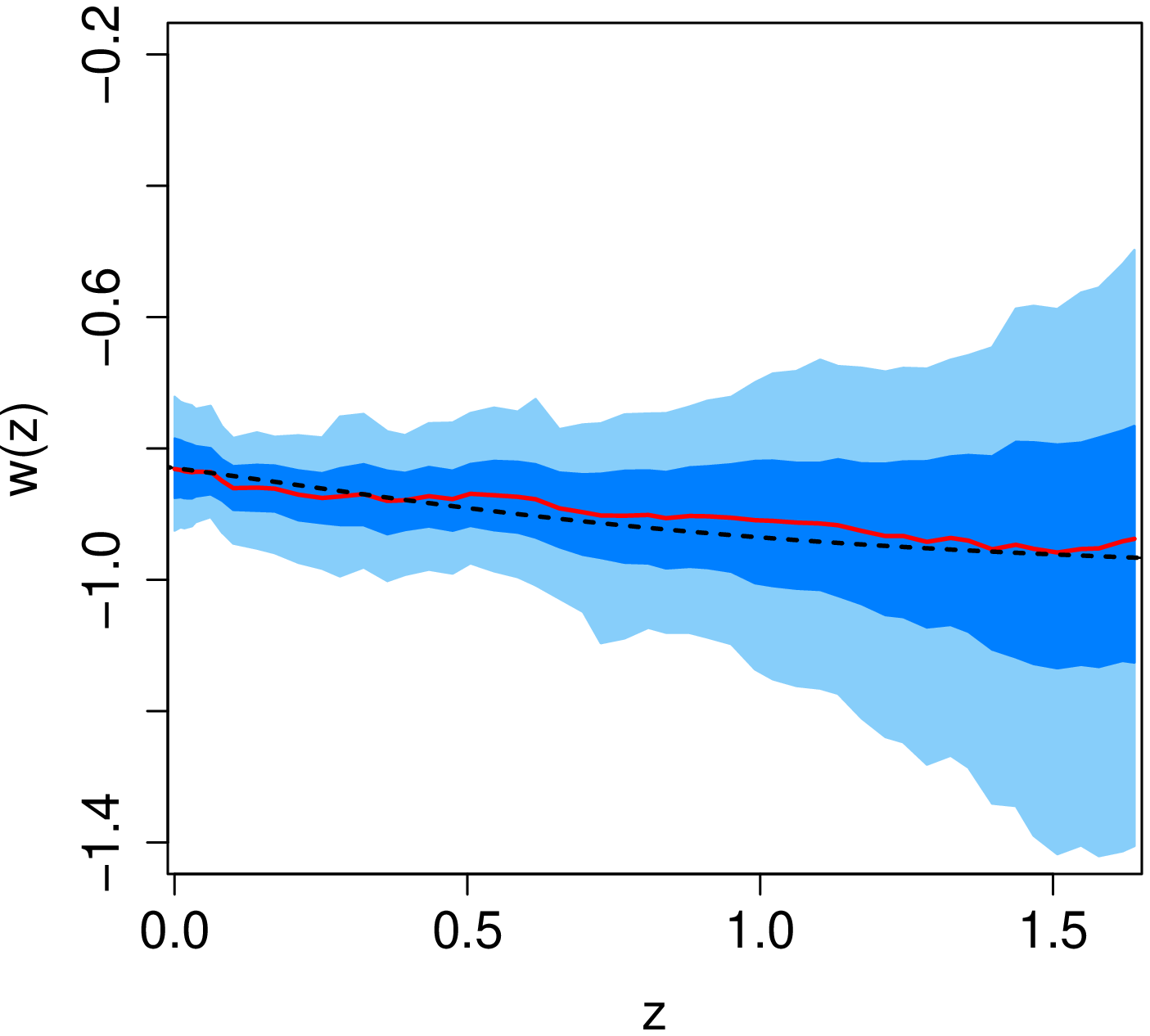}
\includegraphics[width=2.2in]{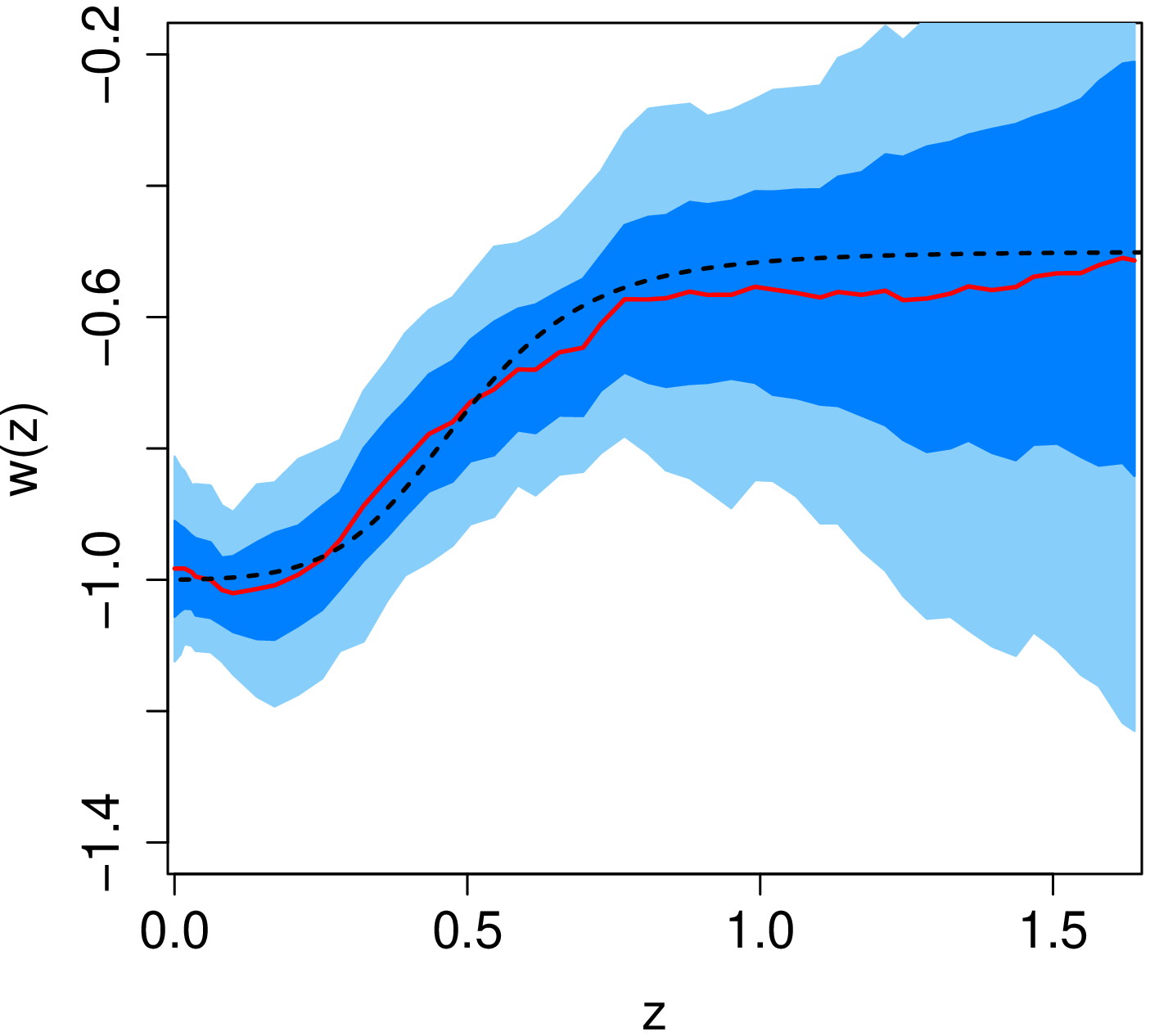}}
\caption{\label{gp} Results displayed as in Figure~\ref{wconst}, but
  with GP model-based reconstruction. For all three datasets the GP
  model succeeds in capturing the true behavior of $w$; the error bars at
  higher $z$ are slightly larger due to the sparser supernova sampling
  beyond $z=1.1$.}  
\centerline{
\includegraphics[width=2.2in]{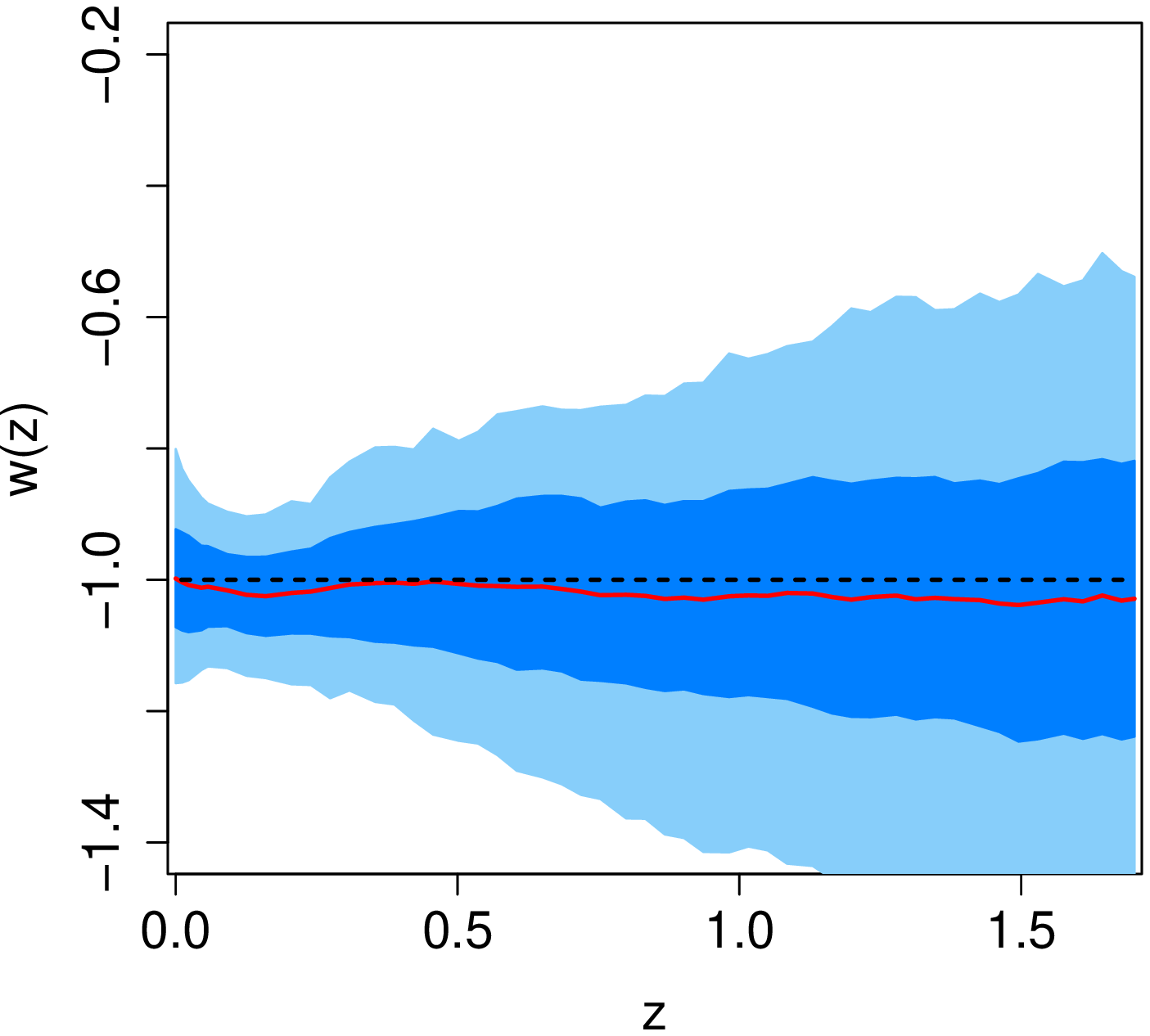}
\includegraphics[width=2.2in]{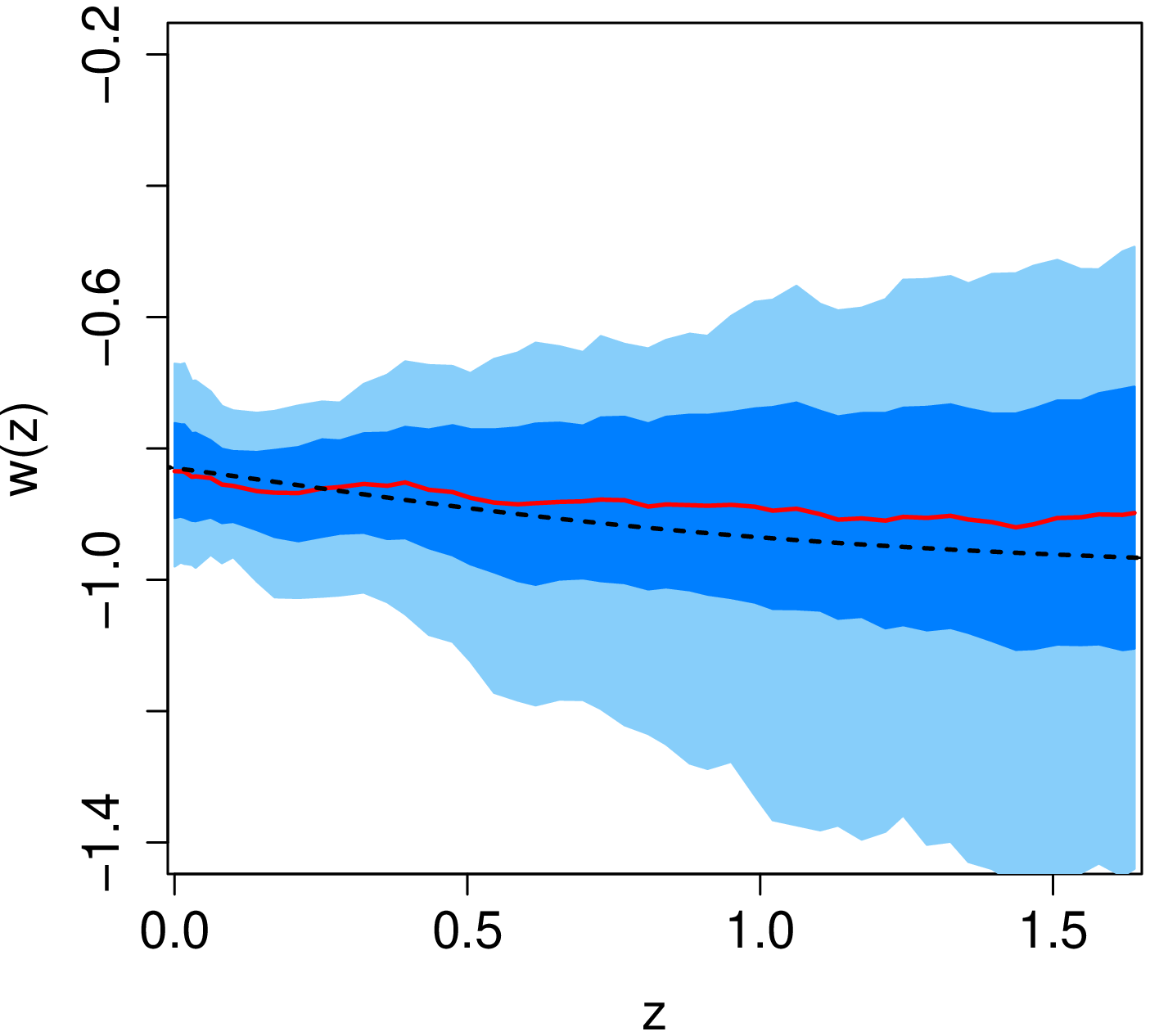}
\includegraphics[width=2.2in]{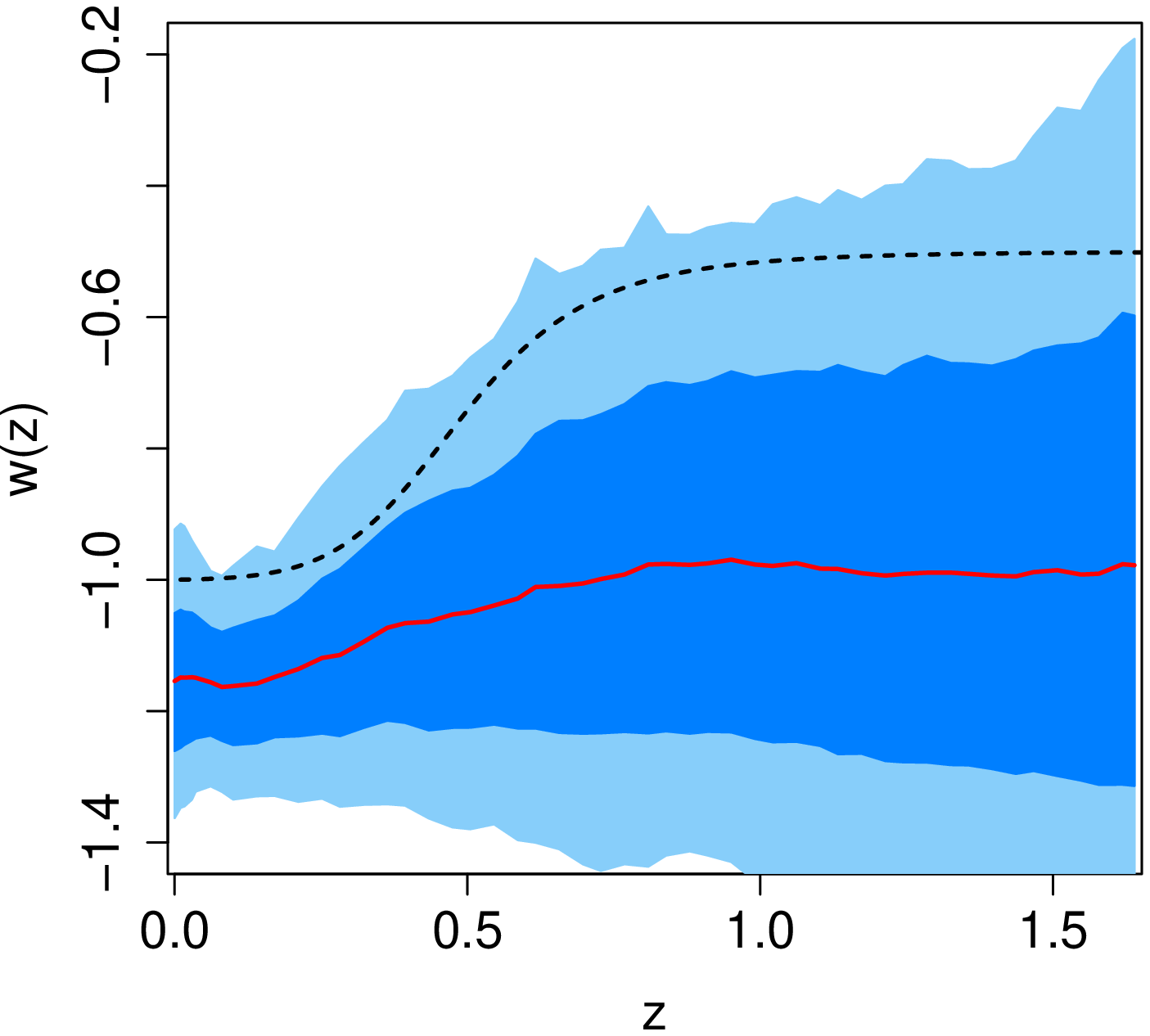}}
\caption{\label{gpoh}As in Figure~\ref{gp} but with
  $\Omega_m$ and $\Delta_\mu$ free to vary. The GP reconstruction performs
  extremely well for all the first two cases and captures the third
  case reasonably well (within error bands).}
\end{figure*}

We set up the following
GP for $w$:
\begin{equation}
w(u)\sim {\rm GP}(-1,K(u,u')).
\end{equation}
Choosing a mean value of -1 is natural, given current observational
constraints on $w$. Even though the mean is fixed, each GP
realization will actually have a different mean with a spread
controlled by $\kappa$. For the second and third dataset we 
adjusted the means during the analysis to slightly different values suggested by
preliminary runs. This adjustment is purely informed by the data and
demonstrates the flexibility of the approach. In principle, the mean
could also be left as a free parameter. After the adjustment we
measured the posterior mean and ensured that it was close to the prior
mean. Table~\ref{table:mean} summarizes the prior and posterior means for
the final analysis. 

\begin{table}[b]
        \caption{Prior and posterior means for $w(u)$ for the GP models.}
        \centering
                \begin{tabular}{l c c }
                        \hline
        Set &  Prior Mean &  Posterior Mean \\ [0.9ex]
\hline
1 ($\Omega_m$, $\Delta_\mu$) fixed & -1.00&  -1.01 \\
1 ($\Omega_m$, $\Delta_\mu$) free  & -1.00 & -1.02 \\
2 ($\Omega_m$, $\Delta_\mu$) fixed & -0.94&  -0.90 \\
2 ($\Omega_m$, $\Delta_\mu$) free  & -0.87 & -0.88 \\
3 ($\Omega_m$, $\Delta_\mu$) fixed & -0.7&  -0.71 \\
3 ($\Omega_m$, $\Delta_\mu$) free  & -1.00 & -1.04 \\
\hline
                \end{tabular}
        \label{table:mean}
\end{table}

Next, recall that we have to integrate over $w(u)$ (Eqn.~\ref{mu}):
\begin{equation}
\label{int}
y(s)=\int_0^s\frac{w(u)}{1+u}du.
\end{equation}
We use the fact that the integral of a GP is also a GP with mean and
correlation dependent on the original GP~\cite{Banerjee}. Therefore
$y(s)$ results in a second GP defined as:
\begin{equation}
y(s)\sim {\rm GP}\left(-\ln (1+s),\kappa^2
\int_0^s\int_0^{s'}\frac{\rho^{|u-u'|^\alpha}dudu'}{(1+u)(1+u')}\right),
\end{equation}
where we choose $\alpha=1$.
The mean value for this GP is simply obtained by integrating
Eqn.~(\ref{int}) for the mean value of the GP for $w(u)$.
We can now construct a joint GP for $y(s)$ and $w(u)$:
\begin{equation}
\left[\begin{array}{c}y(s)\\w(u)\end{array}\right]\sim{\rm
GP}\left[\left[
\begin{array}{c}-\ln(1+s)\\-1\end{array}\right],
\left[\begin{array}{cc}\Sigma_{11} & \Sigma_{12}\\\Sigma_{21}&
\Sigma_{22}\end{array}\right]\right],
\end{equation}
with
\begin{eqnarray}
\Sigma_{11}&=&\kappa^2\int_0^s\int_0^{s'}\frac{\rho^{|u-u'|}dudu'}{(1+u)(1+u')},\\
\Sigma_{22}&=&\kappa^2\rho^{|u-u'|},\\
\Sigma_{12}&=& \Sigma_{21} = \kappa^2\int_0^s\frac{\rho^{|u-u'|}du}{(1+u)}.
\end{eqnarray}
The mean for $y(s)$ given $w(u)$ can be found through the following
relation:
\begin{equation}
\langle
y(s)|w(u)\rangle=-\ln(1+s)+\Sigma_{12}\Sigma_{22}^{-1}\left[w(u)-(-1)\right].
\end{equation}
Note that we never have to calculate the double integral in
$\Sigma_{11}$ which would be numerically expensive. More details about
each step in the GP model algorithm are given in Appendix~\ref{appa}.

\begin{table}
         \caption{GP model - 95\% PIs}
        \centering
                 \begin{tabular}{c c c c c c}
                        \hline
        Set &  $\Omega_{m}$ & $\Delta_\mu$ & $\sigma^2$ & $\rho$ & 
 $\kappa^2$\\ [0.9ex]
  \hline
 1 & 0.27 &0& $0.97^{+0.06}_{-0.05}$ & $0.910 ^{+0.088}_{-0.262}$ & 
 $0.330^{+0.357}_{-0.177}$\\
 2 & 0.27& 0& $0.97^{+0.06}_{-0.05}$ & $0.915^{+0.083}_{-0.258}$ & 
 $0.338^{+0.339}_{-0.179}$\\
 3 & 0.27& 0&$0.97^{+0.06}_{-0.05}$ & $0.802^{+0.141}_{-0.284}$ & 
 $0.406^{+0.426}_{-0.216}$\\
 1 & $0.270^{+0.032}_{-0.043}$ & $-0.003^{+0.058}_{-0.054}$& 
 $0.97^{+0.06}_{-0.05}$ & $0.897^{+0.100}_{-0.266}$ & 
 $0.343^{+0.374}_{-0.183}$ \\
 2 & $0.263^{+0.046}_{-0.051}$ & 
 $-0.004^{+0.018}_{-0.018}$&$0.97^{+0.06}_{-0.05}$
 & $0.903^{+0.095}_{-0.273}$ & $0.343^{+0.394}_{-0.183}$\\
 3 & $0.327^{+0.040}_{-0.070}$ & 
 $-0.007^{+0.019}_{-0.019}$&$0.97^{+0.06}_{-0.05}$
 & $0.852^{+0.143}_{-0.319}$ & $0.351^{+0.403}_{-0.194}$\\
 \hline
                 \end{tabular}
         \label{table:m3}  
\end{table}

Following our practice for the parameterized reconstruction methods,
we first apply the GP-based technique fixing the values for $\Omega_m$
and $\Delta_\mu$. The results are shown in Figure~\ref{gp}.
Reconstruction from the GP model for $w(z)$ is remarkably accurate for
all three data sets. (The noise in the predictions is due to the
choice of the functional form of the covariance function.) In
particular, the reconstruction for the third dataset, where the
parameterized model did not fare well, is very good.
Table~\ref{table:m3} gives the results for the GP model
hyperparameters $\rho$ and $\kappa$ for all three models. Larger
values for $\rho$ indicate a smoother reconstructed function. For a
model with more variation in the data and $w$ crossing the mean
several times, the correlation length would be smaller than for the
models investigated here. Our analysis shows that the $\rho$ is the
smallest for the dataset that varies the most (dataset 3), as
expected. Since even the third dataset is not varying strongly, $\rho$
is still close to one (note that $\rho=1$ is not allowed). In
addition, the interplay between $\rho$ and $\kappa$ which determine
the overall covariance function is non-trivial -- the simultaneous
fitting of $\kappa$ and $\rho$ is therefore an important aspect of our
approach.

Last, we study the results from the GP model, letting $\Omega_m$ and
$\Delta_\mu$ vary. The results are shown in Figure~\ref{gpoh}. As for
the two parameterized models, degeneracies degrade the results in the
cases of varying $w$. For the case of $w=-1$ the prediction for
$\Omega_m$ is very close to the input value (see Table~\ref{table:m3}
for the best-fit values for $\Omega_m$ and $\Delta_\mu$ and the GP
model parameters), and the reconstruction for $w$, albeit somewhat
noisy, is very close to the truth. For the second model, the best-fit
value for $\Omega_m$ is slightly low, but correct within the error
bars. The reconstruction is also in this case very accurate. For the
third dataset the best fit value for $\Omega_m$ is above the true
value, leading to a lower result for $w(z)$. The GP model approach
captures the overall behavior of the true $w(z)$ within the error
bands. However, the problem due to the degeneracy between $w$ and
$\Omega_m$ becomes very apparent in this case. The GP model approach
finds a solution that overestimates $\Omega_m$ with a rather flat $w$.
As we show in the next section, this is a good fit to the data but
does not give a particular good match to the true $w(z)$. These
results will certainly improve if we have stronger constraints on
$\Omega_m$ from different datasets, e.g., CMB or baryon acoustic
oscillation measurements to break the degeneracies between $\Omega_m$
and $w(z)$.

\subsection{Comparison of the Different Approaches }

In order to summarize our findings, we now provide a brief comparison
of the parametric and nonparametric reconstruction results. We
consider two metrics for this comparison: (i) the accuracy of the
reconstructed form of $w(z)$ given that the exact answer is known, and
(ii) the accuracy with which the predicted $w(z)$ fits the data. As
mentioned before, more involved statistical techniques such as BIC did
not lead to satisfying results, as the simplest parametrization was
always identified as the best -- which is obviously not the case for
datasets 2 and 3.

\begin{table}[b]
        \caption{Mean squared error for the reconstructed $w(z)$ w.r.t. 
the exact $w(z)$.}
        \centering
                \begin{tabular}{l c c c }
                        \hline
        \hfill &  Dataset 1 &  Dataset 2 & Dataset 3 \\ [0.9ex]
\hline
$w_0$ ($\Omega_m$, $\Delta_\mu$ fixed) & 0.00001 & 0.0044 & 0.1000 \\
$w_0$ ($\Omega_m$, $\Delta_\mu$ free)  & 0.00004 & 0.0058 & 0.3050 \\
\hline
$w_0-w_a$ ($\Omega_m$, $\Delta_\mu$ fixed) & 0.00003 & 0.0001 & 0.0034 \\
$w_0-w_a$ ($\Omega_m$, $\Delta_\mu$ free) &  0.0175 & 0.0019 &   0.0109\\
\hline
GP ($\Omega_m$, $\Delta_\mu$ fixed) &   0.0006 &  0.0003 &      0.0014\\
GP ($\Omega_m$, $\Delta_\mu$ free) &  0.0006 & 0.0012 &  0.1439 \\
\hline
                \end{tabular}
        \label{table:mse}
\end{table}

\begin{figure}
\centerline{
 \includegraphics[width=2.in]{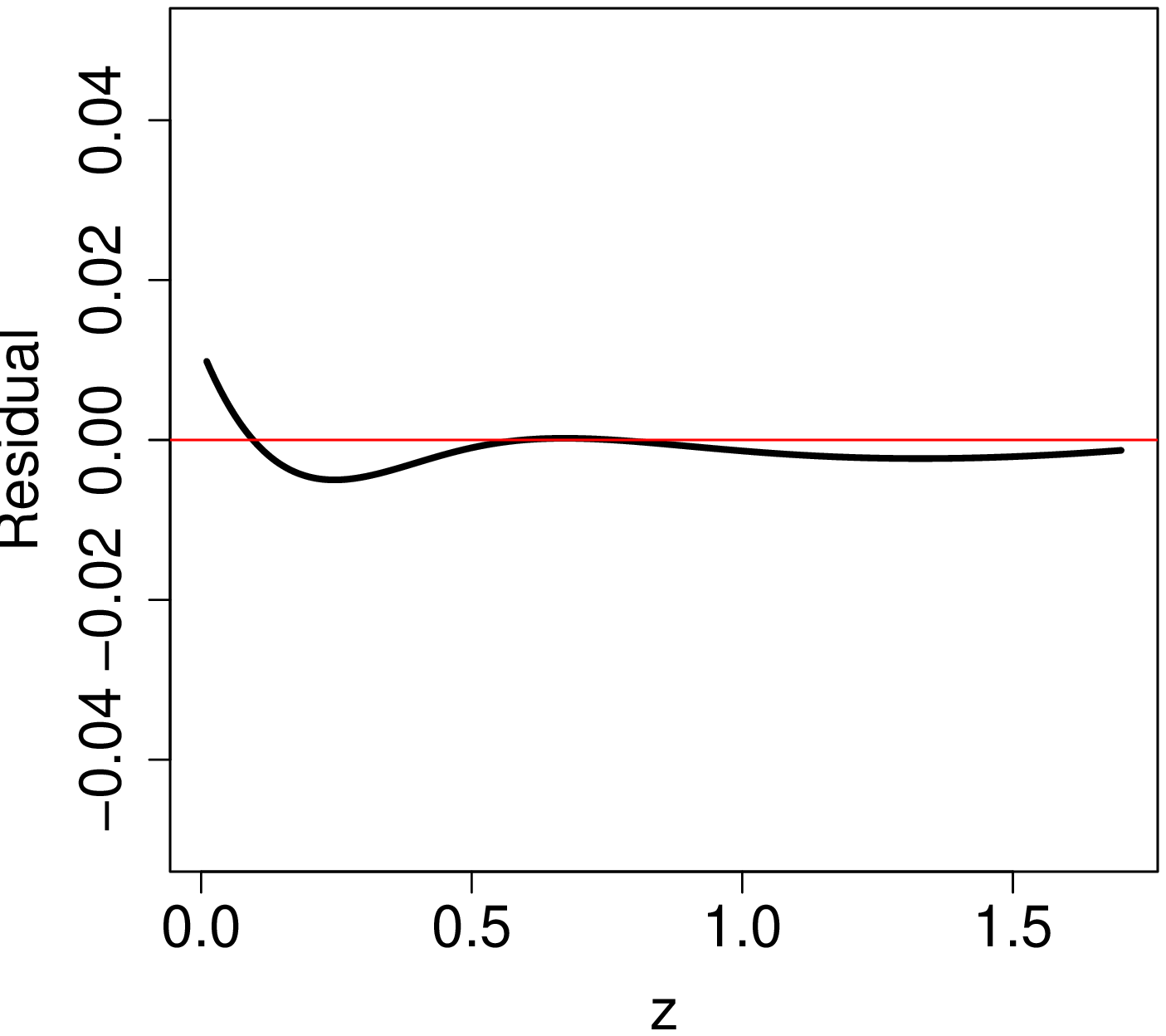}
 \hspace{-1.28cm}
\includegraphics[width=2.in]{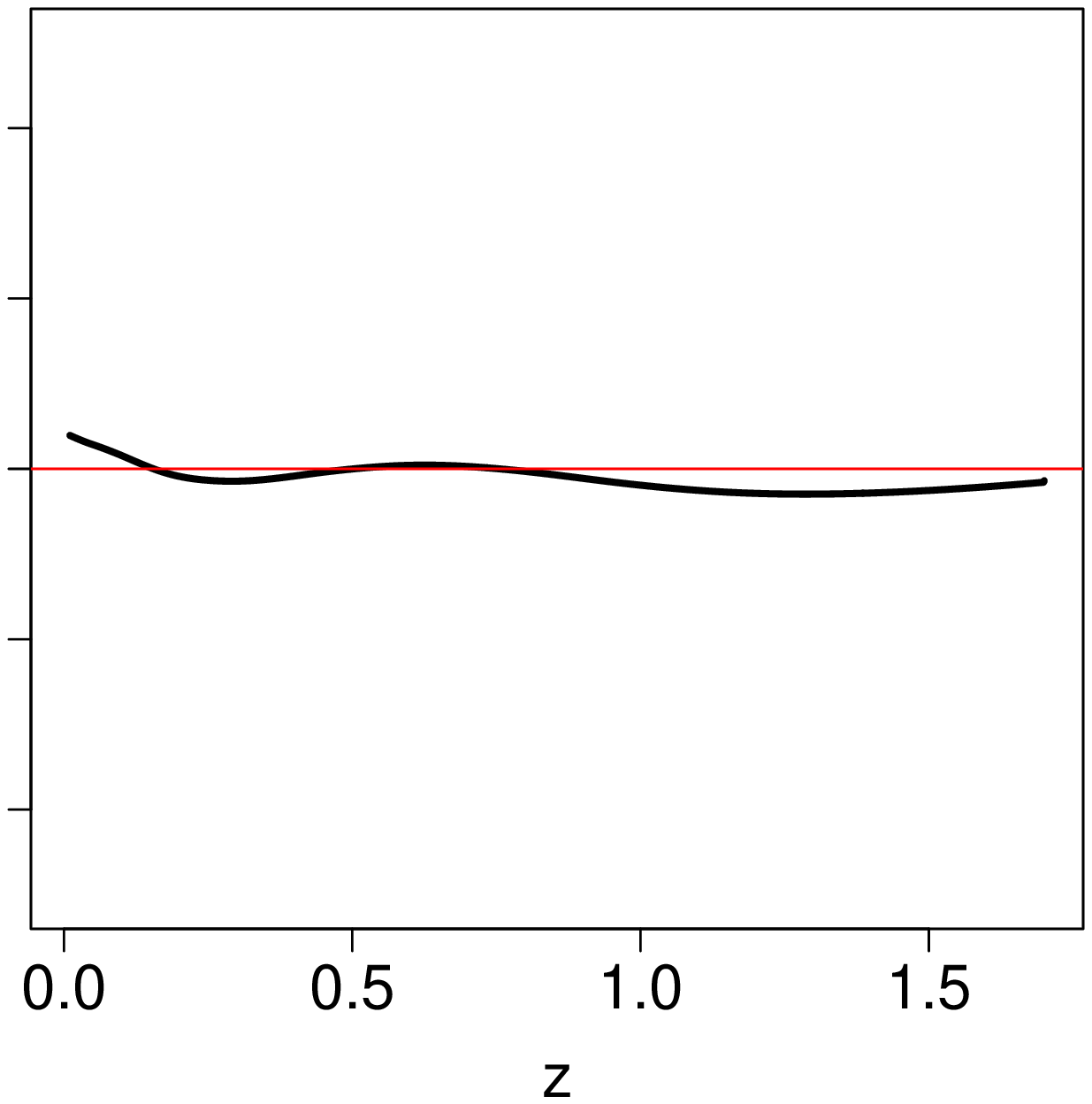}}
\caption{\label{residuals} Residuals for $\mu$ considering the predictions for $w$ for dataset 3 with $\Omega_m$ and $\Delta_{\mu}$ free. The left plot shows the result for the $w_0-w_a$ parametrization, the right plot shows the results for the GP approach.}  
\end{figure}

Table~\ref{table:mse} shows a simple measure of how well the exact
functional form of $w(z)$ has been captured by the three different
approaches. We calculate the mean square error of the reconstructed
history for $w(z)$ with respect to the perfect input $w(z)$ as shown
in the lower panels of Fig.~\ref{data} -- the smaller the error the
better the reconstruction result. This simple test has two minor
shortcomings -- first, it does not account for the realization noise
in the history of $w(z)$ underlying the simulated data so the error
will never be zero. In order to obtain the true $w(z)$ we would have
to perform two derivatives of the noisy simulated data which would
render this test basically meaningless. Second, the error bands of the
predictions are not taken into account. Nevertheless, this comparison
should provide some information about how well the different methods
perform compared to each other.

For the first dataset $w=const$. the parametric reconstruction ansatz
$w=w_0$ provides -- not surprisingly -- the best results; the history
for $w(z)$ is captured extremely well as is $\Omega_m$ with small
errors. For dataset 1, the GP model provides a more accurate answer
than the $w_0-w_a$ parameterization in the case of $\Omega_m$ free.
This is mainly due to the fact that once the parameterized form has
picked up some curvature in $w(z)$, the reconstructed $w(z)$ will
depart more and more from $w=const$ at higher $z$. The GP model
however is flexible enough to avoid such a behavior and stays close to
$w=-1$ over the whole redshift range. For dataset 2, the GP model
performs slightly better than the two parametric reconstruction
approaches for similar reasons as for dataset 1. The $w_0-w_a$
parametrization picks up some time-dependence in the low-$z$ regime
which overestimates the curvature of $w(z)$ at higher $z$ while the GP
approach reconstructs $w(z)$ reasonably well over the whole redshift
range and therefore has a smaller mean square error. For the third
dataset, the mean square error for the GP model is smallest in the
case of $\Omega_m$ fixed but for $\Omega_m$ free kept it is worse than
the result from the $w_0-w_a$ parametrization. For this last case
(dataset 3 and $\Omega_m$ and $\Delta_{\mu}$ free) we employ another
assessment of the accuracy of the prediction which highlights the
well-known degeneracy between $w$ and $\Omega_m$. We only consider the
$w_0-w_a$ parametrization and the GP model approach for this test,
since the $w=const.$ parametrization has obvious shortcomings in this
case.

For each case we fit the $w(z)$ result and then we find the associated
fit for $\mu$. Then we determine the difference between the predicted
$\mu$ and the input $\mu$ for our simulated data. We show the
residuals in Figure~\ref{residuals}. The left figure shows the
residuals for the $w_0-w_a$ parametrization, the right figure for the
GP reconstruction. The solution for $w(z)$ found with the GP model is
clearly a good fit to the data -- it performs slightly better than the
parametrized form in the low redshift range. Due to the the degeneracy
between $w$ and $\Omega_m$ the overall reconstruction of $w(z)$ is on
the other hand worse for the GP model -- this result will improve with
tighter constraints on $\Omega_m$ and complementary datasets such as
BAO measurements which help to break the degeneracy.

\section{Conclusions}
\label{sec:concl}

Characterizing the behavior of the dark energy equation of state is a
first step in understanding the nature and origin of dark energy.
Although a simple cosmological constant model is consistent with
current observations, the implied numerical value has no theoretical
explanation. Alternative dynamical models of dark energy generically
predict time variations in $w$ and a robust detection of such a time
dependence is one of the first targets in dark energy studies.
Supernova measurements remain a very promising probe of $w(z)$ and
future sky surveys can in principle measure $w(z)$ with high accuracy.

In order to fully exploit the power of future measurements, a reliable
and robust reconstruction method is required. In this paper we have
introduced a new reconstruction approach based on GP modeling. The
approach is nonparametric with modeling hyperparameters constrained
directly from the data. We have demonstrated that we can extract
nontrivial behavior of $w$ as a function of redshift with data of the
quality expected from future surveys. We have contrasted our new
method against two approaches, an assumed cosmological constant, and
one with a simple two-parameter model of the variation of $w(z)$. Both
of these models are effective descriptions for only a limited class of
possible behaviors of $w(z)$. In contrast, the generality of the GP
approach results in accurate reconstruction of potentially complex
variability in $w(z)$.

The GP model approach makes only mild smoothness assumptions about
$w(z)$ which are reasonable if we expect that the accelerated
expansion of the Universe is due to a physically well motivated
reason. The major ingredient for the GP model is specified by the
covariance function $K(z,z')$. While the choice of the specific form
for $K(z,z')$ is up to the modeler, the GP approach is rather robust
to this choice and the major hyperparameters influencing $K(z,z')$ are
informed by the data themselves. In addition to choosing a covariance
function, we have to specify a set of priors for cosmological and
model parameters. These priors have to be broad enough to include the
truth but should not be so broad that the Bayesian approach does not
converge. Both model and cosmological parameters are then jointly
determined from the data. While the Bayesian approach is
computationaly rather intensive, it has the great advantage that it
provides robust error bands. The approach outlined here for the
analysis of supernova measurements can easily be extended to include
different cosmological probes such as data from CMB and BAO
observations; work in this direction is currently in progress.
Moreover, the GP-based MCMC procedure can be integrated within
supernova analysis frameworks, e.g., SNANA~\cite{snana} as a cosmology
fitter, following the general methodology presented in
Ref.~\cite{habib07}.

\begin{widetext}

\appendix

\section{GP model algorithm}
\label{appa}

In this appendix we provide implementation details of the GP algorithm
used to reconstruct $w(z)$. The GP model approach requires the
estimation of several variables: the correlation hyperparameters
($\rho$ and $\kappa^2$), the Gaussian process points [$w(u)$ with
$u=(u_1,...,u_m)$, or rather $y(s)$ with $s=(s_1,...s_{m\cdot h})$,
the variance parameter ($\sigma^2$), along with the physical
parameters of interest ($\Omega_m$ and $\Delta_\mu$). $\Omega_m$ and
$\Delta_\mu$ can be added as extra steps; for simplicity we do not
include them in the discussion here (we did include them in our
analysis presented in the main body of the paper.)

A Gaussian process is defined by its mean and correlation function. We
set the prior mean of the Gaussian process to $-1$ for stability;
other values are found to be equally acceptable when the true mean of
$w(z)$ is near $-1$. Even though the mean is fixed, the posterior mean
will not be exactly $-1$ but will have a distribution spread around
$-1$. In the case when the true mean is not $-1$ the posterior value
of the mean of $w(z)$ will indicate this. Preliminary exploration of
the posterior of $w(z)$ can then be used to set the prior mean for
subsequent runs, which provides more stable results.

In the correlation function, the parameter $\alpha$ should be set as a
constant beforehand in the range of 1 to 1.9999, setting it exactly
equal to 2 can cause numerical instability in the covariance matrix,
and values above 2 are not mathematically valid values for the
process. The parameter $\alpha$ controls the smoothness of the overall
GP: $\alpha\leq 2$ will produce a flexible continuous GP but it will
not be differentiable anywhere, while $\alpha=2$ produces a much
smoother continuous GP that is infinitely differentiable everywhere.
In order to allow for maximum flexibility we choose $\alpha=1$
throughout the paper. The correlation length $\rho$ is a free
parameter in the GP model and its value is informed by the data. It is
highly correlated to $\alpha$ and $\kappa^2$ which makes the
interpretation of its final value not straightforward. It is strictly
limited to the region of [0,1) and the GP is not defined for the
limiting case of $\rho=1$. After investigation we found that $\rho$
has a much smaller role in the determination of the nature of the GP
in comparison to $\alpha$ and $\kappa^2$.

Integration of the Gaussian process requires a grid for numerical
integration. Let $n$ be the number of supernova data points in our
dataset, and $m$ be a finite number of Gaussian process points over
this region for evaluation. During the integration we add $h-1$
partition points between each GP point. Our resulting integrated
process $y(s)$ has $m\cdot h$ points. This provides a dense enough
grid to carry out an accurate numerical integration for the outer
integral without slowing down the computations. We find that an $m$
around 50 to 100 is sufficient, and an $h$ between 3 and 5 is a good
balance between accuracy and speed.

As with the parametric models, we employ Bayesian methods where we use
our priors and likelihood to obtain a posterior that can then be
sampled with an MCMC algorithm.  In our case, with our given
likelihood function (given in Eqn.~(\ref{L})), this leads to the
following posterior for the parameters of interest:
\begin{equation}\label{A1}
\sigma^2, \rho, \kappa^2 | \mu, \tau^2, z \propto L(z, \mu,
\tau|y(v),\sigma^2)MVN(y(v)|\rho,\kappa^2)\pi(\rho)\pi(\kappa^2)\pi(\sigma^2).
\end{equation}
$y(v)$ here denotes an arbitrary GP with parameters $\kappa^2$ and
$\rho$.  $\sigma^2$ is the unknown variance parameter from the
likelihood equation and $z$, $\mu$, and $\tau$ are ``observed'' values
from the simulated dataset.

Instead of the usual Gaussian process formulation shown in Eqn.~(\ref{A1}),
we choose an altered form to allow for slower changes in the GP and a
more localized search.  We let $y(v)\sim MVN(-1,\Sigma)$ and
$\Sigma^{-1/2}(y(v)-(-1)) = y^o(v) \sim MVN(0,I)$.  This leads to the
following posterior:
\begin{equation}
L(z, \mu, \tau|y^o(v),\rho,\kappa^2,\sigma^2)MVN(y^o(v);0,I)
\pi(\rho)\pi(\kappa^2)\pi(\sigma^2).
\end{equation}
In this setup, we propose a GP $y^o(v)$ and transform it to $y(v)$.
We then keep track of $y^o(v)$ and make our next proposal based on
these values and not on the $y(v)$.  Thus we are allowing the proposal
to make finer changes each time to boost the acceptance rate, which
tends to be problematic if we were to propose $y(v)$ directly.

The explicit procedure for estimating the process parameters
is as follows and employs standard MCMC techniques.
\begin{enumerate}
\item Initialize all variables: $\rho = \rho_1$,
  $\kappa^2=\kappa^2_1$, and $w^o(u) = w^o_{m,1}(u)$.  $w(u)$ will be
  a vector with $m$ points in our GP and $y(s)$ has $m\cdot h$ points.
  We run this algorithm $q=1,...,Q$ times. Set all tuning parameters,
  $\delta_{1,2,3}$, which need to be tuned until good mixing occurs.

\item Propose $\rho^* = U(\rho_{q-1}-\delta_1,\rho_{q-1}+\delta_1)$
  \begin{enumerate}
  \item Compute the covariance matrix $K_{22*}=
  \rho^{*|u_j-u_i|^{\alpha}}$
  \item Compute the Cholesky decomposition
  for $K_{22*}=U_{*}'U_{*}$
  \item Compute the special
  $K_{12*}= \int_0^{s'}du \rho^{*|{u-s}|^\alpha}/(1+u)$ with
  Chebyshev-Gauss quadrature.
  \item We want $y_{\rho^*}(s)=-\ln(1+s)+[\kappa_{q-1}^2
  K_{12*}][\kappa_{q-1}^2 K_{22*}^{-1}](w_{\rho^*}(u)-(-1))$ where
  $w_{\rho^*}(u)= [\kappa_{q-1} U_{*}']w^o_{m,q-1} +(-1)$ \\
               \begin{align*}
               y_{\rho^*}(s)&=
-\ln(1+s)+[\kappa_{q-1}^2K_{12*}][\kappa_{q-1}^2K_{22*}]^{-1}(\left(\kappa_{q-1}U_{\rho^*}'w^o_{m,q-1}
+(-1)\right)-(-1))\\
              &=-\ln(1+s)+\kappa_{q-1}K_{12*}[(U_{\rho^*}'U_{\rho^*})^{-1}U_{\rho^*}']
w^o_{m,q-1}\\
             &=-\ln(1+s)+\kappa_{q-1}K_{12*}[U_{\rho^*}^{-1}]w^o_{m,q-1}
             \end{align*}
        \item  $L(z, \mu,\tau|y_{\rho^*},\sigma^2_{q-1})
=\exp\left(-\frac{1}{2}\sum{\left(\frac{\mu_i-T(z_i,y_{\rho^*}(u))}{\tau_i\sigma_i}\right)^2}\right)$
 where  the definite integrations in $T(z_i,y_{\rho^*}(u))$ are done
numerically through summations of the trapezoid algorithm.
        \item Accept the proposal $\rho^*$ with probability
$\frac{L_{\rho^*}\pi(\rho^*)}{L_{\rho_{q-1}}\pi(\rho_{q-1})}$
let $\rho_q=\rho^*$, otherwise $\rho_q=\rho_{q-1}$.
   \end{enumerate}

\item Draw $\kappa^{2*} = U(\kappa^2_{q-1}-\delta_2,\kappa^2_{q-1}+\delta_2)$
   \begin{enumerate}
   \item  Compute $y_{\kappa^{2*}}(s)=(-1) \ln(1+s)+\kappa^*
K_{12q}[U_{q-1}^{-1}]w^o_{m,q-1}$\\
   \item $L(z, \mu,\tau|y_{\kappa^{2*}},\sigma^2_{q-1})
     =\exp\left({-\frac{1}{2}\sum{\left(\frac{\mu_i-T(z_i,y_{\kappa^{2*}}(u)
           )}{\tau_i\sigma_i}\right)^2}}\right)$ where the definite
     integrations in $T(z_i,y_{\kappa^{2*}}(u))$ are done numerically
     through summations of the trapezoid algorithm.
   \item Accept with probability
     $\frac{L_{\kappa^{2*}}\pi(\kappa^{2*})}{L_{\kappa^2_{q-1}}\pi(\kappa^2_{q-1})}$.
   \end{enumerate}

\item We propose a non-standard $w_m^*$ for the GP.  We start by
  drawing a proposal for $w^{o*} \sim MVN(w^o_{q-1}, \delta_3
  I_{mxm})$, where $I_{mxm}$ is the identity matrix.
   \begin{enumerate}
    \item  Compute $y^*(s)=(-1) \ln(1+s)+\kappa_{q}
K_{12q}[U_{q}^{-1}]w^{o*}_{m,}$\\
   \item $L_{z,
\mu,\tau|y^*(s),\sigma^2_{q-1}}=e^{-\frac{1}{2}\sum{\frac{\mu_i-T(z_i,y^*(s))}{\tau_i\sigma}^2}}$
   \item Accept with probability $\frac{L_{y^*}MVN(y^{*}|0,I)
}{L_{y_{q-1}}MVN(y_{q-1}|0,I)}$ letting $y_{q}(s)=y^{*}(s)$ and the
corresponding GP realization is $w_{m,q}(u)=w^*_m(u)$
   \end{enumerate}

\item  $\sigma^2_{q}|... \sim
IG\left(\frac{n}{2}+10,\frac{1}{2}\sum{\left(\frac{\mu-T(z|...)}{\tau}
\right)}^2+9
\right)$\\

\item Repeat steps 2-6, $Q$ times and rerun the entire algorithm as
  needed after resetting the tuning parameters
\end{enumerate}

\end{widetext}

\begin{acknowledgments}

  The authors acknowledge support from the LANL Institute for Scalable
  Scientific Data Management. Part of this research was supported by
  the DOE under contract W-7405-ENG-36. UA, SH, KH, and DH acknowledge
  support from the LDRD program at Los Alamos National Laboratory. KH
  was supported in part by NASA. SH and KH acknowledge the hospitality
  of the Aspen Center for Physics, where part of this work was carried
  out. We are indebted to Andreas Albrecht, Eric Linder, Adrian Pope, 
  Martin White, and Michael Wood-Vasey for several useful discussions.

\end{acknowledgments}


\begin{thebibliography}{99}

\bibitem{perlmutter}
  S.~Perlmutter {\it et al.}  [Supernova Cosmology Project Collaboration],
  Astrophys.\ J.\  {\bf 517}, 565 (1999)
  [arXiv:astro-ph/9812133].
  
\bibitem{riess}
  A.G.~Riess {\it et al.}  [Supernova Search Team Collaboration],
  Astron.\ J.\  {\bf 116}, 1009 (1998)
  [arXiv:astro-ph/9805201].

\bibitem{boss}
D.~Schlegel, M.~White, and D.~Eisenstein, arXiv:0902.4680
[astro-ph.CO] 

\bibitem{des}
https://www.darkenergysurvey.org/ 

\bibitem{jdem}
http://jdem.gsfc.nasa.gov/

\bibitem{lsst}
http://www.lsst.org/lsst

\bibitem{frieman08}
 J.~Frieman, M.~Turner and D.~Huterer,
 Ann.\ Rev.\ Astron.\ Astrophys.\  {\bf 46}, 385 (2008)
 [arXiv:0803.0982 [astro-ph]].

\bibitem{turnwhite97}
  M.S.~Turner and M.J.~White,
  Phys.\ Rev.\  D {\bf 56}, 4439 (1997)
  [arXiv:astro-ph/9701138].

\bibitem{quint}C.~Wetterich,
  Nucl.\ Phys.\ B {\bf 302}, 668 (1988); B.~Ratra and
  P.J.E.~Peebles, Phys.\ Rev.\ D {\bf 37}, 3406 (1988);
  P.J~E.~Peebles and B.~Ratra, Astrophys.\ J.\ {\bf 325}, L17
  (1988); R.~R.~Caldwell, R.~Dave and P.~J.~Steinhardt, Phys.\ Rev.\
  Lett.\ {\bf 80}, 1582 (1998) .

\bibitem{hicken}
  M.~Hicken {\it et al.},
  Astrophys.\ J.\ {\bf 700}, 1097 (2009) [arXiv:0901.4804
  [astro-ph.CO]].

\bibitem{amanullah}R.~Amanullah {\it et al.},
  Astrophys.\ J.\  {\bf 716}, 712 (2010)
  [arXiv:1004.1711 [astro-ph.CO]].


\bibitem{huterer98} 
  D. Huterer and M.S. Turner, Phys. Rev. D {\bf 60},
  081301 (1999)

\bibitem{starobinsky98}
  A.A.~Starobinsky,
    JETP Lett.\  {\bf 68}, 757 (1998)
  [Pisma Zh.\ Eksp.\ Teor.\ Fiz.\  {\bf 68}, 721 (1998)]
  [arXiv:astro-ph/9810431].

\bibitem{wang03}
  Y.~Wang and P.~Mukherjee, Astrophys.\ J.\ {\bf 606}, 654 (2004)
  [arXiv:astro-ph/0312192].

\bibitem{daly03} 
  R.A.~Daly and S.G.~Djorgovski,
  Astrophys.\ J.\ {\bf 597}, 9 (2003) [arXiv:astro-ph/0305197].

\bibitem{huterer04} 
  D.~Huterer and A.~Cooray, Phys.\ Rev.\ D {\bf 71},
  023506 (2005) [arXiv:astro-ph/0404062].

\bibitem{gerke04}
  B.F.~Gerke and G.~Efstathiou,
  Mon.\ Not.\ Roy.\ Astron.\ Soc.\  {\bf 335}, 33 (2002)
  [arXiv:astro-ph/0201336].

\bibitem{shaf06}
  A.~Shafieloo, U.~Alam, V.~Sahni and A.A.~Starobinsky,
  Mon.\ Not.\ Roy.\ Astron.\ Soc.\ {\bf 366}, 1081 (2006).

\bibitem{zunckel}
  C.~Zunckel and R.~Trotta, Mon.\ Not.\ Roy.\ Astron.\ Soc.\ {\bf
    380}, 865 (2007) [arXiv:astro-ph/0702695].

\bibitem{hojjati09} 
  A.~Hojjati, L.~Pogosian, and G.-B. Zhao,
  arXiv:0912.4843 [astro-ph.CO].

\bibitem{sahni06}  
  V.~Sahni and A.A.~Starobinsky,
  Int.\ J.\ Mod.\ Phys.\  D {\bf 15}, 2105 (2006)
  [arXiv:astro-ph/0610026].

\bibitem{coorhut99}
  A.R.~Cooray and D.~Huterer, Astrophys.\ J.\ {\bf 513}, L95 (1999)
  [arXiv:astro-ph/9901097].

\bibitem{maor}
  I.~Maor, R.~Brustein and P.J.~Steinhardt,
  Phys.\ Rev.\ Lett.\ {\bf 86}, 6 (2001) [Erratum-ibid.\ {\bf 87},
  049901 (2001)] [arXiv:astro-ph/0007297].

\bibitem{weller}
  J.~Weller and A.J.~Albrecht, Phys.\ Rev.\ Lett.\ {\bf 86}, 1939
  (2001) [arXiv:astro-ph/0008314].
  
\bibitem{chev01}
  M.~Chevallier and D.~Polarski,  Int.\ J.\ Mod.\ Phys.\  D {\bf 10}, 213 (2001)
  [arXiv:gr-qc/0009008].
  
\bibitem{linder03}
  E.V.~Linder,  Phys.\ Rev.\ Lett.\  {\bf 90}, 091301 (2003)
  [arXiv:astro-ph/0208512].

\bibitem{simpson06} F.~Simpson and S.L.~Bridle, Phys. Rev. D {\bf 73},
 083001 (2006), arXiv:astro-ph/0602213. 

\bibitem{genovese}
  C.~Genovese, P.~Freeman, L.~Wasserman, R.~Nichol, and C.~Miller,
  Ann. Appl. Stat. {\bf 3}, 144 (2009) [arXiv:0805.4136 [astro-ph]].

\bibitem{mortonson}  
  M.J.~Mortonson, W.~Hu and D.~Huterer,
  Phys.\ Rev.\  D {\bf 79}, 023004 (2009)
  [arXiv:0810.1744 [astro-ph]].
  
\bibitem{holsclaw}
 T.~Holsclaw, {\em et al.} (in preparation).

\bibitem{Banerjee} 
 S.~Banerjee, B.P.~Carlin, and A.E.~Gelfand, {\em Hierarchical
   Modeling and Analysis for Spatial Data}, New York: Chapman and Hall
 (2004). 
 
 \bibitem{gprefs} 
 C.E.~Rasmussen and K.I.~Williams, {\em Gaussian Processes
   for Machine Learning} (MIT Press, 2006);
 http://www.gaussianprocess.org/gpml/   
  
 \bibitem{heitmann06}
   K.~Heitmann, D.~Higdon, C.~Nakhleh and S.~Habib,
   Astrophys.\ J.\  {\bf 646}, L1 (2006) [arXiv:astro-ph/0606154].

\bibitem{habib07}
  S.~Habib, K.~Heitmann, D.~Higdon, C.~Nakhleh and B.~Williams,
  Phys.\ Rev.\  D {\bf 76}, 083503 (2007)
  [arXiv:astro-ph/0702348].

\bibitem{heitmann09}  
  K.~Heitmann, D.~Higdon, M.~White, S.~Habib, B.J.~Williams and
  C.~Wagner, Astrophys.\ J.\  {\bf 705}, 156 (2009)
  [arXiv:0902.0429 [astro-ph.CO]].
  
\bibitem{lawrence09}  
  E.~Lawrence, K.~Heitmann, M.~White, D.~Higdon, C.~Wagner, S.~Habib
  and B.J.~Williams, Astrophys. J. {\bf 713}, 1322 (2010)
  [arXiv:0912.4490 [astro-ph.CO]].
  
\bibitem{brewer09}
 B.J.~Brewer and D.~Stello, Mon.\ Not.\ Roy.\ Astron.\ Soc.\  {\bf
   395}, 226 (2009) [arXiv:0902.3907].  

\bibitem{way09}
  M.J.~Way, L.V.~Foster, P.R.~Gazis and A.N.~Srivastava, Astrophys.\
  J.\ {\bf 706}, 623 (2009) [arXiv:0905.4081
  [astro-ph.IM]].

\bibitem{bonfield09}
  D.G.~Bonfield, Y.~Sun, N.~Davey, M.J.~Jarvis, F.B.~Abdalla, M.~Banerji and R.G.~Adams,
  Mon.\ Not.\ Roy.\ Astron.\ Soc.\  {\bf
   405}, 987 (2010) 
  [arXiv:0910.4393 [astro-ph.IM]].

\bibitem{alam07}
  U.~Alam, V.~Sahni and A.A.~Starobinsky, JCAP {\bf 0702}, 011 (2007)
  [arXiv:astro-ph/0612381].

\bibitem{flat}
   E.~Komatsu {\em et al.}, Astrophys.~J.~Supp. {\bf 180}, 330 (2009)
  [arXiv:0803.0547 [astro-ph]].
    
\bibitem{sdss} 
  R.~Kessler {\it et al}, Astrophys.\ J.\ Suppl.\ {\bf 185}, 32 (2009)
  [arXiv:0908.4274 [astro-ph.CO]].

\bibitem{lh03} 
  E.V.~Linder and D.~Huterer, Phys. Rev. D {\bf 67},
  081303 (2003).
  
\bibitem{snap}
  G.~Aldering et al. [SNAP Collaboration], arXiv:astro-ph/0405232.
 
\bibitem{coras}
  P.S.~Corasaniti, B.A.~Bassett, C.~Ungarelli and E.J.~Copeland,
  Phys.\ Rev.\ Lett.\ {\bf 90}, 091303 (2003)
  [arXiv:astro-ph/0210209].
 
\bibitem{weller2}J.~Weller and A.~Albrecht,
  Phys.\ Rev.\ D {\bf 65}, 103512 (2002) [arXiv:astro-ph/0106079].

\bibitem{huterer03} 
  D.~Huterer and G.~Starkman, Phys.\ Rev.\ Lett.\ {\bf 90}, 031301
  (2003) [arXiv:astro-ph/0207517].

\bibitem{serra09}
  P.~Serra, A.~Cooray, D.~E.~Holz, A.~Melchiorri, S.~Pandolfi and
  D.~Sarkar, Phys.\ Rev.\ D {\bf 80}, 121302 (2009)
  [arXiv:0908.3186 [astro-ph.CO]].


\bibitem{fomswg}
  A.J.~Albrecht {\it et al.},
  arXiv:0901.0721 [astro-ph.IM].

\bibitem{gelman}
  A.~Gelman, B.~Carlin, H.~Stern, and D.~Rubin, {\it Bayesian Data
    Analysis}, New York: Chapman and Hall (2004).

\bibitem{gamerman}
  D.~Gamerman and H.F.~Lopes, {\it Markov Chain Monte Carlo:
    Stochastic Simulation for Bayesian Inference}, New York: Chapman
  and Hall (2006).

\bibitem{komatsu}
  E.~Komatsu {\it et al.}, arXiv:1001.4538
  [astro-ph.CO].

\bibitem{snana} 
  R.~Kessler {\em et al.} Publ. Astron. Soc. Pac. {\bf
    121}, 1028 (2009), arXiv:0908.4280 [astro-ph.CO].

\end{thebibliography}
\end{document}